\documentclass{a1}
\title{Glueballs in large-$N$ $YM$ by localization on critical points}

\ShortTitle{Glueballs in large-$N$ $YM$ by localization on critical points}

\author{\speaker{Marco Bochicchio} \\INFN-Roma1 and SNS-Pisa \\
Dipartimento di Fisica, Universita' di Roma `Sapienza' \\
Piazzale Aldo Moro 2 , 00185 Roma  \\
       E-mail: \email{marco.bochicchio@roma1.infn.it}}

\abstract{ By exploiting in large-$N$ $YM$ the change of variables from the gauge connection to the $ASD$ part of its curvature by a non-$SUSY$ version of the
Nicolai map, we show that certain twistor Wilson loops supported on a Lagrangian submanifold of twistor space are localized on lattices of surface operators of $Z_N$ holonomy that form translational invariant sectors labelled by the magnetic charge $k=1,2,...,N-1$ at a point.
The localization is obtained reducing the loop equation in the $ASD$ variables in the holomorphic gauge, regularized by analytic continuation to Minkowski space-time, to a critical equation, by exploiting the invariance of the 
v.e.v. of twistor Wilson loops by deformations for the addition of backtracking arcs ending with cusps on the singular divisor of surface operators. Alternatively  the localization is obtained contracting the $YM$ measure in the $ASD$ variables on the fixed points of a semigroup that acts on the fiber of the Lagrangian twistor fibration which twistor Wilson loops are supported on and leaves invariant their v.e.v..
The renormalized effective action induced by the localized $YM$ measure in the $ASD$ variables 
scales according to a large-$N$ beta function of $NSVZ$ type that reproduces the first two universal perturbative coefficients. Because of a non-trivial Jacobian due to the lack of supersymmetry a 
multiplicative renormalization by a $Z$ factor of the $ASD$ field occurs. 
The masses squared of the fluctuations of surface operators in the sectors labelled by $k$, supported on the Lagrangian submanifold analytically continued to Minkowski space-time, form a trajectory linear in $k$ that does not include any massless state. The glueball propagators in the holomorphic/antiholomorphic sector defined by correlators of a complex combination of the $ASD$
curvature and its adjoint saturate at short distances the logarithms of perturbation theory by a sum of pure poles. The anomalous dimensions of long gauge invariant operators belonging to the holomorphic/antiholomorphic sector that are implied by the $Z$ factor coincide with the anomalous dimensions of the scalar operators that occur as the antiferromagnetic ground state of the Hamiltonian spin chain
in the thermodynamic limit, that it is known to provide the anomalous dimensions in the $ASD$ one-loop integrable sector of large-$N$ $YM$. In this framework Regge trajectories of higher spins are related to
fluctuations of surface operators with pole singularities of any order.}

\FullConference{GGI conference "Large-$N$ Gauge Theories"\\
April 6-9 (2011) Florence, Italy}
\begin{document}

\def\beq{\begin{equation}}
\def\eeq{\end{equation}}
\def\bea{\begin{eqnarray}}
\def\eea{\end{eqnarray}}
\def\bq{\begin{quote}}
\def\eq{\end{quote}}

{\bf Contents} \par
{\bf 1. Introduction} \par
\emph{1.1 A purely field theoretical presentation: summing planar diagrams} \par
\emph{ by combining  the Nicolai map with the holomorphic gauge} \par
\emph{1.2  Summary of results} \par
\emph {1.3 A geometric point of view: localization by cohomology and by homology} \par
{\bf 2. Synopsis} \par
{\bf 3. Prologue} \par
\emph{3.1 One-loop beta function of $YM$ by the background field method} \par
\emph{3.2 $NSVZ$ beta function of $\mathcal{N}$ $=1$ $SUSY$ $YM$ by the Nicolai map} \par
\emph{3.3 Non-$SUSY$ Nicolai map from the connection to the $ASD$ curvature in pure $YM$} \par
\emph{3.4 One-loop beta function of $YM$ in the $ASD$ variables by the background field method} \par
{\bf 4. Twistor Wilson loops and non-commutative gauge theories} \par
\emph{4.1 Star algebra / operator algebra correspondence}\par
\emph{Translations / unitary operators correspondence} \par
\emph{Eguchi-Kawai reduction by realization of translations as unitary operators}\par
\emph{and rescaling of the action} \par
\emph{4.2 Twistor Wilson loops }\par
\emph{4.3 Fiber independence of twistor Wilson loops }\par
\emph{4.4 Twistor Wilson loops are supported on a Lagrangian submanifold of twistor space} \par
\emph{4.5 Triviality of twistor Wilson loops in the limit of infinite non-commutativity} \par
{ \bf 5. Quasi-localization lemma} \par
\emph{Quasi-localization lemma for twistor Wilson loops under the action of a semigroup } \par
\emph{of contractions on the fiber, by the fiber independence and  by the contraction of the measure} \par
\emph{in the $ASD$ variables} \par
{ \bf 6. Holomorphic loop equation} \par
\emph{6.1 Holomorphic loop equation for twistor Wilson loops, by the non-$SUSY$ Nicolai map} \par
\emph{in the holomorphic gauge} \par
\emph{6.2 Gauge invariant regularization of the holomorphic loop equation by analytic} \par
\emph{continuation to Minkowski space-time} \par
{\bf 7. Integrating over surface operators} \par
\emph{7.1 Integrating over local systems, by interpreting the non-$SUSY$ Nicolai map as hyper-Kahler} \par
\emph{reduction on a dense set of "local systems in infinite dimension"} \par
\emph{7.2 Reducing to finite-dimension local systems with gauge group $U(N\times \hat N)$,} \par
\emph{i.e. to surface operators, by Morita duality for rational values of non-commutativity} \par
\emph{Recovering the large-$N$ limit by the inductive structure } \par
\emph{7.3 Hyper-Kahler and Lagrangian moduli space of surface operators} \par 
{\bf 8. Localization on fixed points} \par
\emph{Localization on a discrete sum of fixed points, by the quasi-localization} \par
\emph{lemma, by triviality of twistor Wilson loops, and by the local abelianization, } \par
\emph{up to zero modes, due to integrating over surface operators } \par
\emph{The fixed points are surface operators with $Z_N$ holonomy} \par
\emph{The fixed points live inside the closure of the Lagrangian cone} \par
\emph{of the hyper-Kahler moduli space} \par
{\bf 9. Wilsonian beta function by localization} \par
\emph{9.1 Wilsonian beta function by the renormalized effective action in the $ASD$ variables} \par
\emph{and by the localization on surface operators in a neighborhood of the fixed points}\par 
\emph{9.2 Lagrangian moduli of surface operators occur in a neighborhood of the fixed points }\par
\emph{and they determine the zero modes that contribute via Pauli-Villars to the beta function} \par 
{\bf 10. Localization by homology} \par
\emph{Localization by homology, by the holomorphic loop equation,} \par
\emph{by triviality of twistor Wilson loops and by homology theory of arcs ending} \par
\emph{with cusps on the singular divisor of surface operators} \par
{\bf 11. Canonical beta function} \par
\emph{11.1 Canonical beta function, by the localization by homology} \par 
\emph{and by the gluing rules that follow by the localization on local systems,} \par
\emph{i.e. on surface operators. They coincide with the gluing rules of topological strings } \par
\emph{11.2 The canonical normalization involves a multiplicative factor, $Z$, in addition to the }\par
\emph{rescaling by the factor of $g$} \par
{\bf 12. Glueball spectrum} \par
\emph{12.1 Glueball propagators and anomalous dimensions in perturbation theory at large-$N$} \par
\emph{12.2 Localized effective action and holomorphic/antiholomorphic fusion} \par
\emph{12.3 Glueball potential} \par
\emph{The hyper-Kahler reduction on the dense orbits in the neighborhood of the fixed} \par
\emph{points in the unitary gauge requires that the residues of the $ASD$ curvature} \par
\emph{live in an adjoint orbit by the action of the unitary group and that they commute at any } \par
\emph{lattice point. Therefore the $YM$ theory restricted to the lattice hyper-Kahler reduction} \par
\emph{is locally abelian, all the other non-abelian degrees of freedom being zero modes } \par
\emph{of the Jacobian of the non-$SUSY$ Nicolai map associated to the moduli of the local system.} \par
\emph{In the holomorphic gauge they live in an adjoint orbit by the action of the} \par
\emph{complexification of the unitary group.} \par
\emph{The glueball potential arises as the logarithm of the Jacobian from the unitary to the} \par
\emph{holomorphic gauge that is needed to write the holomorphic loop equation} \par
\emph{12.4 All lattices of translational invariant surface operators at the fixed points} \par 
\emph{have degenerate renormalized effective action at the leading large-$N$ limit} \par
\emph{12.5 A kinetic term arises for the fluctuations of Lagrangian supported } \par
\emph{surface operators in the effective action provided the theory is analytically} \par 
\emph{continued to Minkowski space-time } \par 
\emph{12.6 The mass gap arises by the second derivative of the glueball potential} \par
\emph{in the holomorphic/antiholomorphic sector} \par
\emph{12.7 The spectrum is a sum of pure poles by the vanishing as $N^{-\frac{1}{2}}$ of non-quadratic terms} \par 
\emph{in the effective action in the inductive sequence} \par
\emph{12.8 The second derivative of the glueball potential is supported on degenerate eigenvalues, }\par
\emph{confirming that gets contributions by configurations with unbroken gauge group} \par
\emph{12.9 Glueball propagators in the Wilsonian scheme} \par
\emph{12.10 Glueball propagators in the canonical scheme and anomalous dimensions} \par
{\bf 13. Regge trajectories} \par
\emph{Extension to functional integration on surface operators with wild singularities,} \par
\emph{by the same hyper-Kahler reduction due to the non-$SUSY$ Nicolai map extended}  \par
\emph{to twistor connections with wild singularities, i.e. pole singularities of any order}\par
{\bf 14. Conclusions} \par
{\bf 15. Acknowledgments}

\section{Introduction}

A technical presentation of the main ideas and results of this paper is in the synopsis and in a very sketchy way in the list of contents. \par
This section is more of introductory nature, in order to convey in a simpler way part of the meaning of the technical ideas. \par
The problem of the Yang-Mills ($YM$) mass gap as reported in \cite{AJ} has an infrared and an ultraviolet nature at the same time. \par
Indeed the renormalization group ($RG$) requires that every mass scale of the $YM$ theory must depend on the canonical coupling constant, $g_{YM}$,
only through the $RG$ invariant scale, $\Lambda_{YM}$:
\bea
\Lambda_{YM}= \Lambda \exp(-\frac{1}{2\beta_0 g_{YM}^2}) (\beta_0 g_{YM}^2)^{-\frac{\beta_1}{2 \beta_0^2}}(1+...)
\eea
whose dependence on the coupling constant is equivalent to the knowledge of the exact beta function of the theory in some scheme.
The dots refer to the higher loop contributions  irrelevant in the ultraviolet, while the non-analytic two-loop result is explicitly displayed. 
Eq.(1.1) in turn implies that an amazing asymptotic accuracy, as $g_{YM}$ vanishes \footnote{Therefore the mass gap problem is not a strong coupling problem.} when the cutoff, $\Lambda$, diverges, is needed to solve the mass gap problem
and that the mass gap is zero to every order of perturbation theory. \par
One possibility is that such finest asymptotic accuracy may be achieved only by an exact solution both on the ultraviolet side,
for the beta function, and on the infrared side, for the mass gap. \par
While an exact solution of the $YM$ theory, to use just an euphemism, seems completely outside the reach of the present techniques,  
in this paper we propose an exact solution in the large-$N$ limit of the $SU(N)$ $YM$ theory for the beta function and for the glueball spectrum restricted to a special sector of the theory.

\subsection{A purely field theoretical presentation: summing planar diagrams combining the Nicolai map with the holomorphic gauge}

It has been known for a long time that the large-$N$ limit of the pure $YM$ theory can be defined by the celebrated Makeenko-Migdal loop equation \cite{MM, MM1}, Eq.(10.8):
\bea
&&\int_{L_{xx}} dx_{\alpha}<\frac{N}{2 g^2} Tr(\frac{\delta S_{YM}}{\delta A_{\alpha}(x)}\Psi(x,x;A))> \nonumber \\
&&=i \int_{L_{xx}} dx_{\alpha} \int_{L_{xx}} dy_{\alpha} \delta^{(4)}(x-y) <Tr \Psi(x,y;A)>< Tr \Psi(y,x;A)> 
\eea
where:   
\bea
\Psi(x,y;A)=P \exp i\int_{L_{xy}} A_{\alpha} dx_{\alpha}
\eea
is the Wilson loop, i.e. the holonomy of the gauge connection, $A_{\alpha}$.
The left hand side of the Makeenko-Migdal  loop equation contains the critical equation for the $YM$ action and the right hand side contains the contribution
of the "change to loop variables" \footnote{There is a way of deriving the Makeenko-Migdal loop equation as a change of variables from the gauge connection to the trace of
its holonomy, i.e. the trace of a Wilson loop \cite{Mak1}.}. It has been known for long that the Makeenko-Migdal loop equation is a compact way of "summing the planar diagrams" \cite{Hooft}. \par
However, the solution of the Makeenko-Migdal loop equation lives in a von Neumann algebra that is not hyperfinite, i.e. that is not the weak limit of a sequence of matrix algebras. Such non-hyperfinite algebra is too large a mathematical object to handle, sect.(1.3). \par
Nevertheless, the algebra of local single trace operators at next to leading $\frac{1}{N}$ order is, in a sense, the most simple as possible. Correlators of such operators, $O(x)$,
are conjectured to be an infinite sum of free fields, saturating the logarithms of perturbation theory, sect.(12.1):
\bea
G_O(p^2)&=&\int e^{ipx} < O(x) O(0)>_{conn} d^4x \nonumber \\
&=&\sum_{k} \frac{Z_{ O k}}{p^2+m_k^2}
\sim Z^2_O(p^2) p^{2L-4} \log(\frac{p^2}{\mu^2})
\eea
The basic philosophy of this paper is to try to disentangle the more limited, but very interesting information on the spectral side, contained in the free propagators at next to leading
order, from the overwhelming information, but less relevant from a spectral point of view, contained in the algebra of non-local Wilson loops at the leading $\frac{1}{N}$ order. \par
This disentanglement is obtained constructing new kinds of Wilson loops, called twistor Wilson loops for geometrical reasons, built by a connection $B_{\lambda}$, defined in Eq.(4.28):
\bea
\Psi( B_{\lambda};L_{ww})= P \exp i \int_{L_{ww}}( A_z+\lambda D_u) dz+( A_{\bar z}+ \lambda^{-1}D_{\bar u}) d \bar z 
\eea
with the property that their v.e.v. is trivially $1$ in the large-$N$ limit, but non-trivial at next to leading order. \par
The twistor Wilson loops are defined initially in $U(N)$ $YM$ theory on Euclidean space-time, $R^2 \times R^2_{\theta}$, where the first factor is the ordinary Euclidean plane
and the second factor is a non-commutative plane with non-commutative parameter, $\theta$. 
The limit $ \theta \rightarrow \infty $ is the most relevant for us, since this limit is equivalent to the large-$N$ limit of the $SU(N)$ $YM$ theory on commutative space-time, sect.(4). \par
The aim of this paper is to solve exactly a new loop equation in large-$N$ $YM$, that involves the new twistor Wilson loops.
The basic strategy is to solve the new loop equation for twistor Wilson loops by a change of variables \cite{MB1}. We set for simplicity $B=B_1$ in the following. \par
In fact despite the twistor Wilson loops satisfy a Makeenko-Migdal loop equation of standard type, Eq.(10.13): 
\bea
&&\int_{L_{ww}} d\bar z<\frac{N}{2 g^2}Tr(\frac{\delta S_{NC}}{\delta B_z(z,\bar z)}\Psi(z,z;B))>= \nonumber \\
&&i\ \int_{L_{ww}} d\bar z\int_{L_{ww}} dz \delta^{(2)}(z-w) <Tr \Psi(w,z;B)> <Tr \Psi(z,w;B)> \nonumber \\
\eea
the Makeenko-Migdal loop equation is as difficult to solve for twistor Wilson loops, $\Psi(x,x;B)$, as it is for ordinary Wilson loops, $\Psi(x,x;A)$. \par
However, twistor Wilson loops satisfy a new loop equation, called holomorphic loop equation, obtained by a change of variables, actually two changes of variable. 
It is this holomorphic loop equation that is possible to solve exactly \cite{MB1}. \par
The first change of variables, worked out in sect.(3.3), is a non-supersymmetric ($SUSY$) version of the Nicolai map \cite{Nic1, Nic2}, previously known in $\mathcal{N}=1$ $SUSY$ $YM$ theory \cite{V,V1} sect.(3.2), from the gauge connection, $A_{\alpha}$, to the anti-selfdual ($ASD$) part of the curvature, $\mu^-_{\alpha \beta}$.
This change of variables is not one-to-one everywhere in function space even after gauge fixing.
We discuss later in this introduction how this technical difficulty is solved in this paper. \par
There is a further change of variables to a holomorphic gauge, from $\mu=\mu^-_{01}+i \mu^-_{03}$ to $\mu'$ \cite{MB1}, that has the scope of avoiding that the right hand side of the loop equation, i.e. the contribution of the  "change to loop variables"
be field dependent. Indeed this would make the reduction of the loop equation to a critical equation impossible. \par
The holomorphic gauge, sect.(6), is the gauge in which $B_{\bar z}= A_{\bar z}+D_{\bar u}=0$. This gauge can be reached by a gauge transformation in the complexification of the unitary gauge group.
Therefore strictly speaking the choice of a holomorphic gauge defines a change of variables and not a proper gauge transformation of the $YM$ theory. \par 
The result is the new holomorphic loop equation reported in sect.(6). It is obtained following the Makeenko-Migdal technique, as an identity that expresses the fact that the functional integral of a functional derivative vanishes:
\bea
\int  Tr\frac{\delta }{\delta \mu'(z,\bar z)} (e^{-\Gamma}\Psi(B'; L_{zz})) \delta \mu'=0
\eea
The new holomorphic loop equation for twistor loops follows:
\bea
<Tr(\frac{\delta \Gamma}{\delta \mu'(z,\bar z)}\Psi(B'; L_{zz}))>=
\frac{1}{ \pi } \int_{L_{zz}} \frac{ dw}{z-w} <Tr\Psi(B'; L_{zw})> <Tr\Psi(B'; L_{wz})> 
\eea
where $\Psi(B'; L_{zz})$ is the holonomy of $B$ in the gauge $B'_{\bar z}=0$.
The Cauchy kernel arises as the kernel of the operator $\bar \partial^{-1}$ that occurs by functionally differentiating $\Psi(B'; L_{zz})$. \\
Also the holomorphic loop equation contains in the right hand side the contribution of the "change to loop variables" as the Makeenko-Migdal loop equation does.
In fact the right hand side contains the contribution of a certain subclass of planar diagrams that contribute to the
twistor Wilson loops. The other planar diagrams being already included in the logarithm of the Jacobian of the change of variables that occurs in the "effective action", $\Gamma$,
in the left hand side. \par
However, a main point of this paper and of \cite{MB1} is that the right hand side of the holomorphic loop equation
vanishes provided the twistor Wilson loop has a backtracking cusp, i.e. cusp with $0$ cusp angle, $\Omega$, at the point
where the loop equation is evaluated. 
We follow the convention that a straight line has cusp angle $\pi$ while a backtracking cusp has cusp angle $0$. \par
The vanishing of the right hand side implies the localization of the holomorphic loop equation,
since for the cusped twistor Wilson loop the holomorphic loop equation reduces exactly to a critical equation for an effective action:
\bea
<Tr(\frac{\delta \Gamma}{\delta \mu'(z,\bar z)}\Psi(B'; L_{zz}))>= 0
\eea
We can choose the cusped loop with any number of backtracking cusps. In particular we can choose the standard Makeenko-Migdal loop with the shape of the symbol $\infty$, but with a double backtracking cusp
at the non-trivial self-intersection point and the equation of motion for the effective action inserted precisely at the cusp, sect.(6) and sect.(10). \par
We explain in this introduction why the localization of the holomorphic loop equation occurs. \par
The twistor Wilson loops have the following fundamental properties at large $N$. \par
They are non-trivial as operators. Indeed the curvature of the connection, $B$, that occurs in the twistor Wilson loops for $\lambda=1$ is:
\bea
F(B)=\mu
\eea
where $\mu$ is a field of $ASD$ type, Eq.(5.4). If the twistor Wilson loops are chosen in the adjoint representation, their v.e.v. factorizes in the large-$N$ limit in the product of the
fundamental and conjugate representation. The sector generated by the local operators that are polynomial in $(\mu, \bar \mu)$ in the fundamental representation plays a special role in this paper 
and it is referred to as the holomorphic/antiholomorphic sector. \par
The twistor Wilson loops have trivial expectation value at large $N$, i.e. their v.e.v. is exactly 1 at large $\theta$, Eq(4.29):
\bea
\lim_{\theta \rightarrow \infty} <\frac{1}{ N} Tr \Psi(\hat B_{\lambda};L_{ww})  >&=&1  \nonumber \\
<\frac{1}{\cal N} Tr_{\cal N} \Psi(\hat B_{\lambda};L_{ww})     >& = &<\frac{1}{\cal N} Tr_{\cal N} \Psi(\hat B_1;L_{ww}) > \nonumber \\
\eea
and their v.e.v. is $\lambda$-independent for any $\theta$.
This is shown in sect.(4), in particular in sect.(4.5). \par
Therefore their shape can be deformed at will without changing their
v.e.v.. In particular they can be deformed by adding an arc that backtracks, i.e. a path that is oriented in one direction
and then it comes back along the same path in the opposite direction. This invariance property is known as zig-zag symmetry \footnote{The zig-zag symmetry plays an important role
in a string ansatz for the solution of the Makeenko-Migdal loop equation due to Alexander Polyakov \cite{Po1}.}
and it is well known classically for any Wilson loop, but for twistor Wilson loops it holds also at quantum level.
The reason is that in a regularized version the backtracking arc becomes the boundary of a tiny strip ending with a 
cusp. \par
Now, the v.e.v. of an ordinary cusped Wilson loops has an extra divergent logarithmic contribution, with respect to smooth loops, known as cusp anomaly \cite{Py}. 
The cusp anomaly, i.e. the coefficient of the logarithm, actually diverges when the cusp backtracks, i.e. when the cusp angle $\Omega$ tends to $0$ (see for example \cite{Gr, Kr}) :
\bea
 <Tr \Psi(x,x;A)>  \sim \exp(-\frac{\delta}{\Omega} \log(l \Lambda))
 \eea
where $l$ is the length of the cusp and $\delta$ a numerical factor. In the Makeenko-Migdal loop equation of standard type the right hand side
develops extra divergences if a Wilson loop develops a backtracking cusp.
Indeed, performing the two contour integrations along the loop in the right hand side of the Makeenko-Migdal equation:
\bea
&&\int_{L_{xx}} dx_{\alpha} <\frac{N}{2 g^2}Tr(\frac{\delta S_{YM}}{\delta A_{\alpha}(x)}\Psi(x,x;A))> \nonumber \\
&&=i \int_{L_{xx}} dx_{\alpha} \int_{C_{xx}} dy_{\alpha} \delta^{(4)}(x-y) <Tr \Psi(x,x;A)> <Tr1>
\eea
it is obtained \cite{Gr}:
\bea
&&\int_{L_{xx}} dx_{\alpha}<\frac{1}{2 g^2}Tr(\frac{\delta S_{YM}}{\delta A_{\alpha}(x)}\Psi(x,x;A))> \nonumber \\
&&\sim i(P \Lambda^3+\sum_{cusp} \frac{ \cos \Omega_{cusp}}{ \sin \Omega_{cusp}} (\pi -\Omega_{cusp}) \Lambda^2)<Tr \Psi(x,x;A)> <Tr1>
\eea
where $l$ is the perimeter of the loop and $\Omega_{cusp}$ the cusp angle at a cusp. In the limit in which the cusp angle $\Omega_{cusp}$ reaches $0$
the cusp backtracks and the cusp contribution to the
contact term of the Makeenko-Migdal loop equation is divergent. \par
On the contrary, the right hand side of the holomorphic loop equation, after the gauge invariant regularization by analytical continuation to Minkowski described in sect.(6),
vanishes at the backtracking cusp. 
Indeed the right hand side of Eq.(1.8) is a contour integral of a $\delta^{(1)}$ first taken with one orientation and then taken with the opposite
orientation:
\bea
\int dw_+(s)
\delta(z_+(s_{cusp}) -w_+(s))&=& \frac{1}{2}\frac{\dot w_+(s^+_{cusp})}{ |\dot w_+(s_{cusp})|}+ \frac{1}{2}
\frac{\dot w_+(s^-_{cusp})}{|\dot w_+(s^-_{cusp})|} \nonumber \\
&=& \frac{1}{2} - \frac{1}{2} =0
\eea
The delta function arises by analytical continuation to Minkowski, sect.(6).
The light cone Minkowski coordinate is chosen to be parallel to the tangent to the cusp. 
Thus the holomorphic loop equation for a cusped loop reduces exactly to a critical equation for an effective action, provided the equation of motion for the effective action,
$\Gamma$, is restricted to the subalgebra generated by twistor Wilson loops:
\bea
<Tr(\Psi(B; L'_{z_+z_+})\frac{\delta \Gamma}{\delta \mu(z_+,z_-)}\Psi(B; L''_{z_+z_+}))>= 0
\eea
where $L'$ and $L''$ are the two petals in which the Makeenko-Migdal loop with the shape of $\infty$ is decomposed.
All the information about the original "change to loop variables", that is contained in the planar diagrams that contribute to a cusped twistor Wilson loop, is encoded in the logarithm of the Jacobians that enter
the effective action: the Jacobian of the Nicolai map from $A$ to $\mu$ and the Jacobian to the holomorphic gauge from
$\mu$ to $\mu'$. \par
Thus an effective action, $\Gamma$, that represents the "sum of planar diagrams"
for cusped Wilson loops has been obtained. $\Gamma$ contains the interesting information of the localization of the loop equation,
since we can compute in principle the glueball spectrum, in the holomorphic/antiholomorphic sector generated by $(\mu, \bar \mu)$, by fluctuations around the critical points of $\Gamma$. \par
However, there are some technical difficulties. 
To interpret the preceding equation strongly \footnote{If all the matrix elements of an operator vanish, the operator actually vanishes.}:
\bea
\frac{\delta \Gamma}{\delta \mu(z_+,z_-)}= 0
\eea
we must realize explicitly the restriction to the subalgebra of twistor Wilson loops and in addition we must actually find the critical points. \par
Afterward we must be able to check that the fluctuations around the critical points are actually suppressed by $\frac{1}{N}$. \par
Finally, we must be able to understand the loci in function space where the Nicolai map is not one-to-one. \par
Surprisingly, the three problems are actually deeply linked, in such a way that the solution of the third one furnishes the solution of the other two. \par
Indeed the first change of variables, the one from the connection $A_{\alpha}$, to the $ASD$ field, $\mu^-_{\alpha \beta}$, may not be one-to-one even after gauge fixing. \par
Firstly, we  work out the supersymmetric case in sect.(3.2), just as a simpler exercise.
It turns out that the Nicolai map is not one-to-one and at places where multiple solutions of  $A_{\alpha}$ occur for a given $\mu^-_{\alpha \beta}$, there are moduli of  $A_{\alpha}$ and
correspondingly zero modes in the Jacobian of the change of variables.
However, in the $SUSY$ case and for a special observable, the gluino condensate, the only relevant zero modes arise by instantons.
But these zero modes are essential to reproduce the exact supersymmetric beta function of Novikov-Shifman-Vainstein-Zacharov  ($NSVZ$)  \cite{NSVZ}.  \par
Secondly, we work out the non-supersymmetric case, in sect.(3.3). It turns out that, as in the $SUSY$ case, even the first coefficient of the pure $YM$ beta function cannot be reproduced from $\Gamma$
unless zero modes occur, sect.(3.4). But unlike the $SUSY$ case localization on instantons does not reproduce the second coefficient of the $YM$ beta function, sect.(3.4). Thus in the pure $YM$ case we need to understand the moduli of the Nicolai map generically in function space. This is more difficult, since it is known that there is no Hausdorff, i.e. separable, moduli space of all bundles
even in two dimensions. Therefore we have to introduce a dense set, in the sense of distributions,
in function space, on which the moduli problem has a complete answer, by the standards of differential geometry of fiber bundles, sect.(7).
On this dense set also the second change of variables, from $\mu$ to $\mu'$ is well understood, sect.(7). \par
In fact this leads to understanding zero modes also in the pure large-$N$ $YM$ case and to the large-$N$ exact beta function of sect.(9) and sect.(11), via the effective action obtained by the cusped holomorphic loop equation. \par
The aforementioned dense set involves interpreting the non-$SUSY$ Nicolai map as defining a hyper-Kahler reduction on singular instantons,
with singularities of magnetic type:
\bea
F_{\alpha \beta}^- = 
\sum_p \mu^{-}_{\alpha \beta}(p)  \delta^{(2)} (z-z_p)
\eea 
The restriction of the $YM$ measure to the lattice hyper-Kahler reduction, that is dictated by the need of reproducing the beta function at the critical points of $\Gamma$, has two fundamental consequences, sect.(7). It realizes in an explicit and mathematically well defined way
the algebra of twistor Wilson loops, via the equivalence of the hyper-Kahler quotient to the holomorphic quotient. This solves the first problem too. \par
In addition the lattice hyper-Kahler reduction of the $YM$ theory in the Nicolai variables is locally abelian, up to zero modes, because the triple, $ \mu^{-}_{\alpha \beta}(p)$, is commutative at any given lattice point, $p$,
because of fundamental properties of the Hitchin equations, sect.(7).
Thus there is a gauge, referred to as the singular gauge in this paper, in which fluctuations of $ \mu^{-}_{\alpha \beta}(p)$ reduce to fluctuations of eigenvalues, that are suppressed
in the large-$N$ limit, sect.(12). This solves the second problem. 
All the remaining non-abelian degrees of freedom of the theory on the lattice hyper-Kahler quotient are the zero modes associated to the moduli of the local systems determined by the hyper-Kahler reduction.
Therefore the solution of the third problem furnishes the solution to the other two problems as anticipated. \par
Finally, the critical points of the effective action 
can be explicitly found, using the $\lambda$-independence of the v.e.v. of the twistor Wilson loops, Eq(1.11), as fixed points for the action on the Nicolai variables of the semigroup that rescales $\lambda$, sect.(8). They are a "lattice of surface operators" with $Z_N$ holonomy labelled by a magnetic quantum number, $k$, sect.(7).
The surface operators with $Z_N$ holonomy condense because of asymptotic freedom ($AF$), sect.(12).
In turn this leads to the physical interpretation of the magnetic condensate of surface operators in terms of 't Hooft electric/magnetic duality \cite{Super1,Super2,Super3} that requires that, if the $YM$ theory has a mass gap, then either the electric charge condenses (Higgs phase) or the magnetic charge condenses (confinement phase).  \par
The Jacobian of the non-$SUSY$ Nicolai map gives rise at the leading large-$N$ order to an anomalous dimension, sect.(3.4), that contributes to the one-loop exact Wilsonian beta function, sect.(9), and to a canonical beta function of $NSVZ$ type, once the effective action is restricted to surface operators, sect.(11).
At next to leading order this Jacobian gives rise to the glueball kinetic term, sect.(12). \par
Instead the logarithm of the Jacobian of the holomorphic gauge is precisely the glueball potential at leading order, whose second derivative
is the glueball mass matrix at next to leading order. Both are computed in sect.(12). A detailed study shows that a mass gap, proportional to the density, $\rho_k$, of surface operators of magnetic quantum number, $k$, arises, sect.(12). The density scales as $(kN)^{-\frac{1}{2}}$ times the square of the $RG$-invariant scale, but the large-$kN$ vanishing of the density is compensated by large-$kN$ degeneracy factors in the mass matrix and by large-$N$ degeneracy factors in the kinetic term, in such a way that the mass gap survives the large-$N$ limit. \par
Actually the spectrum is of pure poles, because the non-local terms in the effective actions are suppressed by higher powers of the density of surface operators in the expansion
of the effective action in powers of $\rho_k$, that occurs as the usual expansion of a functional determinant in terms of one-loop graphs with multiple insertions of the background field. \par
In fact the spectrum of fluctuations of composite surface operators forms a trajectory linear in $k$ that does not include any massless state. While the spectrum does not depend on the choice
of a particular observable of the $YM$ theory, in our case surface operators, in \cite{MB5} we describe the correspondence between the two-point correlators of long composite surface operators and the glueball propagators. One ingredient of this correspondence is the coincidence of the anomalous dimensions of long surface operators with the anomalous 
dimensions of the ground state of the Hamiltonian spin chain in the thermodynamic limit \cite{Zar1,Zar2}, that provides anomalous dimensions of local operators in the one-loop $ASD$ integrable sector of large-$N$ $YM$, sect.(12). \par
Introducing a lattice of surface operators allows us to write a lattice holomorphic loop equation. 
There is indeed a lattice version of holomorphic loop equation, Eq.(10.1), in which the lattice points are the locations
of singularities of surface operators. This lattice version arises through the discussion presented in sect.(7):
\bea
<Tr(\frac{\delta \Gamma}{\delta \mu(z_p, \bar z_p)'}\Psi'(L_{z_p z_p}))>=
\frac{1}{\pi} \int_{L_{z_p z_p}} \frac{ dw}{z_p -w} <Tr\Psi'(L_{z_p w})> <Tr\Psi'(L_{w z_p })> 
\eea
The lattice holomorphic loop equation equation allows to build more on the geometrical side of the localization, discussed more extensively in sect.(1.3). \par
By homological localization of the loop equation we mean a deformation of the loop that is trivial in homology and for which the  term that arises as the "change to loop variables" vanishes, in such a way that the loop equation is reduced to a critical equation for an effective action \cite{MB1}. Hence the needed homological deformation has to satisfy the following properties.
It has to be trivial in homology. It has to leave the expectation value of the loop invariant. It has to imply the vanishing of the quantum term in the loop equation, i.e. of the term that
contains the contour integral along the loop, sect.(10). \par
Now we deform the twistor loops by adding a backtracking arc ending with a cusp at each lattice point.
The lattice points associated to the divisor of surface operators become the cusps that are the end points, $p$, of the backtracking strings, $b_p$, that
perform the deformation of the loop, $L$. Adding the backtracking strings implies that the contribution of the "change to loop variables" vanishes for the modified loop:
\bea
<Tr(\frac{\delta \Gamma([b_p])}{\delta \mu'(z_p, \bar z_p)}\Psi'(L \cup [b_{p}]))>=0 
\eea
This phenomenon is called homological localization of the holomorphic loop equation
because, in geometrical language, is dual to the cohomological localization
that is described in sect.(1.3) and in sect.(3.2) for the gluino condensate of the $SUSY$ $YM$ theory. \par
In the cohomological localization of sect.(1.3) functional integrals are reduced to critical points by adding "a total differential in function space". \par
In the homological localization of sect.(10) loop equations are reduced to critical points adding to the loop "vanishing boundaries",
i.e. the aforementioned backtracking arcs ending with cusps. \par
The duality is no mystery. It is just the Stokes theorem in the form:
\bea
\int_V d \omega= \int_{\partial V} \omega
\eea
that relates differentials, i.e. the cohomological side, to boundaries, i.e. the homological side. \par
In the holomorphic loop equation of the $YM$ theory the boundaries are the backtracking arcs that are added to the loop to get localization of the holomorphic loop equation.
In the $SUSY$ $YM$ theories the coboundary is the differential furnished by the (twisted) supersymmetric charge. This is not a physical duality. It is purely a mathematical duality of the two
different localizations. \par

\subsection{ Summary of results}

From a purely computational point of view we may present our results in terms of the glueball propagators in a certain sector of the large-$N$ $YM$ theory.
The glueball propagators that we refer to are initially defined in Euclidean signature, constructed by means of fluctuations, $\delta \mu$ \footnote{The precise definition is in Eq.(12.39).}, of certain surface operators \footnote{The idea of integrating in $YM$ theory on local systems associated to an arbitrary parabolic divisor appeared for the first time long ago in some papers by us \cite{MB2, MB3}, by embedding the Hitchin fibration \cite{H2,K1,K4,S3}
in the $YM$ functional integral, and physically corresponds to integrating over surface operators ante litteram \cite{W2}. Explaining how surface operators arise in large-$N$ $YM$ is in fact the subject of this 
paper.} supported on a Lagrangian submanifold of four-dimensional space-time \footnote{The plane $(z, \bar z,z, \bar z)$ is Lagrangian for the symplectic form $dz \wedge d \bar z - du \wedge d \bar u$ in four dimensions.}:
\bea
\Lambda_W^6
\int < \frac{1}{\cal{N }} Tr_{\cal{N}}(\mu \bar \mu)(z, \bar z,z, \bar z)  \frac{1}{\cal{N}}  Tr_{\cal{N}}( \mu \bar \mu)(0,0,0,0)>_{conn} e^{i(p_z \bar z + p_{\bar z} z)} d^2z   \nonumber \\
\eea
where $(z=x_0+i x_1, \bar z=x_0-i x_1, u=x_2+i x_3, \bar u= x_2-i x_3)$  are complex coordinates in Euclidean $R^2 \times R^2 $, $d^2z=dx_0 dx_1$  and
\bea
\mu= \frac{1}{2}(\mu^-_{01}-i \mu^-_{03})
\eea 
with $\bar \mu$ the Hermitian conjugate.
The following identification holds up to (infinite) factors \footnote{The identification extends to composite operators in a certain asymptotic sense, see below and sect.(12). The infinite factors are actually defined and regularized in sect.(12).}:
 \bea
\int \mu(z, \bar z,z, \bar z)e^{i(p_z \bar z + p_{\bar z} z)} d^2z
\sim \frac{1}{2}(F^-_{01}-iF^-_{03})(p_z, p_{\bar z},p_z, p_{\bar z})
\eea
where
\bea
(F^-_{01}-iF^-_{03})(p_z, p_{\bar z},p_u, p_{\bar u})=\int (F^-_{01}-iF^-_{03})(z, \bar z, u, \bar u)  e^{i(p_z \bar z + p_{\bar z} z+p_u \bar u + p_{\bar u} u)} d^2z d^2u 
\eea
is the Fourier transform with
\bea
F_{\alpha \beta}^-=F_{\alpha \beta}- \tilde F_{\alpha \beta} \nonumber \\
 \tilde F_{\alpha \beta}= \frac{1}{2} \epsilon_{\alpha \beta \gamma \delta} F_{\alpha \beta}
\eea
the anti-selfdual ($ASD$) part of the curvature of the gauge connection. The identification holds because the fluctuating field of surface operators, $\delta\mu$, has non-vanishing momenta dual in the
Fourier sense to the Lagrangian support of surface operators and it occurs in the effective action with zero momenta dual in the Fourier sense to the manifold normal to the support. $\mathcal{N}=N \hat N$ is the total rank of the gauge group. The first factor of $N$ is the rank of an $SU(N)$ gauge bundle that is embedded by non-commutative Morita equivalence into $U(N\times \hat N)$. The construction is explained in sect.(7) and sect.(12).
After analytically continuing to Minkowski space-time
in its simplest form of our result reads \footnote{The symbol $\sim$ stays for "equal up to constant irrelevant numerical factors" or "equal up to irrelevant additive terms" depending on the framework.}:
\bea
&&  \Lambda_W^6 
\int < \frac{1}{\cal{N }} Tr_{\cal{N}}(\mu \bar \mu)(x_+, x_-,x_+, x_-)  \frac{1}{\cal{N}}  Tr_{\cal{N}}( \mu \bar \mu)(0,0,0,0)>_{conn} e^{i(p_+x_-+ p_-x_+)} dx_+ dx_-  \nonumber \\
&& \sim   \frac{1}{\mathcal{N}^2}      \sum_{k=1}^{\infty} \frac{ k \rho_k^{-2} \Lambda_W^6}{ - \alpha'  p_+ p_-+ k \Lambda_W^2} \nonumber \\
&& \sim   \frac{1}{\mathcal{N}^2} \sum_{k=1}^{\infty} \frac{ k^2 \Lambda_W^6}{ - \alpha'  p_+ p_-+ k \Lambda_W^2} \nonumber \\
\eea
where, $( x_+=x_4+x_1,x_-=x_4-x_1)$ are light-cone coordinates, $(p_+=p_4+p_1, p_-=p_4-p_1)$ are light-cone momenta, and $\Lambda_W$
is the renormalization group invariant scale in the Wilsonian scheme. \par
$\rho_k$ is the density, in units of $\Lambda_W^{2}$,
\bea
\rho=\sum_p \delta^{(2)} (z-z_p) 
\eea
of surface operators carrying at each lattice point, $p$, magnetic charge $k$ and holonomy valued in the center, $Z_N$, of the gauge group,
i.e. such that:
\bea
e^{2i \mu_p}=e^{i\frac{2 \pi k}{N}}
\eea 
with:
\bea
\mu \sim \frac{1}{2}(F^-_{01}-iF^-_{03}) = \sum_p \mu_p  \delta^{(2)} (z-z_p(u, \bar u))
\eea
and $z_p(u,\bar u)=z_p$.
$\rho_k$ scales with $k$ as:
\bea
\rho_k^{2} \sim \frac{1}{k} 
\eea
and the dimensionless inverse "string tension" in units of $\Lambda_W^{-2}$ is:
\bea
\alpha'=  \frac{10}{3  \pi} n 
\eea
where $n \ge 2$ is a finite positive integer that depends on the choice of the renormalization scheme. In fact $n$ can be reabsorbed into a redefinition of $\Lambda_W$ in any of the
countably many possible inductive sequences that define the $SU(\infty)$ group as a limit of a sequence of finite dimensional $SU(N)$, sect.(7) and sect.(12). \par
The peculiar support, $(x_+, x_-,x_+, x_-)$, of the correlator arises as the projection with Minkowski signature on the base of a Lagrangian submanifold of the twistor space of (complexified) Euclidean space-time that occurs in our approach.
The field $\mu$ is dimensionless and normalized in such a way that the correlator in Eq.(1.7) be renormalization group invariant
\footnote{ In the canonical normalization also an extra factor of the anomalous dimension occurs. This is reported in the formulae below and discussed in sect.(12).}.
The mass spectrum, $ - \alpha'  p_+ p_-+ k \Lambda_W^2=0$, and the multiplicity, $k$, that occurs in the numerator of Eq.(1.7) in the second line, are in fact exact \footnote{Large-$N$ exact linearity of the spectrum, rather than only asymptotic linearity, may look surprising. However, the ratio between the masses of the two lowest scalar states in pure $SU(8)$ $YM$ has been found numerically to be compatible with the value $\sqrt 2$ \cite{Meyer}. 
We would like to thank Michael Teper for a clarifying discussion about this point at the Galileo Galilei Institute workshop on "Large-$N$ Gauge Theories" (2011), hereafter referred to as the GGI workshop. Indeed the best fits in \cite{Meyer} for the continuum limit ratios of the masses (in units of the lattice $RG$-invariant scale, Table (7.14)) of the $J^{PC}$ glueball, $m_{0^{++*}}=4.71(29)$ and  $m_{0^{-+}}=4.72(32)$, to the mass of the lowest scalar, $m_{0^{++}}=3.32(15)$ agree with very good accuracy with $\sqrt2$. Yet, there are larger statistical errors than this agreement may suggest. There is also a $m_{2^{++}}=4.65(19)$ glueball, essentially degenerate in mass with the two aforementioned scalar states, that may in principle couple to our operators. Nevertheless, for reasons explained in the paper, we suggest that the spectrum in Eq.(1.7) is all made by scalars. }
in a certain asymptotic expansion in powers of ${N}^{-\frac{1}{2}} $ of the gauge connection of the surface operators that occur in our approach \footnote{The precise meaning of this statement is clarified in sect.(12).}. \par
At large $N$ there is a Wilsonian scheme in which the Wilsonian beta function is one-loop exact \cite{MB1} and a canonical scheme in which the beta function has a form $NSVZ$ that reproduces the first two universal perturbative coefficients \cite{MB1} 
\footnote{It has been known for some time that the Wilsonian, $g_W$, and the canonical, $g$, coupling constant have different beta functions in general \cite{AA}.}:
\bea
\frac{\partial g_W}{\partial \log \Lambda}=-\beta_0 g_W^3
\eea
and
\bea
\frac{\partial g}{\partial \log \Lambda}=\frac{-\beta_0 g^3+
\frac{1}{(4\pi)^2} g^3 \frac{\partial \log Z}{\partial \log \Lambda} }{1- \frac{4}{(4\pi)^2} g^2 }
\eea
with:
\bea
\beta_0=\frac{1}{(4\pi)^2} \frac{11}{3} \nonumber \\
\eea
where $g=g_{YM}^2 N $ is the 't Hooft canonical coupling constant and $ \frac{\partial \log Z}{\partial \log \Lambda} $ 
is computed to all orders in the 't Hooft Wilsonian coupling constant, $g_W$, by:
\bea
\frac{\partial \log Z}{\partial \log \Lambda} =\frac{ \frac{1}{(4\pi)^2} \frac{10}{3} g_W^2}{1+cg_W^2}
\eea
with $c$ a scheme dependent arbitrary constant.
Indeed since $ \frac{\partial \log Z}{\partial \log \Lambda} $ to the lowest order in the canonical
coupling is:
\bea
\frac{\partial \log Z}{\partial \log \Lambda}=
\frac{1}{(4\pi)^2} \frac{10}{3} g^2 + ...
\eea
the correct value of the first and
second perturbative coefficients of the beta function \cite{AF,AF1,2loop1,2loop2} arise:
\bea
\frac{\partial g}{\partial \log \Lambda}&&=
-\beta_0 g^3+
(\frac{1}{(4\pi)^2} \frac{1}{(4\pi)^2} \frac{10}{3} -\beta_0 \frac{4}{(4\pi)^2} ) g^5 +... \nonumber \\
&&=-\frac{1}{(4\pi)^2}\frac{11}{3} g^3 + \frac{1}{(4\pi)^4} ( \frac{10}{3}
-\frac{44}{3})g^5 +... \nonumber \\
&&=-\frac{1}{(4 \pi)^2} \frac{11}{3} g^3 -\frac{1}{(4 \pi)^4} \frac{34}{3} g^5+...
\eea
In fact a whole family of correlators of the Fourier transform of composite operators of naive dimension $4L$, $O^L(p_+, p_-)$, constructed by surface operators supported on the aforementioned Lagrangian submanifold, are computed in sect.(12). 
In the Wilsonian scheme the result reads:
\bea
&& <Tr_{\mathcal N} O^L(p_+, p_-)  Tr_{\mathcal N}O^L(-p_+, -p_-)>^{(W)}_{conn}  \nonumber \\
&& \sim  \sum_{k=1}^{\infty} \frac{  \Lambda_W^2 k^{2(2L-1)} \Lambda_W^{4(2L-1)}}{ - \alpha'  p_+ p_-+ k \Lambda_W^2} + ...\nonumber \\
&& \sim  (- p_+ p_-)^{4L-2}  \sum_{k=1}^{\infty} \frac{ \Lambda_W^2}{ - \alpha'  p_+ p_-+ k \Lambda_W^2} + ...\nonumber \\
&& \sim  (- p_+ p_-)^{4L-2} \log \frac{ - p_+ p_-}{ \Lambda_W^2} + ...\nonumber \\
\eea
where the dots stand for contact terms, i.e. distributions whose inverse Fourier transform is supported at coinciding points. \par
In the canonical scheme for the same objects anomalous dimensions arise:
\bea
&& <Tr_{\mathcal N} O^L(p_+, p_-)  Tr_{\mathcal N}O^L(-p_+, -p_-)>^{(C)}_{conn}\nonumber \\
&& = g^4(- p_+ p_- ) Z^{-\frac{8L-8}{2}}(- p_+ p_-) <Tr_{\mathcal N} O^L(p_+, p_-)  Tr_{\mathcal N}O^L(-p_+, -p_-)>^{(W)}_{conn}  \nonumber \\
&& \sim g^4(- p_+ p_-) Z^{-\frac{8L-8}{2}}(- p_+ p_-)  (- p_+ p_-)^{4L-2} \log \frac{ - p_+ p_-}{ \Lambda_W^2}
\eea
where $g(- p_+ p_-)$ and $Z(- p_+ p_-)$ are the ($RG$ improved) momentum dependent canonical coupling in Eq.(1.14) and renormalization factor in Eq.(1.16).
A deeper and more complete discussion of the relation between correlators of surface operators and glueball propagators can be found in \cite{MB5}. \par 
For large $L$ the anomalous dimensions agree \footnote{The agreement is at one loop 
since anomalous dimensions are universal, i.e. scheme independent, only at one loop. Actually they agree also for $L=1$, since in this case the anomalous dimension is determined by the beta function via the factor of $g^4$.} with the anomalous dimensions of the ground state \cite{Zar1, Zar2} of the Hamiltonian spin chain in the thermodynamic limit,
furnishing an identification, that is at least asymptotic
for large $L$, between composite surface operators \footnote{We actually mean that we identify the Fourier transform of our composite surface operators with the Fourier transform of composite local operators in the same fashion as in Eq.(1.4). For example, $O^L(p_+, p_-) \sim |(F^-_{01}+iF^-_{03})|^{2L}(p_+, p_-,p_+, p_-)$. } and composite local operators of $YM$ in some regularization scheme \cite{MB5}. Indeed the Hamiltonian spin chain is an integrable model
by which the anomalous dimensions of composite operators of large-$N$ $YM$ in the $ASD$ and $SD$ sector can be computed exactly at one-loop. \par 
The exactness of our formula for the spectrum is not affected by the possibly only
large-$L$ asymptotic identification of the operators, since the spectrum depends only on the occurrence of poles in any correlator of gauge invariant operators. However, the aforementioned asymptotic identification
suggests that all the glueballs in our spectrum are in fact scalar, since this is so for the operators that correspond to the ground state of the Hamiltonian spin chain in the thermodynamic limit \cite{Zar2}.  \par
Yet, there is an extension of our
approach to fluctuations of surface operators defined by connections with wild singularities (i.e. pole singularities of any order), that are naturally associated to Regge trajectories of higher spin (sect.(13)). Their contribution
is not computed in this paper.  \par
Now we re-explain the basic ideas underlying our computations in geometric terms as follows. \par

\subsection{A geometric point of view: localization by cohomology and by homology}

In the last thirty years we witnessed the geometrization of theoretical high energy physics. This geometrization has several faces
but the one that we refer to consists in computing exactly functional integrals by geometrical methods. 
The key idea in solving the analytical problem of performing an integral by geometrical methods lies in the work of Duistermaat and Heckman
\cite{DE} on exact localization of the integral of the exponential of the Hamiltonian of a torus action on a compact symplectic manifold on the fixed points
of the torus action \footnote{For a comprehensive review see \cite{Szabo}.}:
\bea
\frac{1}{n!} \int \omega^{n} \exp{(-\epsilon H)}= \sum_P (\frac{2 \pi}{\epsilon})^n Pf^{-1} (\omega_P^{-1} \partial^2H_P)
\eea
where $Pf$ is the Pfaffian of the skew matrix, $\omega_P^{-1} \partial^2H_P$, at the fixed points, $P$, and $\partial^2H_P$ the Hessian of the Hamiltonian at $P$. Such localization has the following cohomological nature according to Atiyah and Bott \cite{A} and Bismut \cite{Bis1, Bis2}. 
In finite dimension the integral of the exponential of a closed form  $\omega$, $d\omega=0$, on a compact manifold without boundary, $M$, defines a cohomology class
invariant for the addition to the closed form of an exact differential, $d\alpha$, (i.e.  of a coboundary), since $d\alpha$ is trivially closed because $d^2=0$.
Rescaling the exact differential by a large factor, under suitable positivity assumptions, the integral of the cohomology class gets localized on the critical
points of the exact differential and the saddle-point approximation turns out to be exact:
\bea
\int_M \exp(-\omega- t  d\alpha)=\int_M \exp(- \omega)
\eea
Indeed the first $t$-derivative vanishes because it is the integral of a coboundary and $\int_M d\alpha=0$:
\bea
\frac{d}{dt} \int_M \exp(-\omega- t  d\alpha)=-\int_M d(\alpha \exp(-\omega- t  d\alpha))=0
\eea
Therefore the integral is $t$-independent and in the limit $t \rightarrow \infty$ 
can be evaluated by the saddle-point method. Bismut \cite{Bis1, Bis2} was the first one to extend rigorously this kind of argument to infinite dimensions, actually to a functional integral
in one dimension, i.e. to quantum mechanics. In the quantum mechanical setting there are (essentially) no existence
problems for functional integrals and the localization argument is in fact a mathematical proof. \par
However, the subject blossomed in quantum field theory only after Witten paper on localization in two-dimensional
Yang-Mills theory \cite{W} and Witten work on Donaldson invariants \cite{W1} that introduced localization in four-dimensional
supersymmetric gauge theories, by identifying the differential needed to define the cohomology with a twisted super-charge, $Q$,
satisfying $Q^2=0$.
In turn Witten twist of supersymmetry requires to start with at least an $\cal{N}$ $=2$ $SUSY$ $YM$ theory. \par
Thus the infinite-dimensional field theoretical analog of Eq.(1.42) is:
\bea
\int O \exp(-S_{SUSY}- t Q \alpha) 
\eea
with:
\bea
QO=0 \nonumber \\
QS_{SUSY}=0
\eea
There have been a number of applications of the localization idea in four-dimensional $SUSY$ gauge theories,
among which we mention the Nekrasov computation \cite{N} of the prepotential in $\cal{N}$ $=2$ $SUSY$ gauge theories, that
reproduces by localization methods the Seiberg-Witten solution \cite{SW} for the same object, and Pestun \cite{P} computation
of certain twist-$SUSY$ invariant  Wilson loops in $\cal{N}$ $=4,2,2^*$ $SUSY$ gauge theories. \par
From a purely mathematical point of view these exact results state the equality
between a mathematically not well defined object, the original functional integral, and a mathematically well defined and explicit answer, the result of the localization. \par
However, from the point of view of theoretical physics, these results are in fact satisfactory since, waiting for a realization of the constructive program of quantum field theory in four dimensions \cite{GA0},
the explicit answer that is found by localization defines the functional integral by the rules by which it is
computed and contributes to fix the properties that the yet-to-come mathematical construction of the functional
integral has to satisfy: the localization property indeed.   \par  
The aim of this paper is to add, rather surprisingly, a non-supersymmetric chapter to the aforementioned exact results. \par
The simplest way to present our basic result is to compare it with Nekrasov computation of the prepotential.
In the first part of Nekrasov computation the functional integral that evaluates the cohomology of $1$ (i.e. the partition function) is reduced by cohomological localization to a sum of finite dimensional integrals over the instantons moduli spaces:
\bea
Z=\lim_{t \rightarrow \infty} \int 1 \exp(-S_{SUSY}- t Q \alpha)= \sum_k \exp{(-\frac{16 \pi^2 kN}{2 g_W^2})} \Lambda^{2kN} \int_{{\cal M}_k} \wedge \omega
\eea
This depends on the supersymmetry and has no analog in the pure $YM$ case.
On the contrary, in the second part of Nekrasov computation, the finite dimensional integrals over instantons moduli are reduced to a sum over the fixed points for the action of the torus $U(1)^{N-1} \times U(1) \times U(1) $ in $SU(N) \times O(4)$ by applying the Duistermaat-Heckman formula, after a suitable ultraviolet and infrared regularization of the moduli space, by means of a non-commutative deformation parameterized by $\epsilon$:
\bea
\frac{1}{(2kN)!}\int_{{\cal M}^{\epsilon}_k} \wedge \omega \exp{(-\epsilon H)}= (\frac{2 \pi}{\epsilon })^{kN} \sum_P Pf^{-1}(\omega_P^{-1}\partial^2H_P)
\eea
Here $SU(N)$ is the (global) gauge group at infinity and $O(4)$ the group of Euclidean rotations.
These groups are symmetry groups also of the pure $YM$ theory. 
The result of the localization can be resummed into an exact formula for the prepotential \cite{SW}, $\cal { F} $, that is a function of the quantum moduli of the theory, that are related to the v.e.v. of the eigenvalues of the complex scalar field in the adjoint representation of the $\cal{N}$ $=2$ $YM$ theory:
\bea
Z=\exp (\frac{1}{\epsilon^2} \cal { F})
\eea
Despite the prepotential is obtained by the localization of a trivial observable, the cohomology of $1$, from a physical point of view it contains the interesting information of the localization. Indeed, according to Seiberg-Witten \cite{SW}, the prepotential 
contains exact highly non-trivial quantum information. It determines an exact beta function and the low energy effective action in the Coulomb branch of the theory as a function of the translational invariant condensate of the eigenvalues of the scalar field \footnote{In the Coulomb branch the eigenvalues are generically all different in such a way that the unbroken gauge group is $U(1)^{N-1}$.}. Thus the prepotential is used to reach conclusions about the physical theory \cite{SW} that by far exceed the very limited framework of its derivation by localization. \par
A general feature of cohomological localization is that the saddle-point computation can be employed only for the specific observables
that satisfy Eq.(1.45). A fortiori in pure $YM$, that has no $SUSY$, there is no hope that localization may hold, if any, but for very special observables. \par
To say it in a nutshell,
our basic idea for pure $YM$ is to construct special trivial observables, called twistor Wilson loops for geometrical reasons, since they are supported on a Lagrangian submanifold of twistor space of complexified Euclidean space-time.
In a technical sense the trivial twistor Wilson loops are in the homology of 1, rather than in the cohomology of 1, since in pure $YM$ there is no $SUSY$ and thus no interesting cohomology \footnote{There is in fact the Becchi-Rouet-Stora ($BRS$) cohomology associated to gauge-fixing, that leads to localization on gauge-fixed slices of gauge orbits, but it is not relevant for our purposes.}.
The loop equation for twistor Wilson loops can be solved, since they are trivial, in the sense that it can be reduced to a critical equation for an effective action, i.e. it can be localized.
Despite the effective action is obtained by trivial observables, it carries highly non-trivial quantum information, that exceeds by far the framework of localization of $1$. \par
The effective action determines an exact large-$N$ beta function and turns out to be a function of the density, $\rho_k$, in 
units of $\Lambda_W^{2}$,
of the condensate of surface operators of magnetic charge $k$ that occur in the localization of the twistor Wilson loops. \par
In addition the effective action restricted to fluctuations of surface operators supported on a Lagrangian submanifold with Minkowski signature, obtained by a certain Wick rotation from the Lagrangian submanifold which the twistor Wilson loops are supported on, determines the glueball spectrum. \par 
The analytic continuation to Minkowski space-time
is the only way to regularize gauge invariantly the holomorphic loop equation for the twistor Wilson loops, that in turn leads to localization on the critical points of the effective action. \par
We describe now in more detail what the twistor Wilson loops are. \par 
They compute the holonomies along loops of a modified non-Hermitian $YM$ connection, the twistor connection. Its curvature is a non-Hermitian linear combination of the $ASD$ part of the curvature of the ordinary gauge connection. These loops are supported on a Lagrangian submanifold in twistor space of complexified Euclidean space-time, locally the product of a two-dimensional surface immersed in complexified space-time with local (complex) coordinates $(z, \bar z)$ and of a one-dimensional curve immersed in the fiber of the twistor fibration with (not necessarily real) coordinate $\lambda$. \par
The twistor Wilson loops are chosen in the adjoint representation.
The operator definition of twistor Wilson loops involves
the parameter $\lambda$, but their vacuum expectation value (v.e.v.) is $\lambda$-independent, in fact trivially $1$ at large-$N$.
Hence there is a non-compact real version $U(1)_R$, of the complexification $U(1)_C$, of one of the aforementioned $U(1)$ \footnote{ It is a $U(1)$ in the Cartan subgroup of $O(4)$.}
that acts by rescaling $\lambda$ in such a way that the v.e.v. of
twistor Wilson loops is invariant under the aforementioned action. \par
As a consequence we show that twistor Wilson loops in pure $YM$ are localized on the sheaves, defined by the change of variables from the gauge connection to the $ASD$ part of its curvature in the functional integral, fixed by the action of $U(1)_R$. In addition we show that there is a dense set in function space in a neighborhood of the fixed sheaves \footnote{ We refer to the support of the fixed measure as fixed sheaves to imply not any manifold structure for such a locus. However, for a dense set in function space the fixed sheaves at large-$N$ are in fact a manifold that is parameterized by the disjoint union of moduli 
of local systems with fixed conjugacy class of the holonomy of the twistor connection. We refer to fixed points instead when the fixed locus has no moduli and it is a set of disconnected points.}, that at large-$N$ is classified by local systems on a  sphere with a very large number of punctures and with fixed conjugacy class of the holonomy of the twistor connection around the lattice of punctures, with values in the complexification,
$SU(N)_C$, of the gauge group, modulo the global action of $SU(N)_C$. \par
In the physics terminology the local systems are lattices of surface operators satisfying the self-duality ($SD$) equations with singularities:
\bea
F_{\alpha \beta}^- = 
\sum_p \mu^{-}_{\alpha \beta}(p)  \delta^{(2)} (z-z_p)
\eea
In addition the v.e.v. of the aforementioned twistor Wilson loops in the adjoint representation factorizes in the large-$N$ limit in the product of the v.e.v. in the fundamental and conjugate representation.
Then to each factor the following argument applies. \par
On the dense set described by local systems, by translational invariance we can assume that all the conjugacy classes of the holonomies are a copy of the same adjoint orbit, and that the orbit for a holonomy
around one arbitrarily chosen point can be put by the global action of $SU(N)_C$ in canonical form, i.e. either in diagonal or in Jordan form.
Now the global compact $SU(N)$ gauge group acts on such diagonal or Jordan holonomy by conjugation. \par
If the global gauge group is unbroken, as it is believed to be the case for pure $YM$, only the holonomies that are fixed by the entire $SU(N)$ may occur at large-$N$. 
Thus these holonomies  at a preferred point are in fact valued in the center of the gauge group and their orbits reduce to points. But then by translational invariance all the orbits reduce to the center
\footnote{The same conclusion is reached by an inductive argument on the holonomies around each point, without assuming translational invariance, since once the holonomy around a point is shown to
be in the center by the assumption of unbroken gauge group, the global $SU(N)_C$ still acts on the holonomies around each of the remaining points.}.  \par
Besides we show that there is a homological explanation for this
localization on fixed points based on a new localization theory of the loop equation for twistor Wilson loops, such that the actual fixed points that contribute to
the twistor Wilson loops are the critical points of a certain effective action determined by the loop equation \cite{MB1}. The localization by homology of the loop equation, i.e. its reduction to a critical equation, is obtained deforming the loop by adding vanishing boundaries that are backtracking arcs ending with the cusps of the local system \cite{MB1}, an operation allowed by the large-$N$ triviality of twistor Wilson loops, 
by dualizing the idea of deforming a closed form by a coboundary in the cohomological interpretation of the  Duistermaat-Heckman localization. \par
In order to get localization, the first main technical innovation of our approach is a reformulation of the $YM$ theory in terms of a change of variables that in the $\cal{N}$ $=1$ $SUSY$ $YM$ theory has been known as the Nicolai map \cite{Nic1, Nic2}.
The Nicolai map in the $\cal{N}$ $=1$ $SUSY$ $YM$ theory was worked out by De Alfaro, Fubini, Furlan and Veneziano \cite{V, V1} as a change of variables from the gauge connection to the anti-selfdual ($ASD$) part of the gauge curvature, that needs a gauge fixing to be locally invertible, with the property that the Jacobian of the map cancels precisely the fermion determinant in the light-cone gauge. \par
As a preparatory exercise, the Nicolai map allows us to introduce localization also in the pure $\cal{N}$ $=1$ $SUSY$ $YM$ theory \footnote{Witten already observed in his paper \cite{W1} on Donaldson invariants that although most naturally formulated in the $\cal{N}$
$=2$ $SUSY$ theory localization could be extended to certain theories with only $\cal{N}$ $=1$ $SUSY$, called $\cal{N}$ $=2^*$ theories that involve anyway the occurrence of a scalar field.
In pure $\cal{N}$ $=1$ $SUSY$ gauge theory Witten localization does not apply directly since there are no scalars. It is always possible to give the scalars of the $\cal{N}$ $=2^*$ $SUSY$ theories
large masses in order to obtain at low energy pure $\cal{N}$ $=1$ $SUSY$ $YM$ theory. This leads to the modern "weak coupling" approach to the computation of the gluino condensate \cite{Rev, Bian}.} by means of the tautological Parisi-Sourlas
supersymmetry associated to the cancellation of the Jacobian with the fermion determinant \cite{Parisi1, Parisi2}. While it has been known for some time that the Nicolai map can be associated to cohomological localization
\footnote{In the lectures \cite{Dijk} it is worked out the zero dimensional case of the Nicolai map and it is shown indeed that coincides with localization.}, the localization by the Nicolai map has never been worked
out in asymptotically free gauge theories because of the following difficulty.
Naively the Nicolai map maps $\cal{N}$ $=1,2$ $SUSY$ $YM$ into a free theory, that cannot hold true literally. The question arises for example how to reproduce the $NSVZ$ beta function \cite{NSVZ, Shif} by means of the Nicolai map.
Our simple but key observation is that the cancellation of determinants occurs only up to zero modes. Therefore the divergences associated to the Pauli-Villars regulator of the zero modes occur. \par 
In fact understanding how the $NSVZ$ beta function occurs by cohomological localization via the Nicolai map in this paper is only an exercise for understanding localization of the aforementioned twistor Wilson loops in large-$N$ pure $YM$ in the $ASD$ variables. The crucial point is that the localization on the fixed points of the $U(1)_R$ action can be obtained only in the $ASD$ variables. \par
The second main technical innovation consists in interpreting our non-$SUSY$ version of the Nicolai map in the pure $YM$ theory
\cite{MB1,MB2,MB3} as hyper-Kahler reduction \cite{H,HKL} on a dense set in function space, that corresponds to a lattice of
surface operators in the physics terminology. This is an analytical and differential geometric construction that does not need any supersymmetry. It reduces the $YM$ functional integral to a finite dimensional integral with respect to a product measure 
on a lattice and it is the analog of the first part of Nekrasov computation in the supersymmetric case. \par
The physics interpretation is that the localization of the twistor Wilson loops in the large-$N$ $YM$ theory is described in terms of variables that are of purely magnetic type, realizing, in the technical sense of localization of twistor Wilson loops, a new version of 't Hooft long-standing ideas \footnote{In addition to 't Hooft original papers \cite{Super1, Super2, Super3} see also \cite{DW} for a very neat account of 't Hooft duality.} on the $YM$ vacuum as a dual superconductor \cite{Super1, Super2, Super3}. \par
In particular 't Hooft duality in $YM$ theories with fields in the adjoint representation requires that, if the theory has a mass gap, then either the $Z_N$ magnetic charges condense (confining phase) or the  $Z_N$ electric charges condense (Higgs phase). Localization by homology of twistor Wilson loops in pure $YM$ realizes the first alternative, in which the electric charge is unbroken \footnote{The adjoint action of the global gauge group leaves
invariant the center, $Z_N$.} and the magnetic charge is broken in superselection sectors labelled by $k$, the magnetic charge at a (lattice) point, that are degenerate \footnote{The classical action scales as $k$ times the square density of surface operators, $\rho^2$, and the renormalized square density scales as $\frac{1}{k}$. } for the large-$N$ renormalized effective action that occurs in the holomorphic loop equation. From the localized renormalized effective action restricted to fluctuations supported on the aforementioned Lagrangian submanifold it follows also that in each sector there is a mass gap proportional to $\sqrt k$ in units of the common $RG$-invariant scale. \par
Yet, we should stress that
localization is by no means a universal concept, but it applies only to special observables. Therefore, it would be completely wrong to employ surface operators of $Z_N$ holonomy to compute general observables of the $YM$ theory, as it would be completely wrong to employ instantons to compute anything but the gluino condensate in $\mathcal{N}$ $=1$ $SUSY$ $YM$.
 \par
The mathematics interpretation is that we are in fact representing the $YM$ functional integral as an adelic integral over (the moduli space of) local systems. \par
The two aforementioned technical innovations are crucial for our twofold approach to localization in pure $YM$. \par
Firstly, as we just explained, following the spirit of the Duistermaat-Heckman idea our new kind of localization in the large-$N$ pure $YM$ theory involves the action
of a semigroup fixing the v.e.v. of twistor Wilson loops and contracting the support of the functional $YM$ measure, resolved into $ASD$ orbits
by our non-supersymmetric version of the Nicolai map.
At technical level the hyper-Kahler reduction to surface operators furnishes a structure theory of the locus of the fixed-points. \par
Secondly, as well as the Duistermaat-Heckman localization on fixed points has a cohomological explanation, so the new localization on fixed points in pure $YM$ theory
has a homological explanation. 
Indeed there exits a new holomorphic loop equation for twistor Wilson loops that can be localized, i.e. reduced to a critical equation, by deformations
of the loop that are vanishing boundaries (backtracking arcs) in homology, in the dual sense to which a cohomology class represented by an integral of the exponential of
a closed form can be localized by deformations that are coboundaries in cohomology.  At technical level the localization of the holomorphic loop equation for twistor Wilson loops requires that
the backtracking arcs end with cusps supported on the singular divisor of the surface operators. \par
Our new holomorphic loop equation for twistor Wilson loops is derived using the standard technique of the celebrated loop equation of Makeenko and Migdal  \cite{MM,MM1} invented long ago, but 
the crucial difference is that the integration variable that gives origin to the loop equation in our case is not the gauge connection but instead the $ASD$ field of our non-$SUSY$ version of the
Nicolai map, in a holomorphic gauge defined by a further change of variables. The resulting loop equation resembles for the cognoscenti the holomorphic loop equation of Dijkgraaf and Vafa \cite{Vafa, Laz, Kaw} for the holomorphic chiral ring of $\cal{N}$ $=1$ $SUSY$ gauge theories \cite{WD}. \par
The homological localization of the holomorphic loop equation completes the analogy with Nekrasov computation. As well as the prepotential, i.e. the effective action in the low energy sector as a function of the
condensates of the $\cal{N}$ $=2$ $SUSY$ theory, is computed by cohomological localization of $1$, so the large-$N$ effective action of the $YM$ theory in the twistor sector, as a function of the condensates of surface
operators of $Z_N$ holonomy, is computed
by homological localization of $1$, the trivial twistor Wilson loops.
\par
The twistor sector is defined by correlation functions obtained by holomorphic/antiholomorphic fusion a la Cecotti-Vafa \cite{CV,CV1} of the holomorphic/antiholomorphic $ASD$ curvature of surface operators restricted to the aforementioned Lagrangian submanifold of twistor space of complexified space-time with Minkowski signature. \par
On the ultraviolet side, a striking result that follows from the localization on fixed points is that the large-$N$ beta function
for the Wilsonian coupling constant in the $ASD$ variables is one-loop exact, because for twistor Wilson loops, precisely because of the localization, a certain kind of saddle-point
approximation turns out to be exact.  Thus the quantum corrections for these observables are completely accounted by functional determinants whose diagrammatic
expansions contains only one-loop Feynman graphs and possibly the logarithm of the powers of the Pauli-Villars regulator of zero modes. 
Thus in the large-$N$ pure $YM$ theory Eq.(1.1) can be replaced by the much simpler:
\bea
\Lambda_{W}= \Lambda \exp(-\frac{1}{2\beta_0 g_W^2}) 
\eea
where $\Lambda_{W}$ is the $RG$ invariant scale in the Wilsonian scheme in the $ASD$ variables. \par 
At the same time we show that, in the regularization scheme of the homological localization of the holomorphic loop equation, the one-loop exactness for the Wilsonian beta function implies a large-$N$ exact beta function for the canonical coupling of $NSVZ$ type, that reproduces the one- and two-loop perturbative universal coefficients. 
In this scheme the large-$N$ canonical beta function of the pure $YM$ theory in the $ASD$ variables is given by Eq.(1.14). \par
We should stress that the computation of the canonical beta function depends crucially on exploiting the gluing rules for functional integrals in the specific case of the localization on local
systems. \par In particular we show that these gluing rules coincide with the ones of topological strings via the gluing of an associated arc complex \cite{Pen,Pen1,Pen2,Man,Man1}.
Indeed the homological localization is based on "the most local part" of the homology of the essential arc complex of a punctured sphere. \par
The requirement that the homology be essential, i.e. the exclusion of arcs
that can be deformed to a puncture, rules out the local relative homology of compact support around a puncture. In fact "the most local part" of the essential homology that is relevant for us is the essential  homology of the
arc complex with no polygons, that involves only links ending with two different cusps, one in the divisor at the ultraviolet and one in the divisor at the infrared. Naively the Wilsonian beta function in our scheme is a purely ultraviolet concept, and therefore does not distinguish between the local homology and the essential homology of links. Yet, the canonical beta function, that in our scheme involves infrared physics too, does.\par
In addition, while fixed point arguments suffice to display localization directly in large-$N$ $YM$ theory by the change to the $ASD$ variables in the functional integral and the dense hyper-Kahler resolution on local systems, the corresponding effective action, i.e. the logarithm of the density of the localized measure, has intrinsic finite holomorphic ambiguities due to the freedom of making holomorphic changes of variables and possibly holomorphic anomalies at loci where the holomorphic change of variables may develop
singularities. We are able to fix these holomorphic ambiguities only using the localization by homology of the loop equation, via the choice of the holomorphic gauge that is necessary to write down the holomorphic
loop equation. \par
On the infrared side, the holomorphic gauge in the loop equation is essential, because the mass gap and the glueball spectrum occur
precisely because of a non-trivial Jacobian from the unitary to the holomorphic gauge in the effective action. \par 
We end this introduction with some loose heuristic considerations as to why the line of thought of this paper may be able to overcome the main difficulties of the ultraviolet and infrared problem of $YM$
in the restricted sense specified above. \par
While the mass gap problem as formulated in full generality for every correlation function and for every compact gauge group in \cite{AJ} appears presently almost hopeless in our opinion, the program of solving the $SU(N)$ $YM$ theory in the large-$N$ limit has attracted
considerable attention and efforts. \par
A promising avenue is to find an equivalent string theory \cite{Po1, Po2} by effectively resumming 't Hooft perturbative double expansion in powers of $g$ and $N^{-1}$ \cite{Hooft}. In this string theory the v.e.v. of any Wilson loop of the $YM$ theory
in the large-$N$ limit would be computed by a string diagram that is a disk with the loop as boundary. No other interesting observables, but the $RG$-invariant condensates, exist at the leading large-$N$ order
because of the factorization of the v.e.v. of local normalized gauge invariant operators. \par
Now, on the field theory side, the knowledge of all the Wilson loops that would be implied
by the string solution contains a vast information in a mathematical sense. Indeed it has been known for some time that the ambient algebra of the master field \cite{W3} that solves the large-$N$ Makeenko-Migdal loop equation 
for ordinary Wilson loops \cite{MM,MM1} is the Cuntz algebra with four generators \cite{Haa1, Haa2, Cv, Si, Douglas, Gross, Douglas1}
whose Fock space representation is known to be of type $II_1$ but not hyperfinite \cite{Voiculescu}, i.e. not the weak limit of matrix algebras. Indeed such Fock representation is isomorphic to a free group factor with the same number of generators, which is the main explicit example of the "elusive" type $II_1$ non-hyperfinite factors \cite{Voiculescu}. It is clear \cite{MB3} that obtaining the relevant non-hyperfinite information would be extremely difficult in case the von Neumann algebra generated by the actual solution shares with the ambient algebra the non-hyperfinite character.  \par
On the contrary, to the next to leading $\frac{1}{N}$ order, the connected two-points correlation functions of local gauge invariant operators are conjectured to be
the most simple as possible: a sum of an infinite number of propagators of free fields \cite{Pol}, saturating the logarithms of short distance perturbation theory \cite{Narison}:
\bea
\int < \frac{1}{N} \sum_{\alpha \beta} Tr F_{\alpha \beta}^2(x) \sum_{\alpha \beta} \frac{1}{N} Tr F_{\alpha \beta}^2(0)>_{conn}
\ e^{ip x} d^4x= 
\sum_r \frac{Z_r}{p^2+M_r^2}  \sim g^4(p) p^4 \log(\frac{p^2}{\mu^2}) 
\eea
\par
Now any string solution, as it is usually meant, cannot avoid to solve the leading order problem for the Wilson loops, in order to solve the much simpler looking
subleading problem for the free glueball spectrum. This makes such a general, large-$N$ exact, string solution very difficult in our opinion.
We may wonder as to whether we can solve the easy looking subleading problem for the free glueball spectrum avoiding and thus loosing the information about the hard looking problem for the
Wilson loops. Our answer is positive to a certain extent: we construct trivial Wilson loops, the twistor Wilson loops indeed, whose v.e.v. is $1$ in the leading
large-$N$ limit. However, they admit non-trivial $\frac{1}{N}$ corrections and thus morally they couple to a certain non-trivial sector of the large-$N$ theory. \par
By the way, in this restricted sense we believe that there is also an explicitly solvable string theory, that captures the sector of $YM$ accessible to the twistor Wilson loops
defined in this paper. The outlook for this twistor string is described in the conclusions. \par
Coming back to the field theoretical framework, we use our localization theory to localize the twistor Wilson loops, that are in the "homology of $1$", precisely in the dual sense to which Nekrasov
localized the "cohomology of $1$" to get the Seiberg-Witten prepotential. Indeed, although the prepotential is obtained by a "trivial" cohomology, it allows one to reconstruct the low energy effective
theory. Precisely in the same sense, since twistor Wilson loops live in a "trivial" homology, they can be localized by suitable deformations.
Yet, the interesting information is contained in the effective action,
i.e. in the localized measure. The quadratic small fluctuations of the effective action around the localized loci of the measure furnish the glueball spectrum in the twistor sector. \par

\section{Synopsis}

We summarize here the main technical arguments in a logic order and the main results.  \par
In the prologue we describe in some detail the computation of the beta function in the following cases. 
The one-loop beta function in $YM$ by the usual background field method.
The $NSVZ$ beta function in $\cal{N}$ $=1$ $SUSY$ $YM$ by cohomological localization in the Nicolai variables.
The one-loop $YM$ beta function for the Wilsonian coupling in the $ASD$ variables by the usual background field method. \par
These concrete examples are used to furnish a comparison with the computation of the $YM$ beta function by our new localization. 
This section contains many definitions and computational technicalities that are referred to throughout the whole paper. \par
We define also the change of variables from the gauge connection to the $ASD$ curvature in the pure $YM$ case.
In particular we show that, since in pure $YM$ the Jacobian to the $ASD$ variables is not cancelled, as opposed to the $\cal{N}$ $=1$ $SUSY$ $YM$ case in the light-cone gauge,
a multiplicative $Z$ renormalization of the $ASD$ field occurs. In sect.(12) this $Z$ factor is related to the anomalous dimensions of a large class of composite operators that occur
as scalar polynomials in the $ASD$ curvature in the one-loop integrable sector of large-$N$ $YM$.
\par
In sect.(4) we define twistor Wilson loops in non-commutative gauge theories. The twistor Wilson loops are defined on a non-commutative deformation of space-time, that is used as a tool to define the large-$N$ limit much in the way Nekrasov used a non-commutative deformation as a tool to regularize the instantons moduli space. We recall some features of non-commutative gauge theories that we employ in the following sections. \par
We display the following properties of twistor Wilson loops. 
The v.e.v. of twistor Wilson loops is
fiber independent and trivially $1$ in the large-$N$ limit. 
In addition twistor Wilson loops are supported on Lagrangian submanifolds of twistor space. \par
 In sect.(5) we show that the curvature of the twistor connection is of purely $ASD$ type and we describe the localization in large-$N$ pure $YM$ theory of twistor Wilson loops on the fixed sheaves of a semigroup acting on the fiber of the Lagrangian fibration which twistor loops are supported on and contracting the support of the functional measure
in the $ASD$ variables. \par
We refer to this kind of localization as the quasi-localization lemma, since the resulting localized measure is still represented by a residual functional integration on a certain complex path, supported on distribution valued sheaves in fact, rather than by a sum over fixed points.  \par
The quasi-localization lemma is a purely formal computation that involves a quite disputable formal exchange of the order of
limit and integration. However, the exchange of order of limit and integration is justified in sect.(8) in the large-$N$ limit and for a lattice version of the Nicolai map sect.(7), that allows interpreting the Nicolai map as the hyper-Kahler reduction on a dense set in function space in the sense of distributions. \par
We write down the corresponding effective action, i.e. the logarithm of the density of the localized $YM$ measure.
Because of the residual complex integration the effective action has an ambiguity by holomorphic change of variables that we can solve only through the loop equation of sect.(6).
The fixed sheaves in the quasi-localization lemma are characterized by the vanishing of two of the three $ASD$ fields of the non-$SUSY$ Nicolai map.
\par
In sect.(6) we write the holomorphic loop equation for twistor Wilson loops. The holomorphic ambiguity of the effective action of sect.(5) is fixed by a change of variables to the holomorphic gauge, necessary to write down the holomorphic loop equation. 
It is precisely the Jacobian to the holomorphic gauge that generates the glueball potential. \par
This implies that the glueball potential in the holomorphic/antiholomorphic sector defined by twistor loops in the fundamental and conjugate representation must be singular at the fixed points, as it is indeed, for the theory to have a mass gap, since the contribution of the Jacobian to the effective action is formally the logarithm of the square of a holomorphic function.  \par
In sect.(7) we introduce a regularization of the large-$N$ functional integral by integrating on "infinite-dimensional local systems" \footnote{We write it in quotes because these infinite dimensional objects
admit unstable finite dimensional subbundles, thus violating a fundamental property of finite dimensional local systems.} on non-commutative space-time. \par
The idea of integrating on local systems associated to an arbitrary parabolic divisor appeared for the first time long ago in a paper by us \cite{MB2, MB3}, by embedding the Hitchin fibration \cite{H2,S3}
in the $YM$ functional integral, and physically corresponds to integrating over surface operators ante litteram \cite{W2}. 
We employ Morita equivalence \cite{Szabo2} to reduce to the case of ordinary space-time for finite rank bundles.
Thereafter we reconstruct the large-$N$ limit of $YM$ as an inductive limit on the finite rank local systems.  \par
Following the mathematical literature \cite{H,H2,S1,S2,Konno1,Konno2,S3,S4,S5,Moc} we discuss the topological, holomorphic and differential geometric features of the finite rank local systems \cite{S5,S6}.
As topological objects local systems are representations of the fundamental group of a punctured Riemann surface.
As holomorphic objects they are holomorphic connections with regular singularities.
As differential geometric objects they are parabolic harmonic bundles, i.e. parabolic Hitchin bundles \cite{H} equipped with a harmonic metric by a Hitchin-Kobayashi correspondence \cite{S4}.
Remarkably in our setting the harmonic bundles arise as the hyper-Kahler reduction \cite{H} induced by our version of the non-$SUSY$ Nicolai map. \par
Physically the hyper-Kahler reduction \cite{H} is a resolution dense in function space \cite{MB2}, in a neighborhood of the fixed sheaves, of the $ASD$ field \cite{MB3}  as a linear combination 
of two-dimensional delta distributions supported on a lattice of surface operators \cite{W2}. These are local systems that occur in the mathematics and physics literature for completely different reasons, among which we mention the Hitchin-Kobayashi correspondence \cite{S4}, non-abelian Hodge theory \cite{S1,S2}, twistor $D$-modules \cite{Moc}, and last but not least the physics version \cite{W2} of the geometric Langlands correspondence \cite{BD, Frenkel,F,F1,F2,F3,F4,F5}. \par
In sect.(8) we get our localization on fixed points. Indeed we combine the quasi-localization lemma of sect.(5) with the idea of sect.(7) of integrating on local systems to get localization on fixed points. Reducing to finite dimensional local systems needs Morita duality and is allowed implicitly by the triviality of twistor Wilson loops. \par
In fact the quasi-localization lemma depends on the aforementioned disputable formal exchange of the order of
limit and integration. This is justified by showing that, on the set described by the hyper-Kahler reduction, a gauge exists in which the lattice theory is locally abelian,
all the remaining non-abelian degrees of freedom being zero modes associated to the moduli of the local system. Therefore the fluctuation of the eigenvalues of the theory
are suppressed in the large-$N$ limit. Of course the integral of the limit
is the localized measure in the $ASD$ variables, while the limit of the integral is the original $YM$ measure on the gauge connections. \par
In particular the choice of the approximating sequence by finite dimensional  local systems (i.e. stable bundles) is essential for localization on fixed points. 
We show that the fixed manifold restricted to the dense hyper-Kahler locus of local systems is a Lagrangian submanifold of the moduli space of surface operators.
In addition we show that, assuming that the gauge group is unbroken, the fixed manifold is in fact a collection of fixed points represented by surface operators with $Z_N$ holonomy.
This localization is our analog of the Duistermaat-Heckman localization. \par
In sect.(9) we use it to compute the Wilsonian beta function of the large-$N$ $YM$ theory. \par
In sect.(10) we get homological localization of the holomorphic loop equation for the  twistor Wilson loops by means of the lattice version of the holomorphic loop equation obtained integrating over the local systems. \par
The triviality of twistor Wilson loops plays a key role here, since it allows arbitrary deformations of the twistor Wilson loops without changing their expectation value.
In this section the localization on fixed points in not a consequence of the assumption that the gauge group is unbroken, but a consequence of the reduction of
the loop equation to a critical equation. \par
The localization is obtained deforming the twistor Wilson loops by backtracking arcs ending with the cusps of the local system. This is the localization by homology
that combines the holomorphic loop equation of sect.(6) with the idea of integrating over local systems of sect.(7).
This is our analog of cohomological localization by deforming by a coboundary. \par
In particular we derive the glueball potential by computing the Jacobian to the holomorphic gauge of sect.(6) for the local systems of sect.(7). \par
At mathematical level homological localization involves the essential arc complex of a punctured sphere and a combinatorial model of the gluing of the arcs developed in the mathematical literature \cite{Pen,Pen1,Pen2}, in turn inspired by (topological) strings \cite{Man,Man1}. \par
The version of the localization via the loop equation is the one that has, in our opinion, more chances to hold in a strictly mathematical sense. The loop equation for twistor Wilson loops occurs as a formal Schwinger-Dyson or Ward identity derived imposing that the integral of a functional derivative vanishes in function space in the $ASD$ variables in the holomorphic gauge. \par
The left hand side of the loop equation contains the effective action
of the theory that implies a one-loop exact Wilsonian beta function. The right hand side is still divergent, but it can be regularized in a gauge invariant way by analytic continuation to Minkowski space-time
and, by deforming the loop, it can be made to vanish. Thus, despite the loop equation is obtained only as a formal identity (as the Makeenko-Migdal equation is), its solution is defined via its would-be properties,
essentially the fact that it allows analytic continuation to Minkowski space-time. \par
The glueball spectrum occurs only after analytic continuation to Minkowski space-time of the effective action renormalized in Euclidean space for fluctuations of surface operators restricted to the Lagrangian submanifold. \par
The basic idea is that once the large-$N$ localization is obtained for twistor Wilson loops, that are non-local extended objects,
the localized effective action is used to compute physical fluctuations of local operators restricted to certain channels. \par
In sect.(11) we compute the canonical beta function of large-$N$ $YM$ by means of our holomorphic loop equation restricted to the local systems of sect.(7) and we check agreement with the first two universal perturbative coefficients. The result depends crucially on the gluing rules for local systems. \par 
In sect.(12) we display our main result about the mass gap and the glueball spectrum using the effective action of sect.(6) together with its extension to the hyper-Kahler locus
in a neighborhood of the fixed points of sect.(7). \par
To do computations we employ the local model of the singular part of the connections with regular singularities around surface operators of sect.(7). We show that at the (renormalized) critical points the local model is in fact asymptotic for large $N$. We find a trajectory with mass squared exactly linear in $k$ and residues at the poles determined by the multiplicity of the eigenvalues of the $ASD$ curvature at the fixed points and by certain finite counterterms as a function of $k$ in a neighborhood of the fixed points. \par
We display in some detail how the glueball spectrum for the trajectory in the twistor sector follows from the effective
action. We also use our explicit solution to check the long-standing conjecture that the sum of pure poles in the large-$N$ limit on our trajectory saturates the logarithms that occur
in the glueball propagators in perturbation theory. \par
In the Wilsonian scheme we reproduce a factor of a logarithm that occurs in perturbation theory, that arises in our scheme by the $RG$-invariant spectral sum over the glueball. \par
In the perturbative canonical scheme the sum of a logarithm (with a coefficient that can be normalized to $1$) and of its square occurs. The coefficient of the square of the logarithm depends on the operator and 
is related to the anomalous dimension. \par
We observe that in the canonical scheme the multiplicative $Z$ renormalization, that occurs because of the Jacobian to the $ASD$ variables mentioned in sect.(3.4), implies through the localization of composite surface operators the same anomalous dimensions as for the operators associated to the ground state of the Hamiltonian spin chain in the thermodynamic limit, which is known to furnish the one-loop anomalous dimensions of long local gauge invariant operators in the one-loop integrable sector 
of large-$N$ $YM$. This sector is made by $SD$ or $ASD$ fields. \par
This cannot be the whole story, since we get from surface operators just one trajectory. 
In sect.(13) using by now standard results in mathematics \cite{S5,WW}, we extend the hyper-Kahler reduction induced by the non-$SUSY$ Nicolai map to twistor connections with wild singularities, i.e. poles of any order.
We suggest that such an extension corresponds physically to the more realistic case of an infinite family of Regge trajectories of increasing spins. We write the basic definition
of the functional integral on wild surface operators but explicit computations are left for the future. \par
In sect.(14) we summarize our conclusions and we outline some features of the twistor string conjectured to be dual to the $YM$ theory restricted to the sector defined by the twistor Wilson loops of this paper. \par

\section{Prologue}

\subsection{One-loop beta function of $YM$ by the background field method}

This computation is now completely standard, but since it is not easily found in textbooks in the form that we will need in the rest of the paper
we display it here in some detail \footnote{ We would like to thank Luca Lopez for working out a detailed version of this computation during our course at SNS.}. The basic philosophy is as in \cite{Pol}.
The partition function of pure $SU(N)$ $YM$ is:
\bea
Z=\int \delta  A \ e^{-S}
\eea
where:
\bea
S=\frac{N}{2g^2}\int d^4x \ tr_f (F_{\alpha \beta})^2=\frac{N}{4g^2}\int d^4x \ (F^a_{\alpha \beta})^2
\eea
The sum over repeated indices is understood.
The action has been rescaled by a factor of $N$ in such a way that the theory admits a non-trivial large-$N$ limit.
The coupling constant, $g$, is the 't Hooft coupling related to the $YM$ coupling, $g_{YM}$, by $g^2=g_{YM}^2 N$. The normalization of the action is appropriate for a gauge connection
in the fundamental representation of the Lie algebra of $SU(N)$:
\bea
A_{\alpha}=A_{\alpha}^a T^a
\eea
with the Hermitian generators in the fundamental representation normalized as:
\bea
tr_f(T^a T^b)=\frac{1}{2} \ \delta^{ab}
\eea 
The curvature of the $YM$ connection is:
\bea
&&[D_\alpha,  D_\beta] =i F_{\alpha \beta}\\
F_{\alpha \beta}&=&\partial_\alpha  A_\beta -\partial_\beta  A_\alpha +i[A_\alpha ,A_\beta ]\\
F_{\alpha \beta}^a&=&\partial_\alpha  A^a_\beta -\partial_\beta  A^a_\alpha -f^{abc}A_\alpha ^b A_\beta ^c
\eea
where $D_{\alpha}=\partial_{\alpha}+i A_{\alpha}$ is the covariant derivative.
To perform the one-loop computation of the effective action it is convenient to split the gauge connection into a classical background field and a fluctuating quantum field $A_\alpha =\bar{A}_\alpha +\delta A_\alpha $. The Fourier transform of the quantum field is supposed to be supported on momenta much larger than the momenta of the classical background field.
The gauge-fixing is performed by the Faddeev-Popov procedure. It is convenient to choose the Feynman gauge with respect to the background gauge field $\bar A_{\alpha}$:
\bea
\bar D_\alpha ^. \delta A_\alpha -C=0
\eea
where we denote by the dot the adjoint action in the Lie algebra:
\bea
&&D_\alpha (\bar{A})^.\delta A_\beta =\partial_\alpha \delta A_\beta +i[\bar{A}_\alpha ,\delta A_\beta ]\\
&&(D_\alpha (\bar{A})^.\delta A_\beta )^a=\partial_\alpha \delta A_\beta ^a-f^{abc}\bar{A}_\alpha ^b\delta A_\beta ^c=D_\alpha ^{ac}(\bar{A})\delta A_\beta ^c
\eea
where
\bea
D_\alpha ^{ac}(\bar{A})=\partial_\alpha \delta^{ac}-f^{abc}A_\alpha ^b \nonumber \\
=\partial_\alpha \delta^{ac}+iA_\alpha ^{ac} \nonumber
\eea
with $A_\alpha ^{ac} =if^{abc}  A_\alpha ^b $ and $[T^a,T^b] = if^{abc} T^c$.
$C$ is an auxiliary gaussian field whose covariance is chosen in such a way to cancel a longitudinal term in the $YM$ action quadratic in the fluctuating field,
by adding $ \frac{N}{g^2} \int d^4 x tr_f(\bar D_\alpha^.  \delta A_\alpha )^2 $ to the action. Quantities such as $\bar D_{\alpha}$ are evaluated at the background field $\bar A_{\alpha}$.
The gauge fixed partition function reads:
\bea
Z=\int \delta  A  \delta C \exp(-S_{YM}) Det(-\Delta_{\bar{A}}^.) \exp \big(-\frac{N}{g^2} \int d^4 x \ tr_f (C^2) \big)\delta(\bar D_\alpha^. \delta A_\alpha -C)
\eea
where we have inserted the Faddeev-Popov determinant of (minus) the Laplacian in the background field:
\bea
-\Delta_{\bar{A}}^.=-\Delta^2-i\partial_\alpha  \bar A_\alpha^. -2i \bar A_\alpha^. \partial_\alpha + \bar A_\alpha^. \bar A_\alpha^. 
\eea
As a consequence the gauge-fixed action is:
\bea
\int\frac{N}{2g^2} \ tr_f F_{\alpha \beta}^2(\bar{A}+\delta A)+\frac{N}{ g^2} \ tr_f(\bar D_\alpha^.  \delta A_\alpha )^2  d^4 x 
\eea
and the one-loop partition function reads:
\bea
Z_{1-loop}=e^{-\Gamma_{1-loop}(\bar{A})}
=e^{-S_{YM}(\bar{A})}Det^{-1/2}(-\Delta_{\bar{A}}^. \delta_{\alpha \beta}-2i \bar F_{\alpha \beta}^.) Det(-\Delta_{\bar{A}}^.)
\eea
where $\Gamma_{1-loop}(\bar{A})$ is the effective action for the background connection, $\bar{A}_\alpha $, to one-loop order and $\bar F_{\alpha \beta} ^.(...)=[\bar F_{\alpha \beta},..]$. \par
It is very instructive to understand the origin of the spin term, $-2i \bar F_{\alpha \beta}^.$, in the first functional determinant of Eq.(3.14).
By the splitting of the connection into $A_\alpha =\bar{A}_\alpha +\delta A_\alpha $ the curvature decomposes as follows:
\bea
F_{\alpha \beta}(\bar{A}+\delta A)=F_{\alpha \beta}(\bar{A})+D_\alpha (\bar{A})^. \delta A_\beta -D_\beta (\bar{A}) ^.\delta A_\alpha +i[\delta A_\alpha ,\delta A_\beta ]
\eea
Performing the square we keep only up to the quadratic terms in $\delta A_{\alpha}$ in the action, since we are doing a one-loop computation, we understand integration on space-time in the following
and we freely integrate by parts. 
We use the equation of motion, $D_\alpha (\bar{A})^.F_{\alpha \beta}(\bar{A})=0$, to eliminate the linear term in the action. Therefore we get:
\bea
&&F_{\alpha \beta}^2(\bar{A}+\delta A)=F_{\alpha \beta}^2(\bar{A})+(D_\alpha (\bar{A})^.\delta A_\beta -D_\beta (\bar{A}) ^.\delta A_\alpha )^2+2iF_{\alpha \beta}(\bar{A})[\delta A_\alpha ,\delta A_\beta ] \nonumber \\
&&=F_{\alpha \beta}^2(\bar{A})+2(D_\alpha (\bar{A})^.\delta A_\beta )^2-2D_\alpha (\bar{A})^.\delta A_\beta  D_\beta (\bar{A})^.\delta A_\alpha +2iF_{\alpha \beta}(\bar{A})[\delta A_\alpha ,\delta A_\beta ]
\eea
Using
\bea
&&D_\alpha  D_\beta =D_\beta  D_\alpha +i F_{\alpha \beta}\\
&&tr_f(\delta A_\beta [F_{\alpha \beta},\delta A_\alpha ])=tr_f(F_{\alpha \beta}[\delta A_\alpha ,\delta A_\beta ])=-tr_f(\delta A_\alpha [F_{\alpha \beta},\delta A_\beta ])
\eea
the quadratic form in Eq.(3.16) becomes:
\bea
tr_f((D_\alpha ^.\delta A_\beta -D_\beta ^.\delta A_\alpha )^2+2iF_{\alpha \beta}[\delta A_\alpha ,\delta A_\beta ] )\nonumber \\
=tr_f(-2\delta A_\alpha \Delta_{{A}}^. \delta_{\alpha \beta}\delta A_\beta +2\delta A_\beta  D_\alpha^.  D_\beta^.  \delta A_\alpha +2iF_{\alpha \beta}[\delta A_\alpha ,\delta A_\beta ] )\nonumber \\
=tr_f(-2\delta A_\alpha  \Delta_{{A}}^ .\delta_{\alpha \beta}\delta A_\beta +2\delta A_\beta  D_\beta^.  D_\alpha^. \delta A_\alpha +2i\delta A_\beta  F_{\alpha \beta}^.\delta A_\alpha +2iF_{\alpha \beta}[\delta A_\alpha ,\delta A_\beta ]) \nonumber \\
=tr_f(-2\delta A_\alpha  \Delta_{{A}}^.\delta_{\alpha \beta}\delta A_\beta -2 (D_\alpha^. \delta A_\alpha )^2-4i\delta A_\alpha  [F_{\alpha \beta},\delta A_\alpha ])
\eea
where we skip the label of the background field since no confusion can arise.
In the Feynman gauge the second term in the last line is cancelled by the gauge-fixing.
Finally the quadratic form written in components becomes:
\bea
tr_f \big(\delta A_\alpha (-2\Delta_A ^.\delta_{\alpha \beta}-4i \ adF_{\alpha \beta})\delta A_\beta \big) \nonumber \\
=tr_f \big( \delta A_\alpha ^a t^a (-2\Delta_{A}^.\delta_{\alpha \beta}-4i \ adF_{\alpha \beta})\delta A_\beta )^b t^b \big) \nonumber \\
=\delta A_\alpha ^a \big((-\Delta_{{A}}^.\delta_{\alpha \beta}-2i \ adF_{\alpha \beta})\delta A_\beta \big)^a \nonumber \\
=\delta A_\alpha ^a(-\Delta_{{A}} \delta_{\alpha \beta}-2i \ adF_{\alpha \beta})^{ac}\delta A_\beta ^c\nonumber \\
=\delta A_\alpha ^a \big(-(\Delta_{A})^{ac}\delta_{\alpha \beta}+2 f^{abc} F_{\alpha \beta}^b \big)\delta A^c_\beta 
\eea
where $ad F_{\alpha \beta}(...)=[F_{\alpha \beta},...]$, $(ad F_{\alpha \beta})^{ac}=if^{abc}F_{\alpha \beta}^b$ and 
\bea
(\Delta_{A})^{ac}=(D_\alpha )^{ad}(D_\alpha )^{dc}= \Delta^2\delta^{ac}+i\partial_\alpha  A_\alpha ^{ac}+2i A_\alpha ^{ac}\partial_\alpha -A_\alpha ^{ad}A_\alpha ^{dc}
\eea
QED. \par
The following identity holds:
\bea
Det^{-1/2}(-\Delta_{A}\delta_{\alpha \beta}-2i \ adF_{\alpha \beta})=Det^{-1/2}(-\Delta_{A}\delta_{\alpha \beta})Det^{-1/2}(1-2i(-\Delta_{A})^{-1} \ adF_{\alpha \beta})
\eea
The first factor gives:
\bea
Det^{-1/2}(-\Delta_{A}\delta_{\alpha \beta})=Det^{-2}(-\Delta_{A})
\eea
Therefore the one-loop effective action reads:
\bea
e^{-\Gamma_{1-loop}(A)}=e^{-S_{YM}(A)}Det^{-1/2}(1-2i(-\Delta_{A})^{-1} \ adF_{\alpha \beta}) Det^{-1}(-\Delta_{A})
\eea
The first determinant is the spin contribution while the second determinant is the orbital contribution.\\
We can factorize away a trivial infinite constant from the orbital contribution:
\bea
&&Det^{-1}(-\Delta_{A})=Det^{-1}(-\Delta^2-i\partial_\alpha  A_\alpha -2iA_\alpha \partial_\alpha +A_\alpha  A_\alpha )\nonumber \\
&&=Det^{-1}(-\Delta^2)Det^{-1} (1+(-\Delta^2)^{-1}(-i\partial_\alpha  A_\alpha -2iA_\alpha \partial_\alpha +A_\alpha  A_\alpha ))
\eea
where the operators occurring in Eq.(3.25) are now defined by Eq.(3.21).
Using
\bea
Det(1+M)=e^{Tr\log(1+M)}=e^{{Tr M}-Tr(M)^2/2+\ldots}
\eea
at the lowest non-trivial order we get:
\bea
&&Det^{-1}(-\Delta_{A})=Det^{-1}(-\Delta)\exp \big(-Tr((-\Delta)^{-1}(-i\partial_\alpha  A_\alpha -2iA_\alpha \partial_\alpha +A_\alpha  A_\alpha ) )\big) \nonumber \\
&& \exp \big(Tr((-\Delta)^{-1}(-i\partial_\alpha  A_\alpha -2iA_\alpha \partial_\alpha +A_\alpha  A_\alpha )(-\Delta)^{-1}(-i\partial_\alpha  A_\alpha -2iA_\alpha \partial_\alpha +A_\alpha  A_\alpha ))/2 \big)\nonumber \\
\eea
where the trace is over the space-time, the Lie algebra and the vector indices. 
The term $Det^{-1}(-\Delta)$ is an irrelevant constant while the Lie algebra trace of the term  linear in $A_\alpha $ vanishes.
The term $Tr((-\Delta)^{-1}A_\alpha  A_\alpha )$ is a quadratically divergent tadpole that cancels in any gauge invariant regularization scheme, since it would give rise to a mass counterterm for the gauge
connection. Therefore it can be ignored. 
There remains an interesting divergence:
\bea
Det^{-1}(-\Delta_{A})\sim\exp \big(Tr((-\Delta)^{-1}(i\partial_\alpha  A_\alpha +2iA_\alpha \partial_\alpha )(-\Delta)^{-1}(i\partial_\alpha  A_\alpha +2iA_\alpha \partial_\alpha ))/2\big)
\eea
that evaluated in momentum space leads to:
\bea
\exp\big(\frac{1}{2}\int\frac{d^4 k}{(2\pi)^4}\int\frac{d^4 p}{(2\pi)^4}tr(A_\alpha (-k)A_\beta (k))\frac{(2p_\alpha -k_\alpha )(2p_\beta -k_\beta )}{p^2(p-k)^2}\big)
\eea
where the trace $tr$ on the Lie algebra indices refers to the matrices defined in Eq.(3.21).
The logarithmically divergent part of the integral over $d^4p$ has to be transverse in such a way that
\bea
\label{toexp}
\int\frac{d^4 p}{(2\pi)^4}\frac{(2p_\alpha -k_\alpha )(2p_\beta -k_\beta )}{p^2(p-k)^2}=\Pi(k^2)(k^2\delta_{\alpha \beta}-k_\alpha  k_\beta )+\ldots
\eea
where the dots stand for the quadratically divergent part that can be ignored because of the aforementioned reasons.
Taking the trace over the vector indices one gets:
\bea
\int\frac{d^4 p}{(2\pi)^4}\frac{4p^2+k^2-4pk}{p^2(p-k)^2}=3k^2\Pi(k^2)+...
\eea
We are interested in extracting the logarithmic divergencies by expanding the denominator in powers of $k/p$ up to the appropriate order:
\bea
&&\int\frac{d^4 p}{(2\pi)^4}\frac{4p^2+k^2-4pk}{p^2(p-k)^2}=\int\frac{d^4 p}{(2\pi)^4}\frac{4p^2+k^2-4pk}{p^4(1+(k^2-2kp)/p^2)}
\nonumber \\
&&\sim\int\frac{d^4 p}{(2\pi)^4}\frac{-4k^2+k^2+8(pk)^2/p^2}{p^4}=-\int\frac{d^4 p}{(2\pi)^4}\frac{k^2}{p^4}=-\frac{2k^2}{(4\pi)^2}\log\frac{\Lambda}{\mu}=3k^2\Pi(k^2) \nonumber \\
\eea
where we have replaced
\bea 
p_\alpha  p_\beta \rightarrow \frac{1}{4} p^2 \delta_{\alpha \beta}
\eea
into the integral and similarly $(pk)^2\rightarrow p^2 k^2/4$ and we have regularized  
\bea
\int\frac{d^4 p}{p^4}=2\pi^2\int_0^\infty \frac{dp}{p}\rightarrow 2\pi^2\log\frac{\Lambda}{\mu}
\eea
Therefore: 
\bea
\Pi(k^2)=-\frac{2}{3(4\pi)^2} \ \log\frac{\Lambda}{\mu}                                        
\eea
Hence the orbital contribution to the beta function is:
\bea
&&Det^{-1}(-\Delta_{A})\sim \exp\big(-\frac{1}{3(4\pi)^2} \log\frac{\Lambda}{\mu} \int\frac{d^4 k}{(2\pi)^4} tr(A_\alpha (k^2\delta_{\alpha \beta}-k_\alpha  k_\beta )A_\beta)\big)\nonumber \\
&&=\exp\big(-\frac{1}{3(4\pi)^2}\log\frac{\Lambda}{\mu}\int\frac{d^4 k}{(2\pi)^4}A^{ac}_\alpha (k^2\delta_{\alpha \beta}-k_\alpha  k_\beta )A^{ca}_\beta \big)\nonumber \\
&&=\exp\big(-\frac{N}{3(4\pi)^2}  \log (\frac{\Lambda}{\mu})^2\frac{1}{4}\int d^4 x \ (F_{\alpha \beta}^a)^2\big)
\eea
where in the last step we used:
\bea
f^{abc}f^{abd}=N\delta^{cd}
\eea
and 
\bea
\frac{1}{4}\int d^4 x F_{\alpha \beta}^aF_{\alpha \beta}^a\sim\frac{1}{2}\int\frac{d^4 k}{(2\pi)^4} A^a_\alpha (-k)A^a_\beta (k)(k^2\delta_{\alpha \beta}-k_\alpha  k_\beta )
\eea
at quadratic order. \par
Now we have to compute the spin contribution to the effective action. Since $tr F_{\alpha \beta}=0$ up to quadratic order in $F_{\alpha \beta}$ we get:
\bea
Det^{-1/2}(1-2i(-\Delta_{A})^{-1} \ adF_{\alpha \beta})=\exp\big(-Tr((-\Delta_{A})^{-1}adF_{\alpha \beta}(-\Delta_{A})^{-1}adF_{\beta \alpha})\big)
\eea
At lowest order $(-\Delta_{A})\sim(-\Delta)$, therefore
\bea
Tr((-\Delta)^{-1}adF_{\alpha \beta}(-\Delta)^{-1}adF_{\beta \alpha})&=&\int d^4 x\int d^4 y tr(G(x-y)adF_{\alpha \beta}(y) G(y-x) adF_{\beta \alpha}(x))\nonumber \\
&=&-N\int d^4 x\int d^4 y G(x-y)^2 F^a_{\alpha \beta}(y) F^a_{\alpha \beta}(x)
\eea
where in coordinate space
\bea
G(x-y)=\frac{1}{4\pi^2(x-y)^2}
\eea
and
\bea
tr(adF_{\alpha \beta}adF_{\beta \alpha})=(adF_{\alpha \beta})^{ac}(ad F_{\beta \alpha})^{ca}=if^{abc}F_{\alpha \beta}^bif^{cda}F_{\beta \alpha}^d=-NF_{\alpha \beta}^cF_{\alpha \beta}^c
\eea
Assuming that the background field carries momenta much smaller than the fluctuating field we can expand $F_{\alpha \beta}(y)=F_{\alpha \beta}(x)+\ldots$ by Taylor series and keep the first term
since we are interested only in the divergent terms. Thus defining $z=x-y$
\bea
Tr((-\Delta)^{-1}adF_{\alpha \beta}(-\Delta)^{-1}adF_{\beta \alpha})&\sim&-\frac{N}{(4\pi^2)^2} \int \frac{d^4 z}{z^4} \int d^4 x (F^a_{\alpha \beta})^2\nonumber \\
&=&-\frac{2\pi^2N}{(4\pi^2)^2} \log\frac{\Lambda}{\mu}  \int d^4 x (F^a_{\alpha \beta})^2 \nonumber \\
&=&-\frac{4N}{(4\pi)^2}  \log(\frac{\Lambda}{\mu})^2 \frac{1}{4} \int d^4 x (F^a_{\alpha \beta})^2 
\eea
Therefore at this order the divergent part reads:
\bea
Det^{-1/2}(1-2i(-\Delta_{\bar{A}})^{-1} \ adF_{\alpha \beta})\sim\exp\big(\frac{4N}{(4\pi)^2} \log(\frac{\Lambda}{\mu})^2 \frac{1}{4}\int d^4 x (F^a_{\alpha \beta})^2 \big)
\eea
Finally the local part of the one-loop effective action reads:
\bea
\Gamma_{1-loop}&=&S_{YM}+(\frac{N}{3(4\pi)^2}-\frac{4N}{(4\pi)^2}) \log(\frac{\Lambda}{\mu})^2 \frac{1}{4}\int d^4 x (F^a_{\alpha \beta})^2 \nonumber \\
&=&\big(\frac{1}{g^2(\Lambda)}-\frac{11N}{3(4\pi)^2}\log(\frac{\Lambda}{\mu})^2\big) \frac{1}{4}\int d^4 x (F^a_{\alpha \beta})^2
\eea
Therefore the bare coupling constant, $g(\Lambda)$, renormalizes as:
\bea
\label{gbeta}
\frac{1}{2g^2(\Lambda)}=\frac{1}{2g^2(\mu)}+\frac{11}{3}\frac{1}{(4\pi)^2}\log\frac{\Lambda}{\mu}
\eea
or
\bea
g^2(\Lambda)=\frac{g^2(\mu)}{1+2\frac{11}{3(4\pi)^2} \ g^2(\mu)\log\frac{\Lambda}{\mu}}
\eea
that is the solution at one loop of the equation that defines the $\beta$ function:
\bea
&&\beta(g)=\frac{\partial g}{\partial\log\Lambda}=-\beta_0 g^3+... \nonumber \\
&&\beta_0=\frac{11}{3(4\pi)^2}
\eea
Eq. (\ref{gbeta}) can be also written as:
\bea
\Lambda \ e^{-\frac{1}{2\beta_0 g^2(\Lambda)}}=\mu \ e^{-\frac{1}{2\beta_0 g^2(\mu)}}
\eea
Thus the combination:
\bea
\Lambda_{YM}=\Lambda \ e^{-\frac{1}{2\beta_0 g^2(\Lambda)}}
\eea
is independent on the cutoff $\Lambda$ and it is a renormalization group invariant at one loop.

\subsection{A $SUSY$ interlude: cohomological localization by the Nicolai map in $\cal{N}$ $=1$ $SUSY$ $YM$ }

Shortly after Nicolai discovered \cite{Nic1, Nic2} that the vanishing of the vacuum energy in an unbroken 
supersymmetric theory implies the existence of a change of variables whose Jacobian formally sets the functional integral
in ultralocal form, De Alfaro, Fubini, Furlan and Veneziano \footnote{We would like to thank Gabriele Veneziano for several discussions about the Nicolai map over the years and at the GGI.} \cite{V, V1}
worked out explicitly the Nicolai map in the case of $\cal{N}$ $=1$ $SUSY$ $YM$. \par
They found that in this case the Nicolai map is actually the change of variables
from the gauge connection to the $ASD$ part of its curvature in the light-cone gauge, with the property that its Jacobian 
cancels the gluino determinant. \par
In this section we reconsider the Nicolai map of $\cal{N}$ $=1$ $SUSY$ $YM$ paying particular attention 
to the fact that, while generically in function space the aforementioned cancellation occurs exactly, in a renormalizable but not finite supersymmetric
quantum field theory such as $\cal{N}$ $=1$ $SUSY$ $YM$ there should exist loci in function
space where the cancellation occurs in fact only up to zero modes.  \par
Indeed if it were not so the theory would be in fact mapped into a theory of free fields with zero beta function. \par
It is quite clear that the Jacobian of the Nicolai map develops zero modes precisely at loci in function
space where the Nicolai map fails to be one-to-one. If these loci are characterized by moduli then there is a continuous
family of zero modes and the Pauli-Villars regularization of these zero modes in the functional integral furnishes in general
some contribution to the beta function of the theory, thus resolving the puzzle that the Nicolai map maps formally the theory into a theory
of Gaussian fields with vanishing beta function. \par
Understanding the distribution of these zero modes as a function of the Gaussian random field 
which the theory is generically mapped on is in fact a non-perturbative problem seemingly
as difficult as performing the functional integral in the original variables. \par
However, we point out in this section that, thanks to the tautological nilpotent Parisi-Sourlas \cite{Parisi1,Parisi2} $BRS$ symmetry \footnote{We may consider
this $BRS$ symmetry as a remnant of the supersymmetry after gauge-fixing.}
associated to the cancellation of the Jacobian with the gluino determinant, the partition function with the insertion
of certain $BRS$ invariant operators necessary to saturate the zero modes of the gluino determinant
is in fact localized by cohomological localization (sect.(1)) on those (Euclidean) instantons \footnote{We refer here to instantons as configurations satisfying the $SD$ equations,
without any implication about being defined on $S^4$. In fact in the present framework the instantons are naturally defined on $S^2 \times S^2$, because of the analytic continuation 
from Euclidean $(4,0)$ to ultra-hyperbolic $(2,2)$ signature, that can be handled by twistor techniques \cite{Mas} (see also the Appendix in \cite{W4}).}
that can be analytically continued
to ultrahyperbolic signature (this constraint arises because the cancellations due to the Nicolai map actually occur only in the light-cone gauge). \par
Thus because of the localization the occurrence of the zero modes for these special $BRS$-invariant observables can be understood semiclassicaly as they coincide with the moduli of the instantons. \par
The immediate consequence of this localization is an exact formula for the beta function of $\cal{N}$ $=1$ $SUSY$ $YM$,
that quite obviously turns out to be the $NSVZ$ beta function \cite{NSVZ,Shif}, by an almost verbatim reproduction of their original computation. \par
We can now start to work out the details. It turns out that to get localization we need only the information that the gluino determinant cancels the Jacobian 
of the Nicolai map. Firstly let us suppose that there are no zero modes.
Using the identity: 
\bea
Tr(F_{\alpha\beta}^2)=Tr(F_{\alpha\beta}^-)^2/2+Tr(F_{\alpha\beta}\tilde{F}_{\alpha\beta})
\eea
the partition function reads:
\bea
Z= \big[\int  \exp (-\frac{N16 \pi^2Q}{2 g^2}-\frac{1}{8 g^2}\int Tr(F_{\alpha\beta}^- F^{-\alpha\beta})d^4 x) Det( \frac {\delta F_{\alpha\beta}^-}{ \delta A_{\gamma}} )\delta A \big]_{A_+=0} 
\eea
where fields live in the adjoint representation of the Lie algebra of $SU(N)$. 
In Eq.(3.52) we have just expressed the existence of the Nicolai map by inserting its inverse Jacobian, $Det( \frac {\delta F_{\alpha\beta}^-}{ \delta A_{\gamma}} )$,
in place of the gluino determinant.
Thus the theory is mapped into a theory of free fields:
\bea
Z= \int  \exp (-\frac{N16 \pi^2Q}{2 g^2}-\frac{1}{8 g^2}\int Tr(F_{\alpha\beta}^- F^{-\alpha\beta})d^4 x)  \delta F_{\alpha\beta}^- 
\eea
that holds generically in function space where zero modes do not occur. In fact, taking into account the zero modes, we get:
\bea
Z= \int   \Lambda^{n_b [ F_{\alpha\beta}^- ]-\frac{n_f}{2} [F_{\alpha\beta}^- ]}  \int_{\cal M} \frac{Pf <\frac{\delta A[F_{\alpha\beta}^-]}{\delta m} ,\frac{\delta A[F_{\alpha\beta}^-]}{\delta m}>}{Pf< \eta[F_{\alpha\beta}^-] ,\eta[F_{\alpha\beta}^-]>} \nonumber \\
\exp (-\frac{N16 \pi^2Q[F_{\alpha\beta}^-]}{2 g^2}-\frac{1}{8 g^2}\int Tr(F_{\alpha\beta}^- F^{-\alpha\beta})d^4 x)  \delta F_{\alpha\beta}^-
\eea
where the extra factor is the contribution of the Pfaffians of the bosonic and fermionic zero modes \cite{Peter} and of the associated Pauli-Villars regulator, whose origin is explained below.
Going back to Eq.(3.52) we can write it in a more suggestive form introducing anticommuting fields, ($\rho_{\alpha \beta},\eta_{\gamma}$):
\bea
Z= \big[\int  \exp (-\frac{N16 \pi^2Q}{2 g^2}-\frac{1}{8 g^2}\int d^4 x Tr(F_{\alpha\beta}^- F^{-\alpha\beta} + \rho_{\alpha \beta} \frac {\delta F_{\alpha\beta}^-}{ \delta A_{\gamma}}  \eta_{\gamma} ))\delta A \delta \rho \delta \eta\big]_{A_+=0}
\eea
The non-topological term can be rewritten as:
\bea
\big[\exp \int( -\frac{1}{8 g^2}\int d^4 x Tr(E_{\alpha\beta}^- E^{-\alpha\beta}+i E_{\alpha\beta}^-F^{-\alpha\beta} -i \rho_{\alpha \beta} \frac {\delta F_{\alpha\beta}^-}{ \delta A_{\gamma}}  \eta_{\gamma} ))\delta E^- \delta A \delta \rho \delta \eta\big]_{A_+=0}
\eea
The functional integral here has to be interpreted as either in Minkowski space-time $(3,1)$ or in ultrahyperbolic signature $(2,2)$,
since otherwise the light-cone gauge does not exist. In Minkowski signature an overall factor of $i$ in front of the action is understood but not explicitly displayed. In ultrahyperbolic signature the Gaussian integral is defined by analytic continuation. In this form the partition function enjoys the following tautological Parisi-Sourlas $BRS$ symmetry \footnote{To the best of our knowledge the existence of this symmetry in $\mathcal{N}$ $=1$
$SUSY$ $YM$ has never been related to cohomological localization, presumably because it leads to results inconsistent with the $NSVZ$ beta function if zero modes are not taken into account properly. However, the zero dimensional version of the Nicolai map, for which of course no zero modes occur, has been related to cohomological localization in \cite{Dijk}.}:
\bea
Q_{BRS} A_{\gamma} = \eta_{\gamma} \nonumber \\
Q_{BRS} \eta_{\gamma} = 0 \nonumber \\
Q_{BRS} \rho_{\alpha \beta}= E^-_{\alpha \beta} \nonumber \\
Q_{BRS} E^-_{\alpha \beta} = 0
\eea
with $Q_{BRS}^2=0$. The consequence of the existence of this symmetry is that the term $E_{\alpha\beta}^- E^{-\alpha\beta}$ is a coboundary, since $E_{\alpha\beta}^- E^{-\alpha\beta}=
Q(\rho_{\alpha \beta}E^{-\alpha\beta})$. Thus it can be cancelled without changing the cohomology class of the integrand. The resulting functional integral reduces to:
\bea
\big[\exp \int( -\frac{1}{8 g^2}\int d^4 x Tr(i E_{\alpha\beta}^-F^{-\alpha\beta} -i \rho_{\alpha \beta} \frac {\delta F_{\alpha\beta}^-}{ \delta A_{\gamma}}  \eta_{\gamma} ))\delta E^- \delta A \delta \rho \delta \eta\big]_{A_+=0}
\eea
Thus the complete partition function reads:
\bea
Z= \big[\int  \exp (-\frac{N16 \pi^2Q}{2 g^2}) \delta(F_{\alpha\beta}^-)  Det( \frac {\delta F_{\alpha\beta}^-}{ \delta A_{\gamma}} )\delta A \big]_{A_+=0}
\eea
that expresses the fact that the partition function is localized on those instantons that can be analytically continued to Minkoswki \footnote{ In this case the gauge group must be complexified since $SD$ equations exist only in Euclidean or ultrahyperbolic signature
for Hermitian connections \cite{Mas,W4}. } or ultrahyperbolic signature, thus violating the assumption that there are no zero modes.
In case we assume the existence of zero modes from the start we have to insert some fermionic contribution to take into account the fermionic zero modes of the gluino determinant.
This can be done by identifying lexicographically the gluino zero modes, say $\lambda_i$ where $i$ a spinor index, with $\eta_i$ where $i$ is a vector index:
\bea
\big[\int \Lambda^{-\frac{n_f}{2}}  (Tr\epsilon_{ij} \eta_i \eta_j)^{\frac{n_f}{2}}\exp (-\frac{N16 \pi^2Q}{2 g^2}) \nonumber \\
\exp[-\frac{1}{8 g^2}\int d^4 x Tr(E_{\alpha\beta}^- E^{-\alpha\beta}+i E_{\alpha\beta}^-F^{-\alpha\beta} -i \rho_{\alpha \beta} \frac {\delta F_{\alpha\beta}^-}{ \delta A_{\gamma}}  \eta_{\gamma} ))\delta E^- \delta A \delta \rho \delta \eta\big]_{A_+=0}
\eea
Afterwards everything goes through as before and the localized partition function is:
\bea
Z=\exp (-\frac{N16 \pi^2Q(A)}{2 g_W^2}) \Lambda^{n_b-\frac{n_f}{2}}  \int_{{\cal M}_Q} \frac{Pf <\frac{\delta A}{\delta m} ,\frac{\delta A}{\delta m}>}{Pf< \eta ,\eta>}
\eea
where we have explicitly displayed the Pfaffians (i.e. square root of determinants) that occur evaluating the residual integral on the instantons moduli associated to gauge field and gluino zero modes. 
This is precisely the $NSVZ$ result originally found by evaluating the gluino condensate, but for the fact that there is a constraint of analytic continuation for the instantons, say to  
ultrahyperbolic signature, that can be handled by twistor techniques \cite{Mas}. In particular the natural framework for the analytic continuation is to start with twistors on Euclidean $S^2 \times S^2$ and
analytically continue to ultrahyperbolic signature. Doing so the conformal compactification of ultrahyperbolic space-time occurs, that is $S^2 \times S^2/ Z_2$, where $Z_2$ acts by antipodal involution
$\sigma (x,y)=(-x,-y)$ \cite{Mas}. Thus the analytic continuation defines in fact the double cover, $S^2 \times S^2$, of the conformal compactification, $S^2 \times S^2/ Z_2$, of ultrahyperbolic space-time.
As a consequence the possible values of the second Chern class, $Q$, (the topological charge) can only be even for ultrahyperbolic instantons \cite{Mas}.
Thus the gluino condensate cannot be saturated by single instantons. This is perhaps related to the old controversy about the strong versus weak coupling evaluation of the condensate (for reviews see \cite{Rev,Bian}), but it is a matter too far away from our main subject to discuss further. \par
In any case the $NSVZ$ beta function is tautologically reproduced. We show the computation because it is very useful to understand the pure $YM$ case. \par
A striking consequence of the localization on instantons is that the beta function for the Wilsonian coupling constant, $g_W$, is one-loop exact since the only sources of divergences are the zero modes
via the Pauli-Villars regulator:
\bea
\frac{16 \pi^2Q}{2 g_W^2(\mu)}=\frac{16 \pi^2Q}{2 g_W^2(\Lambda)}-(n_b-\frac{n_f}{2}) \log\frac{ \Lambda}{\mu}
\eea  
Now since $n_b=4NQ$ and $n_f=2NQ$ the result for the Wilsonian beta function follows:
\bea
\frac{1}{2 g_W^2(\mu)}=\frac{1}{2 g_W^2(\Lambda)}-\frac{3}{(4 \pi)^2} \log\frac{ \Lambda}{\mu}
\eea  
or differentiating with respect to $ \log \Lambda$:
\bea
\frac{\partial g_W}{\partial \log \Lambda}=-\beta_0 g_W^3
\eea 
with 
\bea
\beta_0=\frac{3}{(4\pi)^2}
\eea
From the one-loop exactness of the Wilsonian beta function it follows the $NSVZ$ formula for the canonical beta function.
Indeed the renormalization of the canonical coupling is obtained rescaling the fields in canonical form, i.e. in such a way that the quadratic part 
of the action is normalized in order to be $g$ independent:
\bea
Z&&=\exp (-\frac{N16 \pi^2Q(gA_c)}{2 g_W^2}) \Lambda^{n_b-\frac{n_f}{2}}  \int_{{\cal M}_Q} \frac{Pf <\frac{\delta g A_c}{\delta m} ,\frac{\delta g A_c}{\delta m}>}{Pf<g \eta_c , g \eta_c>} \nonumber \\
&&=\exp (-\frac{N16 \pi^2Q(gA_c)}{2 g_W^2}) \Lambda^{n_b-\frac{n_f}{2}} g^{n_b-n_f} \int_{{\cal M}_Q} \frac{Pf <\frac{\delta A_c}{\delta m} ,\frac{\delta A_c}{\delta m}>}{Pf<\eta_c ,  \eta_c>} \nonumber \\
&&=\exp (-\frac{N16 \pi^2Q(gA_c)}{2 g^2}) \Lambda^{n_b-\frac{n_f}{2}}  \int_{{\cal M}_Q} \frac{Pf <\frac{\delta A_c}{\delta m} ,\frac{\delta A_c}{\delta m}>}{Pf<\eta_c ,  \eta_c>}
\eea
where we have defined:
\bea
-\frac{N16 \pi^2Q(gA_c)}{2 g^2}=-\frac{N16 \pi^2Q(gA_c)}{2 g_W^2}+(4NQ(gA_c)-2NQ(gA_c)) \log g
\eea 
or
\bea
\frac{1}{2 g_W^2}=\frac{1}{2 g^2} + \frac{2} {(4\pi)^2} \log g
\eea
Differentiating with respect to $ \log \Lambda $ the $NSVZ$ beta function follows:
\bea
\frac{\partial g}{\partial \log \Lambda}=\frac{-\frac{3}{(4\pi)^2} g^3}
{1- \frac{2}{(4\pi)^2}  g^2 }
\eea
QED

\subsection{Non-$SUSY$ Nicolai map in pure $YM$}

Now we have set the stage for the non-$SUSY$ Nicolai map in pure $YM$. \par
In the pure $YM$ case we define our change of variables in a more general way
than in the $SUSY$ case \cite{V,V1}, by allowing arbitrary gauge-fixing. 
Indeed, while in the $SUSY$ theory we need the light-cone gauge to get the cancellations of the determinants, in the pure $YM$ theory
cancellations do not occur at all. Therefore there is no point in choosing a non-covariant gauge.  \par
The $YM$ partition function is (for definitions see sect.(3.1)):
\bea
Z=\int  \exp \big(-\frac{16\pi^2NQ}{2g^2}-\frac{N}{4g^2} \int tr_f (F_{\alpha\beta}^-)^2d^4 x\big)\delta A
\eea
where we used the identity $Tr(F_{\alpha\beta}^2)=Tr(F_{\alpha\beta}^-)^2/2+Tr(F_{\alpha\beta}\tilde{F}_{\alpha\beta})$ as in the $SUSY$ case.
We change variables from the connection to the $ASD$ curvature by introducing in the functional integral the appropriate resolution of the identity:
\bea
1=\int \delta(F_{\alpha\beta}^--\mu_{\alpha\beta}^-) \delta\mu_{\alpha\beta}^-
\eea
Thus
\bea
Z=\int  \exp \big(-\frac{16\pi^2NQ}{g^2}-\frac{N}{4g^2}\int tr_f(\mu_{\alpha\beta}^-)^2d^4 x\big)\delta(F_{\alpha\beta}^--\mu_{\alpha\beta}^-) \delta \mu_{\alpha\beta}^- \delta A
\eea
Exchanging the order of integration we can now perform the integral on the gauge connection because of the delta function.
The easiest way to do this is defining the delta function as:
\bea
\delta(F_{\alpha\beta}^--\mu_{\alpha\beta}^-)= \lim_{\epsilon \rightarrow 0} \ N^{-1}(\epsilon) e^{-\frac{N}{4 \epsilon}\int tr_f(F_{\alpha\beta}^--\mu_{\alpha\beta}^-)^2}
\eea
with $N(\epsilon)$ an irrelevant normalization factor. Thus the Jacobian of the map to the $ASD$ variables can be evaluated as a Gaussian integral for the quadratic form obtained from the expansion to quadratic order of:
\bea
tr_f(F_{\alpha\beta}^--\mu_{\alpha\beta}^-)^2=tr_f(P^-(F_{\alpha\beta}-\mu_{\alpha\beta}))^2 \sim tr_f(P^-(D^._{[\alpha}\delta A_{\beta]}))^2
\eea
where:
\bea
F_{\alpha\beta}=P^-F_{\alpha\beta}+P^+F_{\alpha\beta}=\frac{1}{2}F_{\alpha\beta}^-+\frac{1}{2}F_{\alpha\beta}^+
\eea
and
\bea
P^-_{\alpha\beta\gamma\delta}=\frac{1}{2} \ \delta_{\alpha\gamma}\delta_{\beta\delta}-\frac{1}{4} \ \varepsilon_{\alpha\beta\gamma\delta}
\eea
Therefore, integrating by parts freely and using the same identities as in sect.(3.1), we get:
\bea
&&tr_fP^-(D^._{[\alpha}\delta A_{\beta]})^2=tr_f(D^._{[\alpha}\delta A_{\beta]}-\frac{1}{2}\epsilon_{\alpha\beta\gamma\delta}D^._{[\gamma}\delta A_{\delta]})^2\nonumber \\
&&=tr_f(2( D^._{[\alpha}\delta A_{\beta]})^2-D^._{[\alpha}\delta A_{\beta]} \epsilon_{\alpha \beta \gamma \delta} D^._{[\gamma}\delta A_{\delta]}) \nonumber \\
&&=tr_f(- 4  \delta A_\alpha \Delta_{A}^.  \delta A_\alpha  - 4 (D_{\alpha}^.  \delta A_\alpha)^2  + 4 i F_{\alpha \beta} [\delta A_\alpha, \delta A_\beta] -4 D^._{\alpha}\delta A_{\beta} \epsilon_{\alpha \beta \gamma \delta} D^._{\gamma}\delta A_{\delta}) \nonumber \\
&&=4tr_f(-   \delta A_\alpha \Delta_{A}^.  \delta A_\alpha  -  (D_{\alpha}^.  \delta A_\alpha)^2   + i F_{\alpha \beta} [\delta A_\alpha, \delta A_\beta] +D^._{\gamma} D^._{\alpha} \delta A_{\beta} \epsilon_{\alpha \beta \gamma \delta} \delta A_{\delta}) \nonumber \\
&&=4tr_f(-   \delta A_\alpha \Delta_{A}^.  \delta A_\alpha - (D_{\alpha}^.  \delta A_\alpha)^2   + i F_{\alpha \beta} [\delta A_\alpha, \delta A_\beta] +\frac{1}{2}([D_{\gamma}, D_{\alpha}])^. \delta A_{\beta} \epsilon_{\gamma\alpha \beta \delta} \delta A_{\delta}) \nonumber \\
&&=4tr_f(-   \delta A_\alpha \Delta_{A}^.  \delta A_\alpha  - (D_{\alpha}^.  \delta A_\alpha)^2   + i F_{\alpha \beta} [\delta A_\alpha, \delta A_\beta] + i [\tilde{F}_{\beta \delta}, \delta A_{\beta}]  \delta A_{\delta}) \nonumber \\
&&=4tr_f(-   \delta A_\alpha \Delta_{A}^.  \delta A_\alpha  - (D_{\alpha}^.  \delta A_\alpha)^2   + i F_{\alpha \beta} [\delta A_\alpha, \delta A_\beta] + i \tilde{F}_{\alpha \beta} [\delta A_{\alpha},  \delta A_{\beta}]) \nonumber \\
&&=4tr_f (\delta A_\alpha(-\Delta_{A}^.\delta_{\alpha\beta} + D_\alpha^. D_\beta^.-2i \ ad(P^+F)_{\alpha\beta})\delta A_\beta)
\eea
The partition function becomes:
\bea
Z=\int  \exp \big(-\frac{16\pi^2NQ}{g^2}-\frac{N}{4g^2}\int tr_f (\mu_{\alpha\beta}^-)^2d^4 x\big)  Det^{-1/2}(-\Delta_{{A}}^. \delta_{\alpha \beta}+ D_\alpha^. D_\beta^.-i F_{\alpha \beta}^{+ .})  \delta \mu^- 
\eea
The determinant in Eq.(3.78) does not exist unless the gauge is fixed. This is most conveniently done in a background Feynman gauge:
\bea
Z&&= \lim_{\epsilon \rightarrow 0} \int \delta  A  \delta C \delta \mu^- \exp(-S_{YM})e^{-\frac{N}{4 \epsilon}\int tr_f(F_{\alpha\beta}^--\mu_{\alpha\beta}^-)^2 d^4x} \nonumber\\
&&Det(-\Delta_{{A}}^.) e^{-\frac{N}{\epsilon} \int d^4 x \ tr_f (C^2) }\delta( D_\alpha^. \delta A_\alpha -C)
\eea
As a consequence the gauge-fixed partition function in the $ASD$ variables is:
\bea
Z&&=\int  \exp \big(-\frac{16\pi^2NQ}{g^2}-\frac{N}{4g^2}\int tr_f (\mu_{\alpha\beta}^-)^2d^4 x\big) \nonumber \\
&& Det^{-1/2}(-\Delta_{{A}}^. \delta_{\alpha \beta}-i F_{\alpha \beta}^{+ .}) Det(-\Delta_{{A}}^.) \delta \mu^- 
\eea
that because of the argument displayed below can be rewritten as:
\bea
Z&&=\int  \exp \big(-\frac{16\pi^2NQ}{g^2}-\frac{N}{4g^2}\int tr_f (\mu_{\alpha\beta}^-)^2d^4 x \big) \nonumber \\
&&Det^{-1/2}(-\Delta_{A}^. \delta_{\alpha \beta}-i \mu_{\alpha \beta}^{- .}) Det(-\Delta_{A}^.) \Lambda^{n_b [\mu^-] } \omega^{\frac{n_b[\mu^-]}{2}} \delta \mu^-
\eea
where $\omega$ is a Kahler form on the moduli induced by a Kahler form on the connections. A possible choice for $\omega$ is:
\bea
\omega_I = \frac{1}{2 \pi} \int d^4 x tr_f( \frac{\delta A_z}{\delta m_i} \frac{\delta A_{\bar z}}{\delta \bar m_k}- \frac{\delta A_u}{\delta m_i} \frac{\delta A_{\bar u}}{\delta \bar m_k})\delta m_i \wedge \delta \bar m_k
\eea
As in the $SUSY$ case zero modes have to occur, but in the pure $YM$ case there are also other contributions
to the beta function due to the lack of cancellation of determinants (sect.(3.4)). \par
Let us introduce the matrices $\sigma_{\alpha}=(1,i \tau)$ with $\tau$ the three Hermitian Pauli matrices and the self-dual and anti-selfdual 
matrices (we use the same notation as in \cite{Rev}):
\bea
\sigma_{\alpha\beta}=\frac{1}{4}(\sigma_{\alpha} \bar \sigma_{\beta}-\sigma_{\beta} \bar \sigma_{\alpha}) \nonumber \\
\bar \sigma_{\alpha\beta}=\frac{1}{4}(\bar \sigma_{\alpha}  \sigma_{\beta}-\bar \sigma_{\beta}  \sigma_{\alpha}) \nonumber \\
\eea
The three variations \cite{Rev}:
\bea
P^-(D^._{[\alpha}\delta A_{\beta]})
\eea
can be rewritten as:
\bea
(\tau_i) ^{\dot a} _{b}  {{\not \bar D}^.} ^{\dot b a} \delta A_{a \dot a}=Tr(\tau_i \not \bar D^. \delta \not A)
\eea
and the forth variation:
\bea
D^._{\alpha} \delta A_{\alpha}\eea
as:
\bea
{{\not \bar D}^.} ^{\dot a a} \delta A_{a \dot a}=Tr( \not \bar D^. \delta \not A)
\eea
in such a way that the four variations can be written together as:
\bea
{{\not \bar D}^.} ^{\dot a a} \delta A_{a \dot b}
\eea
where we have defined:
\bea
\not D^.=\sigma_{\alpha} D_{\alpha}^. \nonumber \\
\not \bar D^.= \bar \sigma_{\alpha} D_{\alpha}^.
\eea
Therefore the Euclidean invariant positive semidefinite quadratic form:
\bea
\sum_{\alpha > \beta}  tr_f(P^-(D^._{[\alpha}\delta A_{\beta]}) )^2
+ tr_f ( D_\alpha^. \delta A_\alpha)^2
\eea
can be written as:
\bea
&&\sum_i  tr_f |Tr( \tau_i \not \bar D^. \delta \not A)|^2+  tr_f |Tr( \not \bar D^. \delta \not A)|^2 \nonumber \\
&& =\sum_{\dot a\dot b}   tr_f |{{\not \bar D}^.} ^{\dot a a} \delta A_{a \dot b}|^2 \nonumber \\
&&=- tr_f Tr( \delta \not \bar A \not  D^. \not \bar D^. 1\delta \not A)
\eea
where in the second line we have used the completeness of the $\sigma_{\alpha}$ over the $2 \times 2$ matrices.
The trace, $Tr$, refers to the spin indices and we have freely integrated by parts. Therefore the partition function in the Feynman gauge: 
\bea
Z&&= \lim_{\epsilon \rightarrow 0} \int \delta  A  \delta C \delta \mu^- \exp(-S_{YM})e^{-\frac{N}{4 \epsilon}\int tr_f(F_{\alpha\beta}^--\mu_{\alpha\beta}^-)^2 d^4x} \nonumber \\
&&Det(-\Delta_{{A}}^.) e^{-\frac{N}{\epsilon} \int d^4 x \ tr_f (C^2) }\delta( D_\alpha^. \delta A_\alpha -C) 
\eea
becomes:
\bea
&&\lim_{\epsilon \rightarrow 0} \int \delta  A   \delta \mu^- \exp \big(-\frac{16\pi^2NQ}{g^2}-\frac{N}{4g^2}\int tr_f (\mu_{\alpha\beta}^-)^2d^4 x\big)  Det(-\Delta_{{A}}^.)    \nonumber \\
&&\exp(-\frac{N}{ \epsilon} \sum_{\alpha > \beta} \int tr_f(P^-(D^._{[\alpha}\delta A_{\beta]}) )^2 d^4x) 
\exp(-\frac{N}{\epsilon} \int d^4 x \ tr_f ( D_\alpha^. \delta A_\alpha)^2 ) \nonumber \\
&&= \int  \delta \mu^- \exp \big(-\frac{16\pi^2NQ}{g^2}-\frac{N}{4g^2}\int tr_f (\mu_{\alpha\beta}^-)^2d^4 x\big)  Det(-\Delta_{{A}}^.)   
Det^{-\frac{1}{2}}(-\not  D^. \not \bar D^. \delta_{\dot a \dot b})
\eea 
where:
\bea
- \not  D^. \not \bar D^.=-\Delta_{{A}}^. 1+\sigma_{\alpha\beta} F_{\alpha\beta}^.
\eea
Hence:
\bea
Det^{-\frac{1}{2}}(-\not  D^. \not \bar D^. \delta_{\dot a \dot b})=Det^{-1/2}(-\Delta_{{A}}^. \delta_{\alpha \beta}-i F_{\alpha \beta}^{+ .})
\eea
In addition:
\bea
-\not \bar D^. \not D^.=-\Delta_{{A}}^. 1+\bar \sigma_{\alpha\beta} F_{\alpha\beta}^.
\eea
and symmetrically:
\bea
Det^{-\frac{1}{2}}(-\not \bar D^. \not D^. \delta_{\dot a \dot b})=Det^{-1/2}(-\Delta_{{A}}^. \delta_{\alpha \beta}-i F_{\alpha \beta}^{- .})
\eea
Now, since $\not \bar D^. \not D^.$ and $\not  D^. \not \bar D^.$ have the same spectrum of non-zero modes, it follows that $-\Delta_{{A}}^. \delta_{\alpha \beta}-i F_{\alpha \beta}^{- .}$ has the same non-zero modes as
$-\Delta_{{A}}^. \delta_{\alpha \beta}-i F_{\alpha \beta}^{+ .}$, but it has no zero modes precisely when $-\Delta_{{A}}^. \delta_{\alpha \beta}-i F_{\alpha \beta}^{+.}$ has. 
This is the case when the gauge connection that solves the equation of $ASD$ type, $F_{\alpha\beta}^-(A(m_i))=\mu_{\alpha\beta}^-$, has moduli, $m_i$. Indeed taking the derivative of this equation with respect to the moduli one gets:
\bea
\frac{\delta F_{\alpha\beta}^-}{\delta A_\gamma}\frac{\delta A_\gamma}{\delta m_i}=0
\eea
that implies that the operator: 
\bea
\frac{\delta F_{\alpha\beta}^-}{\delta A_\gamma}\delta A_\gamma =\frac{\delta (P^-F_{\alpha\beta})}{\delta A_\gamma}\delta A_\gamma= P^-(D^._{[\alpha}\delta A_{\beta]})
\eea
has zero modes, and therefore $\not \bar D^.$ and $(-\Delta_{A}^.\delta_{\alpha\beta}-2i \ ad(P^+F)_{\alpha\beta})$ have zero modes too. QED

\subsection{One-loop beta function of pure $YM$ in the $ASD$ variables}

We can now use Eq.(3.81) as the definition of the partition function of $YM$ in the $ASD$ variables.
We can apply the standard background field method of sect.(3.1) for the computation of the beta function in the $ASD$ variables.
The field $\mu_{\alpha \beta}=\bar \mu_{\alpha \beta}+ \delta \mu_{\alpha\beta}$ can be decomposed in a background, $\bar \mu_{\alpha\beta}$, and a fluctuating field, $\delta \mu_{\alpha\beta}$.
The correlations of the fluctuating field can contribute only starting from order of $g^2$. Therefore the only $O(g^0)$ contributions, relevant for the one-loop beta function, arise from the functional determinants.
To evaluate the effective action in the $ASD$ variables it is most convenient to compare it with the standard one-loop effective action of sect.(3.1).
 In the standard background field method the quadratic form:
\bea
\frac{N}{2g^2}\int d^4 x tr_f (F_{\alpha\beta})^2
\eea
is expanded around a solution of the equation of motion, leading to the one-loop effective action:
\bea
Z_{1-loop}&&=e^{-\Gamma_{1-loop}(\bar{A})} \nonumber \\
&&=e^{-S_{YM}(\bar{A})}Det^{-1/2}(-\Delta_{\bar{A}}^. \delta_{\alpha \beta}-2i \bar F_{\alpha \beta}^.) Det(-\Delta_{\bar{A}}^.)
\eea
In the $ASD$ variables because of the delta function that defines the resolution of identity, the "action"
\bea
\lim_{\epsilon\rightarrow 0}\frac{N}{4\epsilon}\int tr_f(F_{\alpha\beta}^--\mu_{\alpha\beta}^-)^2
\eea
is expanded around the background $F_{\alpha\beta}^-(\bar A+\delta A)=\bar \mu_{\alpha\beta}^- +\delta \mu_{\alpha\beta}^-$, leading to the one-loop effective action in the
$ASD$ variables:
\bea
Z_{1-loop}^{ASD}&&=e^{-\Gamma_{1-loop}^{ASD}(\bar{\mu})} \nonumber \\
&&=e^{-\frac{16\pi^2NQ}{2g^2}-\frac{N}{g^2} \int tr_f (\frac{\bar \mu_{\alpha\beta}^-}{2})^2d^4 x} \nonumber \\
&&\int \Lambda^{n_b [\bar \mu^-] } \omega^{\frac{n_b[\bar \mu^-]}{2}}
Det^{-1/2}(-\Delta_{\bar{A}}^. \delta_{\alpha \beta}-2i ad \frac{\bar \mu_{\alpha \beta}^-}{2}) Det(-\Delta_{\bar{A}}^.) 
\eea
Thus the orbital contribution is the same as in the standard background field method. The difference is the spin term, $2i ad(P^-F)_{\alpha\beta}$, as opposed to $2iad F_{\alpha\beta}$,
and the possible contribution of the zero modes, whose occurrence is not generic but depends on the background. Let us suppose at first that zero modes do not occur.
In sect.(3.1) we have seen that the combination of determinants
\bea
&&Det^{-1/2}(-\Delta_{A}^. \delta_{\alpha\beta}-2i \ adF_{\alpha\beta}) Det(-\Delta_{A}^.)\nonumber \\
\eea
leads to the beta function:
\bea
\frac{1}{2g^2(\mu)}=\frac{1}{2g^2(\Lambda)}+(\frac{1}{3(4\pi)^2}-\frac{4}{(4\pi)^2})\log\frac{\Lambda}{\mu}
\eea
where the first term in the brackets is the orbital contribution and the second one is the spin contribution. \par
In the $ASD$ case the orbital part is the same one, while the spin part differs because of the substitution $F_{\alpha\beta}\rightarrow P^-F_{\alpha\beta}=\frac{\mu^-_{\alpha\beta}}{2}$.
Thus both in the standard one-loop effective action and in the $ASD$ variables the orbital contribution is:
\bea
&&N \int tr_f(F_{\alpha\beta})^2  \frac{1}{3(4\pi)^2}\log\frac{\Lambda}{\mu} \nonumber \\
&&=N(4\pi)^2Q \frac{1}{3(4\pi)^2}\log\frac{\Lambda}{\mu}+\frac{N}{2} \int tr_f(F^-_{\alpha\beta})^2  \frac{1}{3(4\pi)^2}\log\frac{\Lambda}{\mu}
\nonumber \\
&&=N(4\pi)^2Q \frac{1}{3(4\pi)^2}\log\frac{\Lambda}{\mu}+ N \int tr_f(P^-F_{\alpha\beta})^2  \frac{2}{3(4\pi)^2}\log\frac{\Lambda}{\mu}
\eea
On the contrary, while the spin contribution for the standard one-loop effective action is:
\bea
- N \int tr_f (F_{\alpha\beta})^2  \frac{4}{(4\pi)^2}\log\frac{\Lambda}{\mu}
\eea
in the $ASD$ variables, because of the substitution $F_{\alpha\beta}\rightarrow P^-F_{\alpha\beta}=\frac{\mu^-_{\alpha\beta}}{2}$, is:
\bea
-N \int tr_f (P^-F_{\alpha\beta})^2  \frac{4}{(4\pi)^2}\log\frac{\Lambda}{\mu}
\eea
Hence
\bea
&&\Gamma_{1-loop}^{ASD}=(\frac{1}{2g^2}+\frac{1}{3(4\pi)^2}\log\frac{\Lambda}{\mu})N(4\pi)^2Q\nonumber \\
&&+N\int tr_f (P^-F_{\alpha\beta})^2(\frac{1}{g^2}+\frac{2}{3(4\pi)^2}\log\frac{\Lambda}{\mu}-\frac{4}{(4\pi)^2}\log\frac{\Lambda}{\mu})\nonumber \\
&&=(\frac{1}{2g^2}+\frac{1}{3(4\pi)^2}\log\frac{\Lambda}{\mu})N(4\pi)^2Q+2N\int tr_f (P^-F_{\alpha\beta})^2(\frac{1}{2g^2}-\frac{5}{3(4\pi)^2}\log\frac{\Lambda}{\mu})\nonumber \\
&&=\frac{N(4\pi)^2QZ_Q^{-1}}{2g^2}+\frac{2NZ^{-1}}{2g^2}\int tr_f (P^-F_{\alpha\beta})^2
\eea
where: 
\bea
Z_Q^{-1}&&=1+\frac{2}{3}\frac{g^2}{(4\pi)^2}\log\frac{\Lambda}{\mu} \nonumber \\
Z^{-1}&&=1-\frac{10}{3}\frac{g^2}{(4\pi)^2}\log\frac{\Lambda}{\mu}
\eea
Thus generically the $YM$ beta function is not reproduced in absence of zero modes as in the $SUSY$ case. Moreover generically the renormalizations of $Q$ and of $\mu_{\alpha\beta}$ are different.
However, if the background field satisfies the equation of motion at leading order, $\mu_{\alpha\beta}=0$, the corresponding $YM$ connection is $SD$ and therefore instantons occur.
In this case the zero modes have to be included and the one-loop beta function is reproduced in the $ASD$ variables:
\bea
Z_{1-loop}^{ASD}
&&=e^{-\frac{(4\pi)^2 NQ}{2g^2}}
Det^{-1/2}(-\Delta_{\bar{A}}^. \delta_{\alpha \beta}) Det(-\Delta_{\bar{A}}^.)     \Lambda^{4NQ } \omega^{2NQ}             \nonumber \\
&&=e^{-\frac{(4\pi)^2 NQ Z^{-1}_Q}{2g^2}}  \Lambda^{4NQ } \omega^{2NQ}
\eea
Yet, we may wonder as to whether the $YM$ theory can be exactly localized on instantons as in the $SUSY$ case.
It is very instructive to check that it cannot be so, otherwise the two-loop beta function is not reproduced.
Rescaling fields in canonical form, as in the $SUSY$ case, we get:
\bea
Z_{1-loop}^{ASD}
&&=e^{-\frac{(4\pi)^2 N Q(g Z^{\frac{1}{2}}_Q A_c)  Z^{-1}_Q }{2g_W^2} }  \Lambda^{4NQ} \omega(g Z^{\frac{1}{2}}_Q A_c)^{2NQ} \nonumber \\
&&=e^{-\frac{(4\pi)^2 N Q(g Z^{\frac{1}{2}}_Q A_c)  Z^{-1}_Q }{2g_W^2} }  \Lambda^{4NQ } (g Z^{\frac{1}{2}}_Q)^{4NQ} \omega( A_c)^{2NQ} \nonumber \\
\eea
Now we can define as in Eq.(3.67)
\bea
-\frac{N16 \pi^2Q(g Z^{\frac{1}{2}}_Q A_c)Z^{-1}_Q}{2 g^2}=-\frac{N16 \pi^2Q(g Z^{\frac{1}{2}}_Q A_c)Z^{-1}_Q}{2 g_W^2}+ 4NQ(g Z^{\frac{1}{2}}_Q A_c) \log (gZ^{\frac{1}{2}}_Q)
\eea 
that implies:
\bea
\frac{1}{2 g_W^2}=\frac{1}{2 g^2}+ \frac{4 Z_Q}{(4 \pi)^2} \log (gZ^{\frac{1}{2}}_Q)
\eea 
and with two-loop accuracy
\bea
\frac{1}{2 g_W^2}\sim \frac{1}{2 g^2}+ \frac{4}{(4 \pi)^2} \log (gZ^{\frac{1}{2}}_Q)
\eea 
Taking the derivative with respect to $\log\Lambda$ and assuming by the localization hypothesis that $g_W$ is one-loop exact, we get with two-loop accuracy:
\bea
\frac{1}{g^3}\frac{\partial g}{\partial\log\Lambda}=-\beta_0+\frac{4}{(4\pi)^2}\frac{1}{g}\frac{\partial g}{\partial\log\Lambda}+\frac{2}{(4\pi)^2}\frac{\partial \log Z_Q}{\partial\log\Lambda}
\eea
with $\beta_0=\frac{11}{3} \frac{1}{(4\pi)^2}$. Therefore
\bea
\frac{\partial g}{\partial\log\Lambda}=\frac{-\beta_0 g^3+\frac{2g^3}{(4\pi)^2}\frac{\partial \log Z_Q}{\partial\log\Lambda}}{1-\frac{4}{(4\pi)^2} \ g^2}
\eea
Since
\bea
\frac{\partial\log Z_Q}{\partial\log \Lambda}\sim -\frac{2}{3}\frac{g_W^2}{(4\pi)^2}\sim -\frac{2}{3}\frac{g^2}{(4\pi)^2}
\eea
it follows that:
\bea
\frac{\partial g}{\partial\log\Lambda}\sim-\frac{11}{3}\frac{g^3}{(4\pi)^2}-\frac{4}{3}\frac{g^5}{(4\pi)^4}-\frac{4}{(4\pi)^2} \beta_0 g^5=-\frac{11}{3}\frac{g^3}{(4\pi)^2}-\frac{48}{3}\frac{g^5}{(4\pi)^4}
\eea
Therefore the second coefficient of the beta function, $\beta_1$, differs from the perturbative result: 
\bea
\beta_1=\frac{48}{3}\frac{1}{(4\pi)^4}\neq \frac{34}{3}\frac{1}{(4\pi)^4}
\eea
Thus  it is not possible to localize the $YM$ partition function on instantons.  QED \par
On the contrary, we will see in the following sections that twistor Wilson loops can be localized on surface operators with $Z_N$ holonomy.

\section{Twistor loops and non-commutative $YM$}

\subsection{Non-commutative Eguchi-Kawai reduction}

We recall some fundamental facts about the non-commutative $YM$ theory \cite{Szabo2,DN} that will be used throughout the whole paper.
These results will allow us to construct the twistor Wilson loops which our approach is entirely based on. \par
The non-commutative $R^d$ is defined by:
\bea
[\hat x^{\alpha}, \hat x^{\beta}]=i \theta^{\alpha \beta} 1
\eea
Let $\hat \Delta(x)$ be:
\bea
\hat \Delta(x) = \int \frac{d^d k}{(2 \pi)^d} e^{ik  \hat x} e^{-ikx}
\eea
and
\bea
\hat f= \int d^dx f(x) \hat \Delta(x)
\eea
for complex functions of rapid decrease in both coordinates and momenta (Schwartz space).
This defines an operator/function correspondence such that:
\bea
\hat f \hat g= \hat {f \star g}
\eea
with:
\bea
(f \star g)(x)= f(x) \exp(\frac{i}{2} \partial^{\alpha}_x \theta^{\alpha \beta} \partial^{\beta}_{y}) g(y)|_{y=x}     \eea
that can be extended to multiple $ \star $ products:
\bea
f_1(x_1) \star...\star f_n(x_n)= \prod_{i<k} \exp(\frac{i}{2} \partial^{\alpha}_{x^i} \theta^{\alpha \beta} \partial^{\beta}_{x^k})
f_1(x_1) ... f_n(x_n)
\eea 
needed in the evaluation of Wilson loops of the non-commutative theory in the function representation: $P_{\star}\exp i \int_{L_{yz}}  A_{\alpha} (x) dx_{\alpha}$.\par
By the operator/function correspondence translations are represented by unitary operators:
\bea
e^{a  \hat \partial}    \hat \Delta(x)    e^{-a  \hat \partial}= \hat \Delta(x+a) 
\eea
where:
\bea
\hat \partial^i(\hat x^j)= \delta^{ij}1
\eea
Thus non-commutative derivations can be represented via Eq.(4.1) and satisfy:
\bea
[\hat \partial_{\alpha}, \hat \partial_{\beta}]=i \theta^{-1}_{\alpha \beta} 1
\eea
In addition the integration on functions coincides with the operator trace up to a factor:
\bea
(2 \pi)^{\frac{d}{2}} Pf(\theta )\hat Tr \hat f && = \int d^dx f(x) \nonumber \\
(2 \pi)^{\frac{d}{2}} Pf(\theta ) \hat Tr(\hat \Delta(x)\hat \Delta(y)) &&=\delta^d(x-y) \nonumber \\
\int d^dx (f \star g)(x)&& =\int d^dx f(x) g(x) 
\eea
The $YM$ action of the $U(N)$ non-commutative gauge theory has the function/:
\bea
\frac{N}{2 g^2} \int d^dx tr_N (F_{\alpha \beta} \star F_{\alpha \beta})(x)
\eea
/operator representation:
\bea
\frac{N}{2 g^2}  (2 \pi)^{\frac{d}{2}} Pf(\theta )   tr_N \hat Tr(-i [\hat \partial_{\alpha}+i \hat A_{\alpha},\hat \partial_{\beta}+i \hat A_{\beta}]+ \theta^{-1}_{\alpha \beta} 1)^2
\eea
where the non-commutative gauge connection is valued in the tensor product of the Lie algebra, $u(N)$, of $U(N)$ in the fundamental representation and of the field $\star$-algebra.
This leads to the non-commutative \cite{Twc,Twl1} \footnote{We would like to thank Antonio Gonzalez-Arroyo and Chris Korthals-Altes for discussions on non-commutative
Eguchi-Kawai reduction and Antonio Gonzalez-Arroyo for a detailed exam of our work at the GGI.} Eguchi-Kawai reduction \cite{EK,Neu,Twl2}
\bea
\frac{N}{2 g^2}  \hat N (\frac{2 \pi}{\Lambda})^d tr_N Tr_{\hat N} (-i [\hat \partial_{\alpha}+i \hat A_{\alpha},\hat \partial_{\beta}+i \hat A_{\beta}]+ \theta^{-1}_{\alpha \beta} 1)^2
\eea
where the trace $Tr_{\hat N}$ is taken now over a subspace of dimension $\hat N$,
with 
\bea
\hat N (\frac{2 \pi}{\Lambda})^d=(2 \pi)^{\frac{d}{2}} Pf(\theta )
\eea
in the large $\hat N, \theta, \Lambda $ limit.
The simplest way to understand the occurrence of the inverse power of the cutoff \footnote{ We would like to thank Yuri Makeenko for discussing this point with us  at the GGI.} in the reduced non-commutative action \footnote{
See \cite{Kaw} for a modern treatment. Another interesting way to understand the same factor in the quenched version of the Eguchi-Kawai reduction is in \cite{Rt}.} is to study the Makeenko-Migdal \cite{MM,MM1} loop equation after having reabsorbed the two factors of $N , \hat N$  
into a unique factor, $\mathcal{N}$ $=N\hat N$, that computes the rank of the tensor product.
For this we need to write the Wilson loop of the non-commutative theory in the operator notation.
In this version the theory is a matrix model of infinite matrices. Thus the Wilson loop must involve a connection constant in space-time:
\bea
\frac{1}{\mathcal{N}}  tr_N Tr_{\hat N} \Psi(\hat A; L_{ww})=\frac{1}{\mathcal{N}}  tr_N Tr_{\hat N} P \exp  \int_{L_{ww}}(\hat \partial_{\alpha} +i \hat A_{\alpha}) dx_{\alpha}
\eea
Indeed this prescription leads to the correct definition of the Wilson loops of the non-commutative theory in the function representation. The proof is as follows.
In the operator representation we can gauge away the non-commutative derivative that occurs in the definition of the Wilson loop by performing
a local gauge transformation with values in the infinite-dimensional unitary group acting on the Fock representation of the non-commutative theory:
\bea
\hat U(x)= e^{ x_{\alpha} \hat \partial_{\alpha}} 
\eea
where $x_{\alpha}$ is a commutative space-time coordinate.
The operator-valued gauge connection transforms under this gauge transformation in the usual way:
\bea
\hat A_{\alpha}^ {\hat U} = \hat U(x) \hat A_{\alpha} \hat U(x)^{-1} + i \partial_{\alpha} \hat U(x) \hat U(x)^{-1}
\eea 
where the partial derivative is the usual partial derivation with respect to the commutative 
parameter $x_{\alpha}$. The operator $\hat \partial_{\alpha}$ must instead transform as a Higgs field in order for $-i \hat \partial_{\alpha} + \hat A_{\alpha}$ to be a connection:
\bea
\hat \partial_{\alpha}^ {\hat U} = \hat U(x) \hat \partial_{\alpha} \hat U(x)^{-1} 
\eea
Correspondingly the Wilson line:
\bea
 \Psi(\hat A; L_{yz})= P \exp i \int_{L_{yz}}(- i \hat \partial_{\alpha} + \hat A_{\alpha}) dx_{\alpha}
\eea
transforms as \footnote{We ignore central terms that vanish for large $\theta$.}:
\bea
&& \hat U(y)\Psi( \hat A; L_{yz}) \hat U(z)^{-1} \nonumber \\
&&= P \exp i  \int_{L_{yz}}(- i \hat \partial_{\alpha}^U + \hat A_{\alpha}^U) dx_{\alpha}  \nonumber \\
&& =  P \exp i \int_{L_{yz}}( \hat U(x) \hat A_{\alpha} \hat U(x)^{-1} -i U(x)\hat \partial_{\alpha}U(x)^{-1} + i \partial_{\alpha} \hat U(x) \hat U(x)^{-1} )dx_{\alpha}  \nonumber \\
&&=  P \exp i \int_{L_{yz}} \hat U(x) \hat A_{\alpha} \hat U(x)^{-1} dx_{\alpha}  \nonumber \\
&& = P_{\star} \exp i \int_{L_{yz}}  A_{\alpha} (x) dx_{\alpha}
\eea
that is the function version of the non-commutative Wilson loop by the operator/function correspondence.
Now the Makeenko-Migdal loop equation \cite{MM,MM1} of the large-$N$ commutative theory is:
\bea
&&<\frac{1}{{N}}  tr_N(\frac{\delta S_{YM}} {\delta A_{\alpha}(x)} \Psi(A; L_{xx}))>  \nonumber \\
&&=i \int_{L_{xx} }dy_{\alpha}  \delta ^{ (d)} (x-y) <\frac{1}{{N}}  tr_N \Psi(A; L_{xy})> < \frac{1}{{N}}  tr_N\Psi(A; L_{yx})>
\eea
where the normalized commutative $YM$ action is:
\bea
S_{YM}=\frac{1}{2 g^2} \int tr_N ( F_{\alpha \beta})^2 d^d x
\eea
and the v.e.v. is defined with respect of the unnormalized action:
\bea
< ... >=Z^{-1} \int  ...  \exp(-\frac{N}{2 g^2} \int tr_N ( F_{\alpha \beta})^2 d^d x)  \delta A
\eea
The loop equation of the non-commutative matrix model is instead \cite{Mak1,Mak2}:
\bea
&&<\frac{1}{{ \cal {N}}}  Tr_{ \cal N}(\frac{\delta S_{NC}} {\delta \hat A_{\alpha}} \Psi(\hat A; L_{xx}))>  \nonumber \\
&&=i \int_{L_{xx} }dy_{\alpha}   <\frac{1}{{\cal N}}  Tr_{\cal N} \Psi(\hat A; L_{xy})> < \frac{1}{{\cal N}}  Tr_{\cal N} \Psi(\hat A; L_{yx})>
\eea
where the normalized action, $S_{NC}$, of the non-commutative theory is:
\bea
S_{NC}=\frac{1}{2 g^2} (\frac{2 \pi}{\Lambda})^d Tr_{\cal N} (-i [\hat \partial_{\alpha}+i \hat A_{\alpha},\hat \partial_{\beta}+i \hat A_{\beta}]+ \theta^{-1}_{\alpha \beta} 1)^2
\eea
and the v.e.v. of the non-commutative theory is defined with respect to the unnormalized action:
\bea
< ... >=Z^{-1} \int  ...  \exp(- \frac { \cal{N} } {2 g^2}  (\frac{2 \pi}{\Lambda})^d Tr_{\cal {N}} (-i [\hat \partial_{\alpha}+i \hat A_{\alpha},\hat \partial_{\beta}+i \hat A_{\beta}]+ \theta^{-1}_{\alpha \beta} 1)^2)
\delta A
\eea
At this point we notice that the factor of $(\frac{2 \pi}{\Lambda})^d$ in the normalized non-commutative action is essential to reproduce the loop equation 
of the commutative gauge theory, since its effect is equivalent to the insertion of the missing $\delta^{(d)}(0)$ in the right hand side of the non-commutative loop equation.
As a consequence the  $\delta^{(d)}(x-y)$ of the commutative loop equation is reproduced provided the trace of Wilson lines vanishes for $x \neq y$ \cite{Mak1,Mak2}:
\bea
 < \frac{1}{{\cal N}} Tr_{\cal N}\Psi(\hat A; L_{xy})>=0
\eea
QED \par
The occurrence of the inverse power of the cutoff in the matrix model version of the non-commutative theory opens the way to saddle-point computations
of new kind in which power-like divergences cancel against the $(\frac{2 \pi}{\Lambda})^d$ factor. In particular if the theory is defined
on $R^2 \times R^2 _{\theta}$ quadratic divergences cancel. This will turn out to be the case for the surface operators of the theory on $R^2 \times R^2 _{\theta}$
in the limit of large $\theta$ introduced in sect.(7) and employed in the whole paper. \par

\subsection{Twistor Wilson loops}

We define twistor Wilson loops in the $YM$ theory
with gauge group $U(N)$ on $R^2 \times R^2 _{\theta}$ with complex coordinates $(z=x_0+i x_1, \bar z=x_0-i x_1, \hat u=\hat x_2+i \hat x_3, \hat {\bar u}=\hat x_2-i \hat x_3)$ 
and non-commutative parameter $\theta$, satisfying $ [\hat \partial_u, \hat \partial_{\bar u}]=\theta^{-1} 1$, as follows:
\bea
Tr_{\cal N} \Psi(\hat B_{\lambda};L_{ww})=Tr_{ \cal{N}} P \exp i \int_{L_{ww}}(\hat A_z+\lambda \hat D_u) dz+(\hat A_{\bar z}+ \lambda^{-1} \hat D_{\bar u}) d \bar z 
\eea
where $\hat D_u=\hat \partial_u+i \hat A_u$ is the covariant derivative along the non-commutative direction $\hat u$ and $\lambda$ a complex parameter. For many purposes it is not restrictive to choose $\lambda$ real,
although other choices are possible, for example a phase, $\lambda=e^{i \delta}$. The plane $(z, \bar z)$ is commutative. The loop, $L_{ww}$, starts and ends at the marked point, $w$, and lies in the commutative plane. Thus we regard the twistor connection, $B_{\lambda}$, whose holonomy the twistor Wilson loop computes, as a non-Hermitian connection
in the commutative plane valued in the tensor product of the $U(N)$ Lie algebra and of the infinite-dimensional operators that generate the Fock 
representation of the non-commutative plane $(\hat u, \hat {\bar u})$. $B_{\lambda}$ is indeed a connection in the
commutative plane since the non-commutative covariant derivative transforms as a Higgs field of the commutative plane. The trace is defined accordingly. The limit of infinite non-commutativity in the plane $(\hat u, \hat {\bar u})$ is understood, being equivalent to the large-$N$ limit of the commutative gauge theory \cite{Twc,Szabo2,AG}. The $U(N)$ non-commutative theory for finite $\theta$ has tachyon instabilities that occur
in non-planar diagrams suppressed by powers of $\theta^{-1}$ and $N^{-1}$ \cite{Szabo2}. Therefore non-commutativity is for us just a mean to define the large-$N$ limit, as well as it is for Nekrasov just a mean to compactify the moduli space of instantons \footnote{Once localization is obtained, the glueball spectrum is computed employing the effective action in the large-$N$ commutative theory, around the localized locus (sect.(12)).}. \par

\subsection{Fiber independence of the v.e.v. of twistor Wilson loops}

It easy to show that the v.e.v. of the twistor Wilson loops is independent on the parameter $\lambda$:
\bea
<\frac{1}{\cal N} Tr_{\cal N} \Psi(\hat B_{\lambda};L_{ww})     >=<\frac{1}{\cal N} Tr_{\cal N} \Psi(\hat B_1;L_{ww}) > 
\eea
The proof is obtained changing variables, rescaling covariant derivatives in the usual definition of the functional integral
of the non-commutative $YM$ theory:
\bea
&&\int   Tr_{ \cal{N}} P \exp i \int_{L_{ww}}(\hat A_z+\lambda \hat D_u) dz+(\hat A_{\bar z}+ \lambda^{-1} \hat D_{\bar u}) d \bar z  \nonumber \\
&&\exp(- \frac { \cal{N} } {2 g^2}  (\frac{2 \pi}{\Lambda})^2 \int d^2 x Tr_{\cal {N}} (-i [\hat D_{\alpha},\hat D_{\beta}]+ \theta^{-1}_{\alpha \beta} 1)^2)  
\delta{\hat A} \delta {\hat{\bar A}  } \delta {\hat D}  \hat {\bar D} \nonumber \\
&&= \int   Tr_{ \cal{N}} P \exp i \int_{L_{ww}}(\hat A_z+ \hat D'_u) dz+(\hat A_{\bar z}+  \hat D'_{\bar u}) d \bar z  \nonumber \\
&&\exp(- \frac { \cal{N} } {2 g^2}  (\frac{2 \pi}{\Lambda})^2 \int d^2 x Tr_{\cal {N}} (- [\hat D'_{\alpha},\hat D'_{\beta}]^2+ {(\theta^{-1}_{\alpha \beta})} ^2 1-2i [\hat D'_{\alpha},\hat D'_{\beta}] \theta^{-1}_{\alpha \beta})  
\delta{\hat A} \delta {\hat{\bar A}  } \delta {\hat D'}  \hat {\bar D'} 
\eea
where: 
\bea
\hat D'_{z}&=&\hat D_{z} \nonumber \\
\hat  D'_{ \bar z}&=&\hat D_{ \bar z} \nonumber \\
\hat D'_{u}&=& \lambda \hat D_{u} \nonumber \\
\hat  D'_{\bar u}&=&\lambda^{-1} \hat D_{\bar u} \nonumber \\
\eea
The formal non-commutative integration measure is invariant under such rescaling because of the
pairwise cancellation of the powers of $\lambda$ and $\lambda^{-1}$. The first term in the non-commutative $YM$ action, proportional to
$ Tr_{\cal {N}}[\hat D'_{\alpha},\hat D'_{\beta}]^2$, is invariant because of rotational invariance in the non-commutative plane. \par
Indeed every $u$ must be contracted with a $\bar u$ by rotational
invariance in the non-commutative plane and thus the factors of $\lambda$ cancel. The only possibly dangerous terms couple the non-commutative
parameter to the commutator $Tr_{\cal {N}}( [\hat D'_{\alpha},\hat D'_{\beta}] \theta^{-1}_{\alpha \beta}) $ but only $Tr_{\cal {N}} ([\hat D'_{u},\hat D'_{\bar u}] \theta^{-1}_{u \bar u})$ survives,
because all the other terms are zero for $R^2 \times R^2_{\theta}$. But the commutator is invariant under $\lambda$-rescaling. \par
We notice that, after rescaling, the integration variables 
$(\delta \hat D'_{u}, \delta \hat  D'_{\bar u})$
should be treated as independent.  For $\lambda$ real this is appropriate if we analytically continue the non-commutative plane to Minkowski space-time, after which the $\lambda$ invariance of the loop is simply
invariance under Lorentz boosts \footnote{We would like to thank Konstantin Zarembo for discussing with us the $\lambda$-independence at the GGI.}. The analytic continuation is also connected with the large-$\theta$ triviality (see below). QED \par
In fact the twistor Wilson loops are trivially $1$ at large-$ \theta $ to all orders in the 't Hooft coupling constant $g$:
\bea
\lim_{\theta \rightarrow \infty} <\frac{1}{ \cal N} Tr_{\cal N} \Psi(\hat B_{\lambda};L_{ww})  >=1 
\eea
Firstly, we show that  triviality holds to the lowest non-trivial order in perturbation theory.
We have in the Feynman gauge in the large-$\theta$ limit \footnote{We would like to thank Luca Lopez for working out a detailed version of this computation during our course at SNS.}:
\bea
&&<Tr_{\cal N}  \big( \int_{L_{ww}}(\hat A_z+\lambda \hat D_u) dz+(\hat A_{\bar z}+ \lambda^{-1}\hat D_{\bar u}) d \bar z \int_{L_{ww}}(\hat A_z+\lambda \hat D_u) dz+(\hat A_{\bar z}+ \lambda^{-1} \hat D_{\bar u}) d \bar z\big)> \nonumber \\
&&= 2 \int_{L_{ww}} dz\int_{L_{ww}}d \bar z (<Tr_{\cal N} (\hat A_z  \hat A_{\bar z})>+i^2 <Tr_{\cal N} (\hat A_u  \hat A_{\bar u})>) \nonumber \\
&&=0 
\eea

\subsection{Twistor Wilson loops are supported on Lagrangian submanifolds of twistor space}

Secondly, we show that triviality holds in the large $\theta$-limit to all orders of perturbation theory.
For this aim it is convenient to gauge away the non-commutative derivatives that occur in the definition of twistor Wilson loops.
This can be done by performing a local gauge transformation with values in the complexification of the gauge group. Although this is not a symmetry
of the theory, the trace of the twistor Wilson loops is left invariant because of the cyclicity property of the trace. Let be
\bea
\hat S(z, \bar z)= e^{i \lambda z \hat \partial_{ u}+ i {\lambda}^{-1} \bar z \hat \partial_{\bar u}} 
\eea
where $(z, \bar z)$ are commutative coordinates.
The components of the operator-valued gauge connection, $\hat B_{\lambda}$, transform under this gauge transformation in the usual way:
\bea
\hat B_{\lambda, z}^ {\hat S} = \hat S(z, \bar z) \hat B_{\lambda, z} \hat S(z, \bar z)^{-1} + i \partial_{z} \hat S(z, \bar z) \hat S(z, \bar z)^{-1} \nonumber \\
\hat B_{\lambda, \bar z}^ {\hat S} = \hat S(z, \bar z) \hat B_{\lambda, \bar z} \hat S(z, \bar z)^{-1} + i \partial_{\bar z} \hat S(z, \bar z) \hat S(z, \bar z)^{-1}
\eea 
where the partial derivatives are the usual partial derivations with respect to the commutative 
parameters $(z, \bar z)$. 
Correspondingly the twistor Wilson line transforms as \footnote{We ignore central terms that vanish for large $\theta$.}:
\bea
&& \hat S(w, \bar w)\Psi( \hat B_{\lambda}; L_{wv}) \hat S(v, \bar v)^{-1} \nonumber \\
&&= P \exp i \int_{L_{wv}} \hat B_{\lambda, z}^ {\hat S}  dz+  \hat B_{\lambda, \bar z}^ {\hat S}   d \bar z \nonumber \\
&&=  P \exp i \int_{L_{wv}}( \hat S(z, \bar z)   (\hat A_z+\lambda \hat D_u) \hat S(z, \bar z)^{-1} - \lambda \hat \partial_{u} ) dz  
+(  \hat S(z, \bar z)   (\hat A_{\bar z}+ \lambda^{-1} \hat D_{\bar u})  \hat S(z, \bar z)^{-1} 
 - \lambda^{-1}  \hat \partial_{\bar u} ) d\bar z \nonumber \\
&&= P \exp i \int_{L_{wv}} \hat S(z, \bar z)   (\hat A_z+i \lambda \hat A_u) \hat S(z, \bar z)^{-1}  dz +  \hat S(z, \bar z)   (\hat A_{\bar z}+ i \lambda^{-1} \hat A_{\bar u})  \hat S(z,\bar z)^{-1}  d \bar z 
\eea
Therefore the twistor loop lies effectively on the submanifold of four-dimensional commutative space-time defined by:
\bea
( z,  \bar z, u,  {\bar u})=(z, \bar z, i \lambda  z, i \lambda^{-1} \bar z)
\eea
with tangent vector:
\bea
(\dot z, \dot {\bar z}, \dot u, \dot {\bar u})=(\dot z,\dot {\bar z}, i \lambda \dot z, i \lambda^{-1}\dot {\bar z})
\eea
This is a Lagrangian submanifold of the (complexified) Euclidean space with respect to the Kahler form $dz \wedge d \bar z+ du \wedge d \bar u$ that lifts to a Lagrangian submanifold
of twistor space provided $\lambda$ is either real or a unitary phase. The two cases correspond to Lagrangian submanifolds of antipodal and circle type
respectively.

\subsection{Triviality of twistor Wilson loops in the limit of infinite non-commutativity}

The proof of triviality of twistor Wilson loops to all orders of perturbation theory in the limit $\theta \rightarrow \infty$ follows now almost immediately. \par
Indeed at any order in perturbation theory
a generic contribution to an ordinary Wilson loop of a commutative gauge theory contains a correlator of gauge fields,
i.e. a Green function, with tensor indices contracted with a product of monomials in
$\dot x_{\alpha}(s)$ at generic insertion points on the loop, labeled by $s$:
\bea
\int ds_1 ds_2..... G_{{\alpha_1}{\alpha_2}...}(x_{\beta}(s_1)-x_{\beta}(s_2),...) \dot x_{\alpha_1}(s_1) \dot x_{\alpha_2}(s_2) ...
\eea
Because of the $O(4)$ invariance of the commutative theory $\dot x_{\alpha_1}(s_1)$ is contracted either with another  $\dot x_{\alpha_2}(s_2)$ or with an $ x_{\alpha_2}(s_2)$  to form  polynomials in  $\dot x_{\alpha}(s) \dot x_{\alpha}(s')$ or in $\dot x_{\alpha}(s) x_{\alpha}(s')$. Indeed all these monomials necessarily contain at least one factor of  $\dot x_{\alpha}$ since the gauge field along the loop has the index contracted with the one of  $\dot x_{\alpha}$. The possible factor of $x_{\alpha}$ arises from the dependence of the Green functions on the coordinates.\par
We now specialize to twistor Wilson loops. \par
In the limit $\theta \rightarrow \infty$ of the non-commutative gauge theory $O(4)$ invariance is recovered, because the theory becomes the large-$N$ limit of the commutative theory,
that obviously is $O(4)$ invariant. \par
Therefore all the monomials just mentioned vanish when evaluated on the Lagrangian submanifold which
the twistor Wilson loop lies on, because they are of the form $\dot z(s) \dot{\bar z}(s')- \dot z(s) \dot{\bar z}(s')=0$ or $z(s) \dot{ \bar z}(s')-z(s) \dot{\bar z}(s')=0$. Thus the "effective propagators" that connect
a Feynman graph at any order to the twistor Wilson loop vanish. The only factors that may spoil the triviality occur if singularities due to denominators of Feynman diagrams
arise, since $ x_{\alpha}(s)  x_{\alpha}(s')$ vanishes too on the Lagrangian submanifold for the same reasons. To cure this we analytically continue the correlators that occur in the computation of twistor Wilson loops
from Euclidean to Minkowski space-time in order to get the $i \epsilon$ prescription, $z_+(s)  z_-(s')-z_+(s)  z_-(s')+i \epsilon=i\epsilon$
in the denominators. \par
The gauge invariant prescription of analytic continuation from Euclidean to Minkowski space-time will be used over and over in the paper and it will play a crucial role. QED \par
We describe now the aforementioned analytic continuation of the twistor Wilson loops at the operator level in the functional integral by means of the following sequence. \par
Firstly, we analytically continue to Minkowski space-time only the commutative plane. Then at operator level the twistor Wilson loops become:
\bea
Tr_{\cal N} \Psi(\hat B_{\lambda};L_{ww}) \rightarrow Tr_{ \cal{N}} P \exp i \int_{L_{ww}}(\hat A_{z+}+i \lambda \hat D_u) dz_++(\hat A_{ z_-}+i \lambda^{-1} \hat D_{\bar u}) dz_- 
\eea
since $x_0 \rightarrow i x_4$, $z \rightarrow i z_+$ and $\bar z \rightarrow i z_-$, with $(z_+=x_4+x_1, z_-=x_4-x_1)$ and $A_z \rightarrow -i A_{z_+}$,  $A_{\bar z} \rightarrow -i A_{z_-}$.
The support of the twistor Wilson loops analytically continued in this way becomes:
\bea
(z, \bar z, i \lambda  z, i \lambda^{-1} \bar z)
 \rightarrow (z_+, z_-, -\lambda  z_+, -\lambda^{-1} z_-)
\eea
that is Lagrangian with respect to $- dz_+ \wedge dz_-+ du \wedge d \bar u$ for a real section of the complexified Euclidean space-time. \par
In sect.(6) we write a holomorphic loop equation that the twistor Wilson loops satisfy in Euclidean space-time. The holomorphic loop equation
involves in the left hand side an Euclidean effective action that should be renormalized in Euclidean space-time and in the right hand side a contour integral along the loop that is not well defined in Euclidean space-time and that should be regularized. \par
We will show that there is an essentially unique way of regularizing by analytical continuation to Minkowski space-time.
Thus after renormalization of the effective action the holomorphic loop equation makes sense in Minkowski space-time.
In sect.(11) the renormalized effective action in Minkowski space-time restricted to fluctuations of surface operators supported on the Lagrangian submanifold:
\bea
 (z_+, z_-, -\lambda  z_+, -\lambda^{-1} z_-)
\eea
is used to compute the glueball spectrum. In the effective action this Lagrangian submanifold is obtained first restricting to the Lagrangian submanifold in Euclidean space:
\bea
(z,\bar z,  -\lambda  z,  -\lambda^{-1} \bar z)
\eea
and then analytically continuing.
This can be done in two ways.
The following choice for the analytic continuation $u \rightarrow i u_+ , \bar u \rightarrow i u_-$ leads to trivial twistor Wilson loops:
\bea
Tr_{\cal N} \Psi(\hat B_{\lambda};L_{ww}) \rightarrow Tr_{ \cal{N}} P \exp i \int_{L_{ww}}(\hat A_{z_+} + \lambda \hat D_{u_+}) dz_+ +(\hat A_{ z_-}+\lambda^{-1} \hat D_{ u_-}) dz_- 
\eea
supported on the Lagrangian submanifold:
\bea
 (z_+, z_-, i\lambda  z_+, i\lambda^{-1} z_-)
\eea
for which the correlators that occur in the evaluation of the twistor Wilson loops are vanishing as shown in the triviality proof.
The other choice for the analytic continuation $u \rightarrow  u_+ , \bar u \rightarrow  u_-$ leads to:
\bea
Tr_{\cal N} \Psi(\hat B_{\lambda};L_{ww}) \rightarrow Tr_{ \cal{N}} P \exp i \int_{L_{ww}}(\hat A_{z_+} + i \lambda \hat D_{u_+}) dz_+ + (\hat A_{ z_-}+i\lambda^{-1} \hat D_{ u_-}) dz_- 
\eea
supported on the Lagrangian submanifold:
\bea
 (z_+, z_-, -\lambda  z_+, -\lambda^{-1} z_-)
\eea
and to non-trivial twistor loops that satisfy the same loop equation in Minkowski, but not the localization property.  
The point is that the effective action, $\Gamma$, is naturally defined on the Lagrangian submanifold that is the support of the non-trivial twistor Wilson loops,
and thus it carries the interesting information of the localization. \par 
The trivial twistor loops are obviously finite at  $\theta = \infty$, i.e. they have no cusp and perimeter divergences, in analogy with certain supersymmetric Wilson loops. \par
Indeed the cognoscenti may have noticed that twistor Wilson loops resemble locally $BPS$ Wilson loops of theories with extended supersymmetry.
In fact our triviality proof mimics the argument about a certain non-renormalization property \cite{Gr} 
of locally $BPS$ Wilson loops. Indeed it has been argued in \cite{Gr} that a locally $BPS$ Wilson loop in the four-dimensional $ \cal{N} $ $=4$ $SUSY$ gauge
theory:
\bea
Tr P \exp i\int_{L} A_{\alpha}\dot {x}_{\alpha}(s) ds+i\phi_{b} \dot{y}_{b}(s) ds
\eea
has no perimeter divergence, to all orders in perturbation theory, because of the local $BPS$ constraint:
\bea
\sum_{\alpha} \dot x^2_{\alpha}(s)-\sum_{b} \dot y^2_{b}(s)=0
\eea
At lowest order of perturbation theory this constraint assures the cancellation
of the contribution to the perimeter divergence of the gauge propagator versus the scalar propagator, because of the 
factor of $i^2$ in front of the scalar propagator at that order. \par
As far as the perimeter divergence is concerned, it is argued in \cite{Gr} that this cancellation occurs
to all orders in perturbation theory, when the locally $BPS$ Wilson loop is seen as the dimensional reduction
to four dimensions of the ten-dimensional Wilson loop of the ten-dimensional $ \cal{N} $ $=1$ $SUSY$ $YM$ theory
from which the four-dimensional $ \cal{N} $ $=4$ $SUSY$ theory is obtained. \par
Remarkably, in the argument of \cite{Gr}
$SUSY$ plays no direct role. In fact the argument is based only on $O(10)$ rotational symmetry of the parent $d=10$ $\cal{N}$ $=1$ $SUSY$
gauge theory from which the daughter $d=4$ $\cal{N}$ $=4$ $SUSY$
gauge theory derives by dimensional reduction,
as we show momentarily. \par
 In ten dimensions the coefficient of the perimeter divergence
of an ordinary unitary Wilson loop, at any order in perturbation theory,
must necessarily contain as a factor a polynomial in the $O(10)$ invariant 
quantity $\sum_{M} \dot x^2_{M}(s)$.
Indeed the perimeter divergence arises
when all insertion points coincide, in such a way that all the arguments
of the Green function vanish. \par
In this case the Green function provides
a factor that, by $O(10)$ rotational invariance, must be a polynomial
in ten-dimensional Kronecker delta, since all the difference vectors in the Green function are zero at coinciding points
and thus no other tensorial structure can be produced. \par
This combines with the factors of $\dot x_{M}(s)$ to produce an invariant polynomial
in $\sum_{M} \dot x^2_{M}(s)$ with no constant term, since the lowest order
contribution is zero by direct computation.
But $\sum_{M} \dot x^2_{M}(s)=\sum_{\alpha} \dot x^2_{\alpha}(s)-\sum_{b} \dot y^2_{b}(s)=0$  is zero for a $BPS$ Wilson loop because of
the $BPS$ constraint.
A naive application of this argument to the four-dimensional $BPS$ Wilson loop
would imply the absence of the perimeter divergence for this loop on the basis of the $O(10)$
rotational invariance of the theory before the dimensional reduction. QED \par

\section{The quasi-localization lemma for twistor loops in large-$N$ $YM$}

We use the $\lambda$-independence to show that the v.e.v. of twistor Wilson loops is localized on the sheaves fixed by the semigroup
rescaling $\lambda$. This involves a delicate and subtle interchange between limit and integration, that will be justified in sect.(8), after introducing a lattice regularization of the functional integral
of differential geometric nature in sect.(7). In addition it will be checked by direct computation that the result of the localization agrees with the perturbative triviality. \par
In this section for simplicity we use a notation that does not distinguish between commutative
and non-commutative theories and therefore we do not add hats to operator valued quantities of non-commutative theories.
The framework has been set in the previous section, therefore this use should not generate ambiguities. \par 
It is convenient to choose our twistor Wilson loops in the adjoint representation and to use the fact that in the large-$N$ limit
their v.e.v. factorizes in the product of the v.e.v. of the fundamental representation and of its conjugate.
Then, for the factor in the fundamental representation, localization proceeds as follows.
We write the $YM$ partition function by means of the non-$SUSY$ analog \cite{MB1} of the Nicolai map of $\cal{N}$ $=1$ $SUSY$ $YM$ theory worked out in sect.(3.3),
introducing in the functional integral the appropriate resolution of identity:
\bea
1= \int \delta(F^{-}_{\alpha \beta}-\mu^{-}_{\alpha \beta}) \delta\mu^{-}_{\alpha \beta}  
\eea
\bea
Z=\int \exp(-\frac{N 8 \pi^2 }{g^2} Q-\frac{N}{4g^2} \sum_{\alpha \neq \beta} \int Tr_f(\mu^{-2}_{\alpha \beta}) d^4x)
 \delta(F^{-}_{\alpha \beta}-\mu^{-}_{\alpha \beta}) \delta\mu^{-}_{\alpha \beta} \delta A_{\alpha} 
\eea 
$Q$ is the second Chern class (the topological charge) and $\mu^{-}_{\alpha \beta}$
is a field of $ASD$ type. The equations of $ASD$ type in the resolution of identity,
$F_{01}-F_{23}=\mu^-_{01} ,
F_{02}-F_{31}=\mu^-_{02} ,
F_{03}-F_{12}=\mu^-_{03} $,
can be rewritten in the form of a Hitchin system (taking into account the central extension that occurs in the non-commutative case):
\bea
-i F_A+[D,\bar D] -\theta^{-1}1&&=\mu^0=\frac{1}{2}\mu^-_{01} \nonumber \\
-i\partial_{A}  \bar D&&= n=\frac{1}{4}(\mu^-_{02}+i\mu^-_{03}) \nonumber \\
-i\bar \partial_A D&&=\bar n=\frac{1}{4}(\mu^-_{02}-i\mu^-_{03}) 
\eea
or equivalently in terms of the non-Hermitian
connection whose holonomy is computed by the twistor Wilson loop with parameter $\rho$,
$B_{\rho}=A+\rho D+ \rho^{-1} \bar D=(A_z+ \rho D_u) dz+(A_{\bar z}+ \rho^{-1} D_{\bar u}) d \bar z$,
\bea
-i F_{B_{\rho}} -\theta^{-1}1&&= \mu_{\rho}=\mu^0+\rho^{-1} n- \rho \bar n  \nonumber \\
-i\partial_{A}  \bar D&&= n  \nonumber \\
-i\bar \partial_A D&&=\bar n
\eea
The resolution of identity in the functional integral
then reads:
\bea
1=\int \delta n \delta  \bar n \int_{C_{\rho}} \delta \mu_{\rho} \delta(-i F_{B_{\rho}} - \mu_{\rho}-\theta^{-1}1) \delta(-i\partial_{A} \bar D- n) \delta(-i\bar \partial_A D- \bar n) 
\eea
where the measure, $\delta \mu_{\rho}$, along the path, $C_{\rho}$, is over the non-Hermitian path with fixed $n$ and $\bar n$ and varying $\mu^0$. The resolution
of identity is independent, as $\rho$ varies, on the complex path of integration, $C_{\rho}$. Indeed the constraint implied by the first of Eq.(5.4) on  $C_{\rho}$ is a linear complex combination
of the Hermitian constraint  in the first  of Eq.(5.3) with coefficient $1$  and of the remaining two with coefficients depending on $\rho$.  \par
Let us consider the v.e.v. of twistor Wilson loops:
\bea
&&\int \delta n \delta \bar n \int_{C_{\rho}}\delta  \mu_{\rho} \exp(-\frac{N 8 \pi^2 }{g^2} Q-\frac{N 4}{g^2}  \int Tr_f( \mu^{0})^2  +4Tr_f(n \bar n) d^4x) \nonumber\\
&& Tr_f P \exp i \int_{L_{ww}}(A_z+\lambda D_u) dz+(A_{\bar z}+ \lambda^{-1}D_{\bar u}) d \bar z  \nonumber \\
&&\delta(-i F_{B_{\rho}} - \mu_{\rho}-\theta^{-1}1) \delta(-i\partial_{A} \bar D- n) \delta(-i\bar \partial_A D- \bar n) 
\delta A \delta \bar A \delta D \delta \bar D
\eea
and let us change variables in the functional integral rescaling the non-commutative covariant derivatives:
\bea
&&\int \delta n \delta \bar n \int_{C_{\rho}}\delta  \mu_{\rho} \exp(-\frac{N 8 \pi^2 }{g^2} Q-\frac{N4}{g^2}  \int Tr_f(\mu^{0})^2+4Tr_f(n \bar n) d^4x)\nonumber \\
&& Tr_f P \exp i \int_{L_{ww}}(A_z+D'_u) dz+(A_{\bar z}+ D'_{\bar u}) d \bar z  \nonumber \\
&&\delta( -i  F_A+ [D',\bar D']-\theta^{-1}1-\mu^0
-i\frac{\lambda}{\rho} \partial_{A}  \bar D' +i \frac{\rho}{\lambda} \bar \partial_A D' -\rho^{-1} n+\rho \bar n)\nonumber \\
&&\delta(-i \lambda \partial_{A} \bar D'-n) \delta(-i \lambda ^{-1}\bar \partial_A D'-  \bar n)
\delta A \delta \bar A \delta D' \delta \bar D' 
\eea
Taking the limit $\lambda \rightarrow 0$ inside the functional integral, the last line implies localization on $ n=0$ and 
$\bar \partial_A D' =0$. The $\delta n$ integral is performed by means of the second delta function. $\delta \bar n$ decouples from the third delta function.
The independence on the path $C_{\rho}$ in the neighborhood of $\rho=0$, that we denote, choosing $\rho= \alpha \lambda$, $C_{\alpha {\lambda}^+}$ with $\alpha$ fixed as $\lambda \rightarrow 0^+$, implies that the $\delta \bar n$ integral decouples and that 
$\partial_{A}  \bar D'=0$ as well. \par
Firstly, on $C_{\alpha {\lambda}^+}$ the integral over $\delta \bar n$ decouples because $\delta \bar n$ disappears from the first delta function as well.
Secondly, the first delta function implies  $n=0$ as $\lambda \rightarrow 0$ as well. Therefore, setting $n=0$ from the start in the first delta function because of the second delta function, 
the argument of the first delta function contains the combination $-i  F_A+ [D',\bar D']-\theta^{-1}1-\mu^0 +i \alpha^{-1} \partial_{A}  \bar D' $,
that can be $\alpha$ independent and thus zero for every $\alpha$ only if the two terms are zero separately. Therefore also $\partial_{A}  \bar D'=0$.
Notice that for this argument to hold
it is not necessary to consider the variables $(n, D')$ as Hermitian conjugated to $(\bar n, \bar D')$. Indeed they are not so because of the $\lambda$ rescaling and/or the analytic continuation that is necessary to regularize the twistor loops for the triviality property to hold. \par
We notice that the localized density has a holomorphic ambiguity,
since we can represent the same measure using a different density, performing holomorphic transformations without spoiling the quasi-localization lemma : $
\delta  \mu_{0^{+}}= \frac{\delta  \mu_{0^{+}}} {\delta  \mu'_{0^{+}}} \delta \mu'_{0^{+}} $.
The final result for the localized effective measure is:
\bea
\big[ \int_{C_{0^{+}}}  \delta \mu'_{0^{+}}  \frac{\delta  \mu_{0^{+}} } { \delta  \mu'_{0^{+}} }
\exp(-\frac{N 8 \pi^2 }{g^2} Q-\frac{N}{4g^2} \sum_{\alpha \neq \beta} \int Tr_f(\mu^{-2}_{\alpha \beta}) d^4x)
\delta(F^{-}_{\alpha \beta}-\mu^{-}_{\alpha \beta})  \big]_{n=\bar n=0} \delta A_{\alpha} 
\eea
where we have reintroduced the covariant notation. \par
Thus the twistor loops are localized on the fixed sheaves for which two of the $ASD$ fields vanish: $\mu^-_{02}=\mu^-_{03}=0$. QED \par
The integration on the gauge connection in Eq.(5.8) can be explicitly performed in the Feynman gauge in the way explained in sect.(3.3), to obtain: 
\bea
&& |\int_{C_1}   { \delta  \mu' } e^{-\Gamma}|^2  \nonumber \\
&& =\big [ \int_{C_1}  { \delta  \mu'} \exp(-\frac{N 8 \pi^2 }{g_W^2} Q-\frac{N}{4g_W^2} \sum_{\alpha \neq \beta} \int Tr_f(\mu^{-2}_{\alpha \beta}) d^4x) \nonumber \\
&& Det^{-\frac{1}{2}}(-\Delta_A^. \delta_{\alpha \beta} -i ad_{ \mu^-_{\alpha \beta}} ) Det(-\Delta_A^.)
(\frac{\Lambda}{2 \pi})^{n_b} Det^{\frac{1}{2}} \omega    \frac{\delta  \mu } { \delta  \mu' }     \times c.c. \big] _{n=\bar n=0}
\eea
The complex conjugate factor arises by the conjugate representation, $Det^{\frac{1}{2}} \omega$ is the contribution of the $n_b$ zero modes due to the moduli, and $\Lambda$ the corresponding Pauli-Villars regulator. \par
The volume form on the connections admits several different representations as the Liouville measure associated to a symplectic form, since different symplectic forms
may lead to the same volume form. One possible choice for the symplectic form, $\omega$, is $\omega_I$, displayed in Eq.(3.82). However, in sect.(7) and sect.(12) we will see that there is a different choice of $\omega$,
compatible with holomorphic/antiholomorphic fusion, that is most convenient in the computation of the glueball spectrum.

\section{Holomorphic loop equation for twistor Wilson loops}

\subsection{Holomorphic loop equation}

We now specialize to the case $\rho=1$ for the twistor connection $B_{1}=B$. The partition function reads: 
\bea
Z&&=\int \delta n \delta \bar n \int_{C_{1}}\delta  \mu'  \frac{\delta  \mu } { \delta  \mu' }  \exp(-\frac{N 8 \pi^2 }{g^2} Q-\frac{N 4}{g^2}  \int Tr_f( \mu^{0})^2  +4Tr_f(n \bar n) d^4x) \nonumber\\
&& \delta(-i F_{B} - \mu-\theta^{-1}1) \delta(-i\partial_{A} \bar D- n) \delta(-i\bar \partial_A D- \bar n) 
\delta A \delta \bar A \delta D \delta \bar D
\eea
Writing the holomorphic loop equation for twistor Wilson loops requires that $\mu'$ be chosen in the holomorphic gauge, $B_{\bar z}=0$. 
Strictly speaking this is a change of variables and not a gauge transformation with values in the unitary gauge group, since this gauge can be reached only
by a gauge transformation with values in the complexification of the gauge group. \par
The further change of variables to the holomorphic gauge is a new key feature of our approach to the
large-$N$ $YM$ theory. It is based on the idea that twistor Wilson loops, being holomorphic functionals
of $\mu'$, behave as the chiral (i.e. holomorphic) super-fields of an $\cal{N}$ $=1$ $SUSY$ gauge theory.
In fact the new holomorphic loop equation resembles for the cognoscenti the holomorphic loop equation that occurs
in the Dijkgraaf-Vafa theory \cite{Vafa,Laz,Kaw,WD} of the  glueball superpotential \footnote{The name is perhaps misleading because the glueball superpotential cannot 
be used to compute the mass of any glueball state of the $\cal{N}$ $=1$ $SUSY$ theories.} in $\cal{N}$ $=1$ $SUSY$ gauge theories.
Thus the v.e.v. is taken with respect to the measure:
\bea
<...>&&= Z^{-1} \int \delta n \delta \bar n \int_{C_{1}} \delta  \mu'  ... \exp(-\frac{N 8 \pi^2 }{g^2} Q-\frac{N 4}{g^2}  \int Tr_f( \mu \bar \mu)  +Tr_f(n + \bar n)^2 d^4x) \nonumber\\
&& \delta(-i F_{B} - \mu-\theta^{-1}1) \delta(-i\partial_{A} \bar D- n) \delta(-i\bar \partial_A D- \bar n) \frac{\delta  \mu } { \delta  \mu' }  
\delta A \delta \bar A \delta D \delta \bar D
\eea
The holomorphic loop equation is obtained following the Makeenko-Migdal technique, as an identity that expresses the fact that the functional integral of a functional derivative vanishes:
\bea
\int  Tr\frac{\delta }{\delta \mu'(z,\bar z)} (e^{-\Gamma}\Psi(B'; L_{zz})) \delta \mu'=0
\eea
The new holomorphic loop equation for twistor loops follows:
\bea
<Tr(\frac{\delta \Gamma}{\delta \mu'(z,\bar z)}\Psi(B'; L_{zz}))>=
\frac{1}{ \pi } \int_{L_{zz}} \frac{ dw}{z-w} <Tr\Psi(B'; L_{zw})> <Tr\Psi(B'; L_{wz})> 
\eea
where $\Psi(B'; L_{zz})$ is the holonomy of $B$ in the gauge $B'_{\bar z}=0$.
The Cauchy kernel arises as the kernel of the operator $\bar \partial^{-1}$ that occurs by functionally differentiating $\Psi(B'; L_{zz})$ with respect to $\mu'$ via $i \bar \partial B_z'=\mu'$.
Assuming that the loop $L_{zz}$ is simple, i.e. it has no self-intersections, the holomorphic loop equation linearizes:
\bea
<Tr(\frac{\delta \Gamma}{\delta \mu'(z, \bar z)}\Psi(B'; L_{zz}))>=
\frac{1}{ \pi } \int_{L_{zz}} \frac{ dw}{z-w} <Tr\Psi(B'; L_{zw})><Tr1>
\eea

\subsection{Regularization by analytic continuation to Minkowski space-time}

The contour integration in the right hand side (i.e in the term that accounts for quantum fluctuations) of the loop equation
includes the pole of the Cauchy
kernel. We need therefore a regularization. \par
The natural choice consists in analytically continuing the loop equation
from Euclidean to Minkowski space-time, $z \rightarrow i( z_+ + i \epsilon)$.
It is at the heart of the Euclidean approach to quantum field theory that
this analytic continuation be in fact possible. \par
In the approach to localization by the holomorphic loop equation the analytic continuation
is performed only after functional integration and renormalization, that are performed in Euclidean space. Thus we think that this procedure has chances to work also from the point of view of the constructive
quantum field theory. \par
In fact the approach to localization via the holomorphic loop equation, when combined with the integration on local systems of the next section,
leads to more complete and satisfactory results than the localization on fixed points.  \par
The result of the $i \epsilon$ regularization of the Cauchy kernel is the sum of
two distributions, the principal part of the real Cauchy kernel and
a one-dimensional delta function:
\bea
\frac{1}{z_+ -w_+ +i\epsilon}=  P\frac{1}{z_+ -w_+}- i \pi \delta(z_+ -w_+)
\eea
The loop equation thus regularized is:
\bea
&&<Tr(\frac{\delta \Gamma_M}{\delta \mu'(z_+,  z_-)}\Psi(B'; L_{z_+z_+}))> \nonumber \\
&&=\frac{1}{ \pi} \int_{L_{z_+z_+}} (P\frac{dw_+}{z_+ -w_+} - i \pi  dw_+ \delta(z_+ -w_+)) <Tr\Psi(B'; L_{z_+w_+})> <Tr\Psi(B'; L_{w_+z_+})>
\eea
where now both the distributions on the right hand side are integrable along the loop.
This regularization has the great virtue of being manifestly gauge invariant, an unusual feature 
for loop equations.
In addition this regularization is not loop dependent. \par
The right hand side of the loop equation
contains now two contributions.
A delta-like one dimensional contact term, that is supported on closed
loops and a principal part distribution that is supported 
on open loops. Since by gauge invariance it is consistent to assume
that the expectation value of open loops vanishes, as in Eq.(4.27), the principal part 
does not contribute and the loop equation in the holomorphic gauge reduces to:
\bea
&&<Tr(\frac{\delta \Gamma_M}{\delta \mu'(z_+,  z_-)}\Psi(B'; L_{z_+z_+}))> \nonumber \\
&&= - i \int_{L_{z_+z_+}}  dw_+ \delta(z_+ -w_+) <Tr\Psi(B'; L_{z_+w_+})> <Tr\Psi(B'; L_{w_+z_+})>
\eea
As briefly outlined in sect.(1.1), the holomorphic loop equation for cusped loops with the shape of the symbol $\infty$
and the cusp at the non-trivial self-intersection point reduces to a critical equation for the effective action $\Gamma$
in Minkowski signature:
\bea
<Tr(\frac{\delta \Gamma}{\delta \mu'(z_+, z_-)}\Psi(B'; L_{z_+z_-}))>=0
\eea
since the contour integral in the right hand side of the holomorphic loop equation vanishes
because of the opposite orientation in a neighborhood of the backtracking cusp.
The geometrical side of this localization of the loop equation is described in sect.(10). \par
Taking $w_+=z_+$ and using the transformation properties of $\mu'$ and of the holonomy of $B'$,
the preceding equation can be rewritten in terms
the curvature, $\mu$, and of the connection, $B$, in a unitary gauge:
\bea
&&<Tr(\frac{\delta \Gamma}{\delta \mu(z_+,  z_-)}\Psi(B; L_{z_+z_+}))> \nonumber \\
&&= - i \int_{L_{z_+z_+}}  dw_+ \delta(z_+ -w_+) <Tr\Psi(B; L_{z_+w_+})> <Tr\Psi(B; L_{w_+z_+})>
\eea
where we have used the condition that v.e.v. of the trace of open loops vanishes
to substitute the holonomy of $B'$ with the holonomy of $B$. \par
As a consequence of the localization of the holomorphic loop equation, the equation of motion of $\Gamma$ vanishes
when restricted to the subalgebra of twistor Wilson loops:
\bea
<Tr(\Psi(B; L'_{z_+z_+})\frac{\delta \Gamma}{\delta \mu(z_+,z_-)}\Psi(B; L''_{z_+z_+}))>= 0
\eea
where $L'$ and $L''$ are the two petals of the loop.
In the next section we construct an explicit realization of this sublagebra,
on which we can interpret the localization of the holomorphic loop equation strongly:
\bea
\frac{\delta \Gamma}{\delta \mu(z_+,z_-)}= 0
\eea

\section{Integrating on surface operators in large-$N$ $YM$}

\subsection{Integrating on infinite-dimensional local systems}

We would like to give a precise mathematical meaning to the formal manipulations of the functional measure in sect.(5) and sect.(6).
One possibility would be introducing a lattice regularization of the functional integral according to Wilson \cite{Wilson}.
However, this kind of lattice regularization would spoil completely the geometrical structure, since 
in the Wilson regularization the gauge connection lives on links and the curvature on plaquettes, a fact that makes exploiting the map from the connection
to the $ASD$ curvature problematic, not to mention the understanding of the moduli of the loci 
at which this map is not one-to-one, for which the zero modes necessary to get the correct beta function occur (sect.(3.3) and sect.(3.4)) . \par
Therefore we introduce a new regularization of the $YM$ functional integral
that allows us to maintain the differential geometric structure.
The differential geometric structure is crucial to get a structure theory of the locus of the fixed points of the functional measure 
and to understand the zero modes 
of the determinants, that in turn affect the beta function of the theory. \par
Our new regularization of the $YM$ theory in the large-$N$ limit is performed in two steps.
In the first step the resolution of identity in the Nicolai map on $R^2 \times R^2_{\theta}$ is represented in the operator notation of sect.(4.1) as a functional integral on infinite-dimensional parabolic bundles,
as suggested long ago in \cite{MB2, MB3}:
\bea
1= \int \delta(-i[\hat D_{\alpha},\hat D_{\beta}]^--\sum_p \hat \mu^{-}_{\alpha \beta}(p)  \delta^{(2)} (z-z_p) - \theta^{-1}_{\alpha \beta} \hat 1 ) \prod_{p} \delta \hat \mu^{-}_{\alpha \beta}(p) 
\eea
In this notation all the dependence on the non-commutative coordinates is absorbed into the infinite dimensional nature of the operators that occur in the non-commutative
Eguchi-Kawai reduction. Therefore the base of the infinite-dimensional parabolic bundles is the two-dimensional surface,  $R^2$, labelled by the commutative coordinates $(z, \bar z)$. \par
This amounts to substitute the continuous field, $\hat \mu^{-}_{\alpha \beta}(z, \bar z)$, of the Nicolai map with the lattice field, $\hat \mu^{-}_{\alpha \beta}(p)$, 
by the resolution:
\bea
\hat \mu^{-}_{\alpha \beta}(z, \bar z)=\sum_p \hat \mu^{-}_{\alpha \beta}(p)\delta^{(2)} (z-z_p)
\eea
This resolution is dense in the sense of distributions, since for any smooth test function of compact support:
\bea
N^{-1}_D\sum_p f(z_p ,\bar z_p)  \hat \mu^{-}_{\alpha \beta}(p) \rightarrow \int f(z,\bar z) \hat \mu^{-}_{\alpha \beta}(z, \bar z) d^2z
\eea
On this dense set in function space \footnote{The three operators $  \hat \mu^{-}_{\alpha \beta}(p)$ are reduced in fact to large finite dimensional matrices by Morita equivalence as explained 
momentarily. In the finite dimensional case these matrices all commute as a consequence of the local model of the Hitchin equations \cite{S5, S6}. Therefore the resolution in Eq.(7.2) turns out to be dense
in function space in the sense of the distributions only in a certain neighborhood of the fixed points, $ \hat \mu^{-}_{02}(p)= \hat \mu^{-}_{03}(p)=0$. The missing degrees of freedom occur as moduli (see below).
Nevertheless the local degrees of freedom and the moduli in a neighborhood of the fixed points are enough to reproduce the correct universal one- and two- loop contributions to the beta function (sect.(9) and sect.(11)) . Hence the mentioned degrees of freedom are in fact dense in function space in a neighborhood of the fixed points in the large-$N$ limit (see below).  } the resolution of identity of the Nicolai map can be interpreted as hyper-Kahler reduction \cite{MB2, MB3}.
Indeed the three constraints of $ASD$ type are the Hermitian and the complex moment maps for the Hamiltonian action of the infinite-dimensional unitary gauge group on
the commutative plane $R^2$:
\bea
-i F_{\hat A}+[\hat D,\hat {\bar D}] -\theta^{-1}1&=& \sum_p \hat \mu^0_p \delta^{(2)} (z-z_p)  \nonumber \\
-i\partial_{\hat A}^.  \hat {\bar D}&=&\sum_p  \hat n_p  \delta^{(2)} (z-z_p)  \ \nonumber \\
-i\bar \partial_{\hat A}^. \hat D&=&\sum_p \hat {\bar n}_p \delta^{(2)} (z-z_p)
\eea
with respect to the three symplectic forms \cite{MB2,MB3,MB1} \footnote{We use the same labels $(I,J,K)$ of the symplectic forms as in \cite{W2} for the finite dimensional case.}:
\bea
\omega_I&&=\frac{1}{2 \pi}\int d^2 z  tr_f \hat Tr (\delta {\hat A}_{z} \wedge \delta {\hat A}_{\bar z}+ \delta {\hat D}_u \wedge \delta {\hat D}_{\bar u}) \nonumber \\
\omega_J-i\omega_K&&=\frac{1}{2 \pi i}\int d^2 z  tr_f \hat Tr (\delta {\hat A}_{z} \wedge \delta {\hat D}_{\bar u} ) \nonumber \\
\omega_J+i\omega_K&&=\frac{1}{2 \pi i}\int d^2 z  tr_f \hat Tr (\delta {\hat A}_{\bar z} \wedge \delta {\hat D}_{u}) 
\eea
as it follows immediately from the interpretation as (infinite-dimensional) Hitchin systems \cite{H,HKL}. \par
Thus we come to the remarkable conclusion that, on a dense set associated to a lattice divisor, the Nicolai map can be interpreted, up to gauge equivalence, as a resolution of the gauge connection
into orbits parametrized formally by hyper-Kahler moduli spaces, that arise as the quotient of the manifold defined by Eq.(7.4) for the action of the unitary gauge group \cite{H,HKL}.
Yet, for the moment, this construction is somehow formal, because of the infinite-dimensional nature of the bundles involved.
Thus it is unclear what the moduli theory of these infinite-dimensional bundles is. \par

\subsection{Reducing to finite dimension by Morita duality and inductive structure}

In order to reduce to finite dimensional bundles a possible way out is to compactify the non-commutative plane, $R^2_{\theta}$, on a non-commutative torus of large area, $L^2$.
The corresponding non-commutative $U(N)$ gauge theory enjoys, for rational values of the dimensionless non-commutative parameter, $2 \pi \theta L^{-2}= \frac{\hat M}{\hat N}$, 
Morita duality \cite{Szabo2,AG,GA,K} to a theory on a commutative torus of area  $L^2 \hat N^{-2}$, with gauge group $U(N \times \hat N)$, with the same 't Hooft coupling constant $g$, and 
with twisted boundary conditions corresponding to a 't Hooft flux. \par
Indeed, starting from the infinite-dimensional non-commutative case we can perform an infinite-dimensional unitary gauge transformation, $\hat U(u,\bar u)$, depending on the commutative parameters, $(u, \bar u)$,
in such a way that:
\bea
-i[\hat D_{\alpha},\hat D_{\beta}]^--\sum_p \hat \mu^{-}_{\alpha \beta}(p)  \delta^{(2)} (z-z_p) - \phi_{\alpha \beta} \hat 1 =0
\eea
becomes:
\bea
&&\big[\partial_{\alpha} \hat A _{\beta}(z,\bar z, u, \bar u) -\partial_{\beta} \hat A _{\alpha}(z,\bar z, u, \bar u) +i [\hat A _{\alpha}(z,\bar z, u, \bar u) , \hat A _{\beta}(z,\bar z, u, \bar u)]\big]^- \nonumber \\
&&= \sum_p \hat U(u,\bar u) \hat \mu^{-}_{\alpha \beta}(p) \hat U^{-1}(u,\bar u) \delta^{(2)} (z-z_p) +\phi_{\alpha \beta} \hat 1
\eea
where:
\bea
\hat A _{\beta}(z,\bar z, u, \bar u)= \hat U(u,\bar u) \hat A _{\beta}(z,\bar z) {\hat U}^{-1}(u,\bar u)+i \partial_{\beta}\hat U(u,\bar u) {\hat U}^{-1}(u,\bar u)
\eea
and $ \phi_{\alpha \beta} $ is an antisymmetric field that takes into account a possible $U(1)$ background flux in addition to the central term that arises by the inverse of the non-commutativity. \par
The content of Morita duality is that these seemingly infinite-dimensional equations admit a finite dimensional solution \cite{Szabo2} on a commutative torus with coordinates $(u, \bar u)$, gauge group $U(N \times \hat N)$, twisted boundary conditions corresponding to a 't Hooft flux and a $U(1)$ background flux, $\phi'_{\alpha \beta}$. Despite the structure group is $U(N \times \hat N)$, the Morita equivalent theory does not describe the most general $U(N \times \hat N) $ gauge theory, 
but only the one that satisfies the twisted boundary conditions. In particular the twisted boundary conditions require solutions of the kind:
\bea
&&\big[\partial_{\alpha}  A _{\beta}(z,\bar z, u, \bar u) -\partial_{\beta}  A _{\alpha}(z,\bar z, u, \bar u) +i [ A _{\alpha}(z,\bar z, u, \bar u) ,  A _{\beta}(z,\bar z, u, \bar u)]\big]^- \nonumber \\
&&= \sum_p  U(u,\bar u) \mu^{-}_{\alpha \beta}(p) U^{-1}(u,\bar u) \delta^{(2)} (z-z_p) + \phi'_{\alpha \beta} 1_{N \times \hat N}
\eea
where now $U(u,\bar u)$ are $U(\hat N)$ matrices and $ \mu^{-}_{\alpha \beta}(p) $ lives in the tensor product of Lie algebras, $u(N) \times u(\hat N)$, \cite{Szabo2,AG, GA, K}. 
These are the equations that define surface operators in finite dimension but for the fact that the gauge connection satisfies twisted boundary conditions on the torus. \par
More explicitly any $U(N)$ connection of the $YM$ theory on $R^2 \times T^2_{\theta}$, with coordinates $(y, \hat x)$, with periodic boundary conditions on the non-commutative torus, $T^2_{\theta}$, with $2 \pi \theta L^{-2}= \frac{\hat M}{\hat N}$, admits the expansion \cite{AG}:
\bea
A(\hat x,y)=\sum_{l \in  Z^2} a_l(y) e^{-2 \pi i l^. \hat x/L}
\eea
and the corresponding Morita equivalent $U( N \hat N)$ connection reads \cite{AG}:
\bea
A'(x,y)=\sum_{l \in  Z^2} a_l(y) V^{-\hat M l_1} U^{l_2} \omega^{-\hat M l_1 l_2/2} e^{-2 \pi i l^. x/ \hat N L}
\eea
where the matrices $(U, V)$ are the clock and shift matrices of $SU(\hat N)$ and $\omega= e^{2 \pi i / \hat N}$:
\bea
UV=\omega VU
\eea
The traceless part of $A'$ is a connection on the twisted 't Hooft bundle $SU(N \hat N)/Z_{N \hat N}$, with magnetic flux $\hat M' = r N mod (N \hat N)$ \cite{AG}:
\bea
A'(x_j+L/\hat N)=\Gamma_j A'(x_j) \bar \Gamma_j
\eea
with
\bea
\Gamma_1&=& 1_N \times U^r \nonumber \\
\Gamma_2&=& 1_N \times  V
\eea
where $r$ is an integer that occurs in the definition of the $SL(2,Z)$ matrix that defines the Morita equivalence \cite{AG}. \par
Therefore the theory contains an untwisted sector $SU(N) \times 1_{\hat N}$ that is diagonally embedded in $U(N \times \hat N)$.
This untwisted sector plays a special role, because carries zero momentum on the torus and thus it may condense. \par
We choose in this paper $\frac{2 \pi \theta}{L^2}=\frac{\hat M}{\hat N}= \rightarrow \frac{1}{n}$ for any integer $n \geq 2$, because it allows us to perform the large $\theta$ limit, necessary to reproduce the large-$N$ limit of the commutative $SU(N)$ theory, uniformly for large $\hat N$. It turns out that the choice of $n$ is just the choice of a renormalization scheme
associated to a particular inductive sequence of finite dimensional bundles used to define the large-$\mathcal{N}$ limit (sect.(12)). \par
From a physical point of view we apply 't Hooft duality ideas to the $U(N)$ non-commutative theory.
The structure group of this theory is the tensor product of $U(N)$ and of the group of $\star$-gauge transformations. We assume $Z_N$ magnetic condensation for the untwisted $SU(N)$ factor
diagonally embedded, because, as we have already remarked, the untwisted sector has zero momentum on the torus and therefore it may condense in the vacuum of the localized theory.  \par
However, we will find in sect.(12) that the $RG$-flow in a given $U(\mathcal{N})$ Morita equivalent commutative $YM$ theory must change the rank, $N$, and the degeneracy, $\hat N$, of the
untwisted $SU(N) \times 1_{\hat N}$ sector, keeping the product, $N \hat N=\mathcal{N}$, constant, in order to define non-trivial finite correlation functions of composite surface operators. 
Yet, the limiting bundle in the large $\mathcal{N}$ limit is, independently on $n$, a version of $SU(\infty)$ embedded with $\infty$ multiplicity in $U(\infty)$. \par
Because the Morita equivalent twisted connections live in the tensor product, $u(N) \times u(\hat N)$, there is a natural embedding in the Morita equivalent theory of $SU(N)$ connections with $Z_N$ holonomy in $Z_N \times 1_{\hat N}$. \par
For computational technical reasons it may be convenient to get rid of the twist by means of the thermodynamic limit, to get back (commutative) $R^2$. Going back to the original gauge (that would be singular on the commutative torus
but not on $R^2$) we obtain:
\bea
&&\big[\partial_{\alpha}  A _{\beta}(z,\bar z, u, \bar u) -\partial_{\beta}  A _{\alpha}(z,\bar z, u, \bar u) +i [ A _{\alpha}(z,\bar z, u, \bar u) , A _{\beta}(z,\bar z, u, \bar u)]\big]^- \nonumber \\
&&=\sum_p \mu^{-}_{\alpha \beta}(p)  \delta^{(2)} (z-z_p) +\phi'_{\alpha \beta}  1_{N \times \hat N}
\eea
that is the standard defining equation of a lattice of surface operators in presence of a central magnetic field. 
The central term is now irrelevant in the large-$N$ limit \footnote{In sect.(12.7) it still plays a role.} since without the non-commutativity it does split, at difference of Eq.(7.7).  \par
The solutions of Eq.(7.11) describe surface operators with singularities supported on the $(u, \bar u)$ plane and, because of translational invariance of the vacuum \footnote{We mean in fact the vacuum on which
twistor Wilson loops can be localized on.} in the $(u, \bar u)$ plane, reduce to the standard two-dimensional Hitchin equations in the $(z, \bar z)$ plane. The large-$N$ and large-$\theta$ limit of the non-commutative theory is therefore recovered as a double large-$N$ and large-$\hat N$ limit. \par
For further use we need to know that the second Chern class has an extension to surface operators as a parabolic Chern class. In the notation of $\cite{W2}$:
\bea
\frac{1}{ 16 \pi^2} \int d^4x F_{\alpha\beta}\tilde{F}_{\alpha\beta}=Q+\sum_p tr_f(\alpha_p m_p)+\frac{1}{2} \sum_p  D_p \cap D_p tr_f(\alpha_p^2)
\eea
where $Q$ is the usual second Chern class of the $U(N)$ bundle without the parabolic structure, $\alpha_p$ is the vector of the parabolic weights at the point $p$, i.e.
the vector of the eigenvalues of $F^-_{01}$ divided by $2 \pi$ modulo $1$ in the fundamental representation, $m_p$ the magnetic flux through the surface $D_p$ of the singular divisor $p \times D_p$ of the surface operator and $D_p \cap D_p$ the index of self-intersection of the surface $D_p$.  \par
There is one more symplectic form that plays an important role in this paper. It occurs as the symplectic form associated to the twistor connection:
\bea
\omega_{\rho}
&&=\frac{1}{2 \pi}\int d^2 z  tr_f \hat Tr (\delta {\hat B}_{\rho z} \wedge \delta {\hat B}_{\rho \bar z}) \nonumber \\
&&=\omega_I-i\rho (\omega_J+i\omega_K)-i\rho^{-1}(\omega_J-i\omega_K)
\eea
For $\rho=-1$ it will be employed as an ingredient of the holomorphic/antiholomorphic fusion in sect.(12). Indeed it follows from Eq.(5.4) that $\omega_{-1}=\omega$ depends holomorphically on $\mu_{-1}$. \par
In sect.(12) we employ a modification of $\omega$, $\omega'$, defined over a punctured sphere rather than over its compactification obtained adding the singular divisor. The relation between the two forms is (Eq.(3.30) of  \cite{Malkin}):
\bea
\omega=\omega'+\sum_p Tr(\mu_p(\delta g_p g_p^{-1})^2 )
\eea
where the terms in the sum over $p$ represent the Kirillov forms on the adjoints orbits at $p$. $\omega'$ depends only on the holonomy of the connection (Eq.(3.13) of \cite{Malkin}). \par 
To summarize we have started from the point-like parabolic singularities of the non-commutative Eguchi-Kawai reduced theory and we have ended with surface-like singularities in the Morita equivalent commutative theory.
We can turn the argument around and say that
the aforementioned point-like parabolic singularities of the non-commutative partial large-$N$ Eguchi-Kawai reduction
are daughters of codimension-two singularities of the four-dimensional
parent gauge theory. Codimension-two singularities of this kind have been introduced in \cite{MB2, MB3} in the pure $YM$ theory as an "elliptic fibration of parabolic bundles" for the purpose of getting control over the large-$N$ limit of the pure $YM$ theory exploiting the integrability
of the Hitchin fibration.
In \cite{W2} they have been introduced in the $\cal{N}$ $=4$ $SUSY$ $YM$ theory for the study of the geometric Langlands correspondence, under the name of "surface operators", and this is now the name universally used in the physical literature. \par

\subsection{Moduli of surface operators}

To study the moduli space of surface operators in Eq.(7.11) it is convenient to compactify the $(z, \bar z)$ plane on a sphere.
The moduli space has three different equivalent descriptions that are all employed in this paper. There is a vast mathematics \cite{S1,S2,KM,Konno1,Konno2,S4,S5,S6,Moc} and physics literature \cite{W2,F1} on parabolic Hitchin bundles \footnote{These references are by no means a complete list.}. Thus we summarize briefly the 
essential results \cite{S1,S2,S5,S6}. \par
The first description of the moduli space is of differential geometric nature as a Hitchin system and hyper-Kahler quotient, that in our approach follows by the the non-$SUSY$ non-commutative Nicolai map on a dense set,
as we just discussed.
This is the description that occurs combining the quasi-localization lemma with the idea of integrating on surface operators.
In the hyper-Kahler description the structure group of the bundles involved is compact. In our case $U(N)$ or $SU(N)$. Thus we refer to the gauge fixing in this framework as the unitary gauge.
It is convenient to write Eq.(7.11) in non-covariant notation exactly as Hitchin equations (we disregard for the moment the central $U(1)$ extension that splits):
\bea
-i F_A-[A_u, A_{\bar u}]&&=\sum_p \mu^0_p \delta^{(2)}(z-z_p)\nonumber \\
-i\partial_{A} ^.A_{\bar u}&&= \sum_p n_p \delta^{(2)}(z-z_p)\nonumber \\
-i\bar \partial_A ^.A_{u}&&=\sum_p \bar n_p \delta^{(2)}(z-z_p)
\eea
Because of the delta function at $p$ in general the gauge connection has a pole singularity. The triple $(\mu^0_p, n_p,\bar n_p)$ determines the coefficients of the leading behavior of the gauge connection around the pole. The local model arises by restricting to such leading behavior \cite{S5,S6}. Since $\mu^0_p$ is Hermitian
it is always diagonalizable.
Let us consider first the semisimple case for which by definition all the eigenvalues of $\mu^0_p$ \footnote{We have normalized $\mu_p$ 
in such a way that the holonomy around $p$ is $e^{2i \mu_p}$.} are different modulo $\pi$. \par
A study of the local model implies that in this case also $n_p$ and $\bar n_p$ can be diagonalized 
simultaneously with $\mu^0_p$ by a compact gauge transformation, $g_p$ \cite{S5}. This is a quite remarkable fact, referred to in this paper as local abelianization \footnote{A proof in the physicists style of the result in \cite{S5} about the commutativity of the triple $(\mu^0_p, n_p,\bar n_p)$ can be found in \cite{W2}.}
and it is the ultimate reason that allows the explicit computations of sect.(12) in the large-$N \hat N$ limit. \par
Thus the matrix, $\mu_p=\mu^0_p+n_p - \bar n_p$, commutes with its adjoint, $\bar \mu_p$, i.e. it is normal. Hence the compact adjoint orbits at a point, $g_p \lambda_p g_p^{-1}$, where $\lambda_p$ are the (in general complex) eigenvalues
of $\mu_p$, label some moduli of the solution of Eq.(7.15). There are other moduli that are not immediately manifest in the unitary gauge. They arise as moduli of the metric of the Hitchin bundle that is implicit in the definition of the unitary structure \cite{S6}. \par
When some eigenvalues of $\mu^0_p$ are degenerate modulo $\pi$ the asymptotic behavior of the connection changes. This is recalled below after describing the other representations of the moduli space. \par
The second description of the moduli space is of holomorphic nature. It arises by a meromorphic connection in a holomorphic gauge. 
Indeed the Hitchin equations imply the flatness equation:
\bea
-i F(B)&&=\sum \mu_p \delta^{(2)}(z-z_p) \nonumber \\
F(B)&&=\partial_z B_{\bar z}-\partial_{\bar z } B_{ z}+i[B_z, B_{\bar z}]
\eea
for the non-Hermitian connection:
\bea
B_z=A_z+i A_u \nonumber\\
B_{\bar z}=A_{\bar z}+i A_{\bar u} 
\eea
The moduli space arises as the Kahler quotient of the space of solutions of the flatness equation, Eq.(7.16), with respect to the action of the complexification of the gauge group. Because of  a well known result \cite{H,HKL} it coincides with the hyper-Kahler quotient of the three equations, Eq.(7.15), with respect
to the action of the compact gauge group \cite{S5}.
The structure of the moduli space is particularly transparent in a holomorphic gauge:
\bea
B_{\bar z}=0
\eea
In this gauge $B_z$ is a meromorphic connection:
\bea
i \partial_{\bar z }B_{ z}=\sum_p \mu'_p \delta^{(2)}(z-z_p)
\eea
with residue at $p$ determined by $\mu'_p$, that is conjugate to $\mu_p$ by a gauge transformation in the complexification of the gauge group.
This description is the most transparent
to understand the moduli space because all the local moduli are labelled by the adjoint orbit in the complexification of the gauge group, $\mu_p'=G_p \lambda_p G_p^{-1}$. \par
The holomorphic description arises in the holomorphic loop equation.  \par
It implies also that Hitchin equations are associated to local systems, i.e. to fiber bundles with locally constant transition functions \cite{S1,Local}.  
A local system on a complex curve is the same as a representation of the fundamental group of a Riemann surface with punctures \cite{S1,Local}.
This is the topological description of the moduli, and it is also the easiest to understand globally.
Indeed the residues of the meromorphic connection, $B_z$, determine its holonomy around $p$:
\bea
M_p&=& Pe^{{i \int_{L_p}} B_z dz} \nonumber \\
       &=& e^{2 i \mu_p'}
\eea
The global moduli space on a punctured sphere is therefore the quotient of the algebraic variety:
\bea
\prod_p M_p=1
\eea
modulo the adjoint action of the complexification of the global gauge group (this description has been employed in sect.(1)). \par
We come now to the non-semisimple case \cite{S1,S5,S6}. \par
If the eigenvalues of the holonomy around $p$, $e^{2i \lambda_p}$, are not all different, the holonomy cannot be diagonalized in general but it can be put in Jordan form.
In this case in the unitary gauge some eigenvalues of $n_p$ and of its Hermitian conjugate  $\bar n_p$ are degenerate as well and the Higgs field, $A_u$, in some directions in color space has not anymore a pole singularity but only a milder one, a pole divided by powers of a logarithm. The power of the logarithm and the coefficient of the pole are determined by the off-diagonal parameters in the Jordan form of the holonomy \cite{S6}. \par
Perhaps the most important property of the hyper-Kahler construction from a physical point of view in the semisimple case is the fact that in a unitary gauge the coefficients of the delta function at a point
$(\mu^0_p,n_p,\bar n_p)$ commute and thus can be diagonalized at the same time by a unitary gauge transformation. \par
On the opposite in a holomorphic gauge the residues of the meromorphic
connection and the local holonomies of the twistor connection cannot be diagonalized in general by a unitary transformation but generically only by a transformation in the complexification of the gauge group. 
Thus there is mismatch in the number of local degrees of freedom between the holomorphic and unitary descriptions. \par
This is explained by the fact that in the unitary description the missing degrees of freedom arise as moduli of the
Hermitian metric associated to the Higgs field \cite{S6}. Physically this means that the dimension of local fluctuations of the $ASD$ curvature, $\mu_p$, of semisimple type that occur by the hyper-Kahler construction in a unitary gauge is just one-half
of the dimension that occurs in the holomorphic description of local fluctuations, $\mu'_p$. This leads to a non-trivial Jacobian from the unitary to the holomorphic gauge whose logarithm turns out to be the glueball potential.  \par
In the next section we consider solutions of the Hitchin equations for connections with $Z_N$ holonomy. These connections have no (local) moduli since the adjoint orbit of the center is the center.
They turn out to be the Hitchin bundles that occur at the fixed points 
of the quasi-localization lemma. 
They satisfy the Hitchin equations:
\bea
-i F_A-[A_u,A_{\bar u}]&&= \sum_p \lambda_p \delta^{(2)}(z-z_p) \nonumber \\
\partial_{A}^.  A_{\bar u}&&= 0 \nonumber \\
\bar \partial_A ^. A_u&&=0 
\eea
with
 $e^{2i\lambda_p}\in Z_N$. Therefore:
\bea 
\label{vorticesZN}
2\lambda_p=diag(\underbrace{2\pi(k-N)/N}_{k},\underbrace{2\pi k/N}_{N-k})
\eea
These equations are invariant for the following $U(1)$ action:
\bea
A_u &&\rightarrow e^{i\theta} A_u \nonumber \\
A_{\bar u}&& \rightarrow e^{-i\theta} A_{\bar u} 
\eea
Since there are no moduli this $U(1)$ must act by gauge transformations:
\bea
g _{\theta}A_u g _{\theta}^{-1} &&= e^{i\theta} A_u \nonumber \\
g _{\theta} A_{\bar u}g _{\theta}^{-1} &&= e^{-i\theta} A_{\bar u}  \nonumber \\
g _{\theta}A_z g _{\theta}^{-1} &&=  A_z \nonumber \\
g _{\theta} A_{\bar z}g _{\theta}^{-1}&& =  A_{\bar z}    
\eea
More generally the set of moduli fixed by this $U(1)$ action is the Lagrangian submanifold of the hyper-Kahler moduli space for which the Higgs field, $A_u$, is nilpotent \cite{SL,Konno2,BD}. \par
There are fundamentally two interesting types of orbits in the Lagrangian cone of the hyper-Kahler moduli space \cite{SL}:
the orbits with unitary holonomies, for which the Higgs field vanishes identically;
the orbits that correspond to Hodge bundles, for which the holonomies are valued in a real version of the complexification of the gauge group \cite{SL}. \par
In the first case the holonomies  can always be diagonalized, despite the eigenvalues may not be all different. 
In the second case the holonomies cannot be diagonalized, but can be set in Jordan form. 
Both the orbits play a role in the computation of the Wilsonian beta function (sect.(9)).

\section{Localization on fixed points in large-$N$ $YM$} 

We use the description of surface operators as local systems (i.e. as representations of the fundamental group) to obtain localization on fixed points. \par
Indeed for the lattice theory of sect.(7) we can justify the exchange of the order of integration and limits that occurs in the quasi-localization lemma of sect.(5),
taking advantage of the existence of the singular gauge to reduce the lattice theory to a locally abelian theory, whose eigenvalues label the gauge orbits of the hyper-Kahler reduction
and do not fluctuate in the large-$N \hat N$ limit. \par
In the aforementioned singular gauge the partition function reduces to an integral over the eigenvalues and the zero modes of the locally abelian theory:
\bea
Z&&= \big| \int e^{-\Gamma} \prod_p \delta \lambda_p \delta \nu_p \delta \bar \nu_p \big |^2   \nonumber \\
 &&=\big| \int  \delta' A   \delta' \bar{A} \delta'  D   \delta'   { \bar D} \exp \big(-\frac{4N \hat N}{g_W^2}  \sum_p  tr_N    Tr_{\hat N } ( \lambda_p  {\bar \lambda}_p + 4 \nu_p  {\bar \nu}_p)  \big) \nonumber \\
&&\delta(-i F_{ B} - \sum_p  (\lambda_p - \nu_p+\bar \nu_p)  \delta^{(2)} (z-z_p)) \nonumber \\
&&\delta(- i \partial_{ A}{\bar D}- \sum_p  {\nu}_p  \delta^{(2)} (z-z_p))\delta(- i\bar \partial_{A}  D - \sum_p  {\bar \nu}_p  \delta^{(2)} (z-z_p))  \nonumber \\ 
&& \frac{\Delta(\nu_p+ \bar \nu_p) }{\Delta(\lambda_p-\nu_p+\bar \nu_p)} \Lambda^{n_b} \omega^{\frac{n_b}{2}} \prod_p \delta \lambda_p \delta \nu_p \delta \bar \nu_p\big|^2
\eea
where $(\lambda_p, \nu_p)$ are the eigenvalues of $(\mu^0_p, n_p)$ and $\Delta(\lambda)$ is the Vandermonde determinant of the eigenvalues, Eq.(12.23).
The Vandermonde determinant that occurs in the numerator is due to gauge fixing $n_p$ in triangular form, by the action of the unitary gauge group in the singular
gauge. As a consequence $n_p$ is  automatically diagonal since it arises from the solution of the Hitchin equations, sect.(7) and sect.(12). The Vandermonde determinant in the denominator arises
because of the combination of gauge fixing in the singular gauge and of the change of variables to the holomorphic gauge, sect.(12). \par
At large-$N \hat N$ it is not restrictive to assume that only one set of eigenvalues $(\tilde \lambda_p, \tilde \nu_p)$ or a discrete sum of them actually contribute to the partition function:
\bea
Z&&=\big| e^{-\Gamma}  \big |^2   \nonumber \\
 &&=\big| \int  \delta' A   \delta' \bar{A} \delta'  D   \delta'   { \bar D} \exp \big(-\frac{4N \hat N}{g_W^2}  \sum_p  tr_N    Tr_{\hat N } ( \tilde \lambda_p  {\bar {\tilde\lambda}_p} + 4\tilde \nu_p  {\bar {\tilde\nu}_p})\big) \nonumber \\
&&\delta(-i F_{ B} - \sum_p  (\tilde \lambda_p -\tilde \nu_p+ \bar{\tilde \nu}_p) \delta^{(2)} (z-z_p)) \nonumber \\
&&\delta(- i \partial_{ A}{\bar D}- \sum_p  \tilde {\nu}_p  \delta^{(2)} (z-z_p))\delta(- i\bar \partial_{A}  D - \sum_p  \bar {\tilde\nu}_p  \delta^{(2)} (z-z_p))  \nonumber \\ 
&& \frac{\Delta(\tilde \nu_p+\bar {\tilde \nu}_p)}{\Delta(\tilde \lambda_p -\tilde \nu_p+\bar{\tilde \nu}_p)} \Lambda^{n_b} \omega^{\frac{n_b}{2}} \big|^2
\eea
Therefore we can now justify the interchange of order of limits and integration in the quasi-localization lemma (sect.(5)), since in fact at large-$N \hat N$ there is no integration variable 
in the arguments of the delta functions that depend on the $ASD$ lattice variables:
\bea
&&\big| \int  \delta' A   \delta' \bar{A} \delta'  D'   \delta'   { \bar D'} \exp \big(-\frac{4N \hat N}{g_W^2}  \sum_p  tr_N    Tr_{\hat N } ( \tilde \lambda_p  {\bar {\tilde\lambda}_p} + 4\tilde \nu_p  {\bar {\tilde\nu}_p})\big) \nonumber \\
&& Tr_f P \exp i \int_{L_{ww}}(A_z+D'_u) dz+(A_{\bar z}+ D'_{\bar u}) d \bar z  \nonumber \\
&&\delta(-i  F_A+ [D',\bar D'] -i\frac{\lambda}{\rho} \partial_{A}  \bar D' +i \frac{\rho}{\lambda} \bar \partial_A D'- \sum_p  (\tilde \lambda_p  \delta^{(2)} (z-z_p) -\rho^{-1}   \tilde {\nu}_p  \delta^{(2)} (z-z_p)  +\rho   \bar {\tilde\nu}_p  \delta^{(2)} (z-z_p)))
\nonumber \\
&&\delta(- i \lambda \partial_{ A}{\bar D'}- \sum_p  \tilde {\nu}_p  \delta^{(2)} (z-z_p))\delta(- i \lambda^{-1}\bar \partial_{A}  D' - \sum_p  \bar {\tilde\nu}_p  \delta^{(2)} (z-z_p))  \nonumber \\ 
&& \frac{\Delta(\rho^{-1} \tilde \nu_p + \rho \bar{\tilde \nu}_p)}{\Delta(\tilde \lambda_p- \rho^{-1}\tilde \nu_p+ \rho \bar{\tilde \nu}_p)} \Lambda^{n_b} \omega^{\frac{n_b}{2}} \big|^2
\eea
Thus we get at the fixed points:
\bea
Z&&=\big| e^{-\Gamma}  \big |^2   \nonumber \\
 &&=\big| \int  \delta A'   \delta \bar{A'} \delta  D'   \delta   { \bar D'} \exp \big(-\frac{4N \hat N}{g_W^2}  \sum_p  tr_N    Tr_{\hat N } ( \tilde \lambda_p  {\bar {\tilde\lambda}_p}) \big) \nonumber \\
&&\delta(-i F_{ B} - \sum_p  \tilde \lambda_p  \delta^{(2)} (z-z_p)) \delta( \partial_{ A}{\bar D})\delta(\bar \partial_{A}  D)  \nonumber \\ 
&& \frac{\Delta(\epsilon_p)}{\Delta(\tilde \lambda_p)} \Lambda^{n_b} \omega^{\frac{n_b}{2}} \big|^2
\eea
It remains to integrate on the moduli and determining the eigenvalues, $ \tilde \lambda_p$, that is resolved requiring that the global gauge group be
unbroken at the critical points. Indeed
physically the basic idea is that  the twistor Wilson loops in the adjoint representation, of both the non-commutative theory with gauge group $U(N)$ and of its Morita equivalent counterpart, satisfy the general criteria of 't Hooft duality. Thus if the gauge group is unbroken the twistor Wilson loops in the adjoint representation must be localized, by large-$N$ factorization, on a condensate of the tensor product of $Z_N$ magnetic vortices with the conjugate representation. We have noticed in sect.(7)
that  the correct ansatz for the localized locus is in fact $Z_N \times 1_{\hat N}$, i.e. $Z_N$ occurs with degeneracy  $\hat N$. \par
The formal argument is as follows. The complexification of the global gauge group acts on the holonomy at one point, $p_1$, by the adjoint action, in such a way that $M_{p_1}$ can be put in canonical form.
$M_{p_1}$ can be diagonalized if it has distinct eigenvalues, while in general it can be put in Jordan form. \par
In the large-$N$ limit it is possible to restrict the integration measure, $d\mu_p$, to orbits whose holonomies have fixed eigenvalues, since this restriction
implies an error of subleading order in $\frac{1}{N}$.
In addition by translational invariance the conjugacy class of the orbits at all the points $p$ must be the same. 
Finally the global $SU(N)$ gauge group (and not only the $U(1)^{N-1}$ torus as in the Nekrasov case) must fix $M_{p_1}$, i.e. $g M_{p_1} g^{-1}=M_{p_1}$, since otherwise $M_{p_1}$ would break spontaneously the global gauge symmetry. Therefore $M_{p_1}$
must be central and thus must be in $Z_N$. But then all the orbits collapse to a point and there are no moduli at the fixed points \footnote{As we have seen in the introduction the requirement of translational invariance
can be relaxed.}. \par
However, in any neighborhood of the fixed points of the global gauge group, the orbits are non-trivial and moduli there exist. 
There are essentially two different ways to describe these neighborhoods of the fixed points.
One possibility is deforming infinitesimally the unitary part of the eigenvalues of the holonomy, the other possibility is deforming the holonomy along nilpotent directions.
Both possibilities are discussed in the next section. \par
Thus if we first compute the effective measure in a neighborhood of the fixed points and then we sit on the fixed points
the induced measure will contain the powers of the Pauli-Villars regulator due to the moduli.
This has an analog in the localization of the $\cal{N}$ $=2$ $SUSY$ $YM$ partition function, where
generically instantons have moduli (this is essential to get the correct beta function in that case too), but the instantons at the fixed points of
the torus action have not. \par 
Thus at the fixed points the contour integral over $\mu_{0^{+}}$ in Eq.(5.8) collapses to a discrete sum over sectors with $Z_N$ holonomy.
The reduced Eguchi-Kawai effective action of the localized theory is now:
\bea
&&\sum_{Z_N}  [\exp(-\frac{N 8 \pi^2 }{N_2 g_W^2} Q-\frac{N}{N_2 4g_W^2} \sum_{\alpha \neq \beta} \int tr_f(\mu^{-2}_{\alpha \beta}) d^4x)\nonumber \\
&&Det^{-\frac{1}{2}}(-\Delta_A \delta_{\alpha \beta} -i ad_{ \mu^-_{\alpha \beta}} ) Det(-\Delta_A)
(\frac{\Lambda}{2 \pi})^{n_b} Det^{\frac{1}{2}} \omega    \frac{\delta  \mu_{0^{+}} } { \delta  \mu'_{0^{+}} }     \times c.c. \big] _{n=\bar n=0}
\eea
The connection, $A$, denotes the solution of the equation
$[F^-_{\alpha \beta}- \sum_{p} \mu ^-_{\alpha \beta}(p) \delta^2(z-z_{p_{(u, \bar u)}}) =0]_{n=\bar n=0} $ in each $Z_N$ sector.
$Det^{\frac{1}{2}} \omega$ is the contribution of the $n_b$ zero modes due to the moduli and $\Lambda$ the corresponding Pauli-Villars regulator. 
The complex conjugate factor arises by the conjugate representation. \par
Thus the holonomy of adjoint twistor Wilson loops at the fixed points is trivial, because of the cancellation of $Z_N$ factors 
between the fundamental and the conjugate representations. Hence for twistor Wilson loops the same result is obtained performing the limit $\lambda \rightarrow 0$ outside the functional integral, leading to triviality
via $\lambda$-independence and the all order argument of sect.(4.5), and inside the
functional integral, leading to localization on the tensor product of $Z_N$ surface operators and the conjugate representation, and to triviality as well.

\section{Wilsonian beta function}

\subsection{Beta function by restricting the non-$SUSY$ Nicolai map to surface operators}

We now proceed to the computation of the beta function.
In order to define the renormalization of the coupling constant it is necessary to compute the classical $YM$ action of surface operators.
As we have seen in sect.(7) surface operators are defined by connections on parabolic bundles.
In a mathematical sense we can think of parabolic bundles in two different ways. 
Either parabolic bundles occur on space-time with no boundary and with a divisor and a parabolic structure that belong
to the space-time. This is the point of view in this paper and in some mathematical literature. \par
Or they arise on space-time with boundary, where the boundary is the parabolic divisor.
This is the point of view of \cite{W2}.
In the latter case the insertion of a surface operator keeps the finiteness of the action, since the singular parabolic locus is not included in the
space-time integral that computes the action. This justifies also the term operators, since their occurrence is the analog of operator insertions \footnote{We would like to thank Edward Witten for a discussion about this point.}. \par
However, our point of view is that the surface operators are dynamical objects and therefore their singular divisor is included in the 
path integral in space-time.  \par
Despite the parabolic singularity, the topological term in the action has a well defined mathematical extension to parabolic bundles, as parabolic Chern class (Eq.(7.16)). \par
This is not the case for the term involving the $ASD$ field.
As a consequence the classical $YM$ action is quadratically divergent on each singular divisor, $p \times D_p$, of a surface operator,
with a divergence proportional
to the area of the singular locus of each surface operator.
Therefore we need a way to handle this classical divergence. \par
We have already recalled in sect.(4.1) that
when the codimension-two surface is non-commutative, as in our case, the $YM$ action of the corresponding non-commutative
reduced Eguchi-Kawai ($EK$) model is rescaled by a power of the inverse cutoff, that cancels precisely \cite{MB1} the quadratic
divergence that occurs evaluating the classical $YM$ action on surface operators. 
This allows us to define a new kind of semi-classical computation for which the classical $YM$ action is finite on parabolic bundles.
In our case the $EK$ reduction is only partial in such a way that the action is:
\bea
\frac{N}{2 g^2}  \hat N (\frac{2 \pi}{\Lambda})^2 \int d^2x tr_N Tr_{\hat N} (-i [\hat \partial_{\alpha}+i \hat A_{\alpha},\hat \partial_{\beta}+i \hat A_{\beta}]+ \theta^{-1}_{\alpha \beta} 1)^2
\eea
where the trace $Tr_{\hat N}$ is taken over a subspace of dimension $\hat N$,
with 
\bea
\hat N (\frac{2 \pi}{\Lambda})^2=2 \pi \theta
\eea
in the large $\hat N, \theta, \Lambda $ limit.
The contribution of each parabolic singularity in the action reads:
\bea
&&\int d^2x \delta^{(2)}(x-x_p)^2 \nonumber \\
&&= \delta^{(2)}(0) \int d^2x \delta^{(2)}(x-x_p) \nonumber \\
&&=  (\frac{\Lambda}{2 \pi})^2
\eea
and it is cancelled by its inverse in the $EK$ reduced action.
It is convenient to describe the same result in terms of the action of a commutative gauge theory.
In this case the action reads:
\bea
\frac{N}{2 g^2}  \hat N  \int d^4x tr_N Tr_{\hat N} (-i [ \partial_{\alpha}+i  A_{\alpha}, \partial_{\beta}+i  A_{\beta}]+ \theta^{-1}_{\alpha \beta} 1)^2
\eea
and it is divergent on a surface operator as:
\bea
&&\int d^4x \delta^{(2)}(x-x_p)^2 \nonumber \\
&&= \delta^{(2)}(0) V_2\int d^2x \delta^{(2)}(x-x_p) \nonumber \\
&&=  (\frac{\Lambda}{2 \pi})^2 V_2 \nonumber \\
&&= N_2
\eea
where $V_2$ is the area of the singular divisor of the surface operator and the last equality is the definition of $N_2$.
Thus the reduced action is related to the one of a commutative gauge theory by the factor of $N_2^{-1}$ \footnote{Precisely the same factor arises in the quenched version of the $EK$ reduction \cite{Rt}.}:
\bea
\frac{N}{2 g^2}  \hat N (\frac{2 \pi}{\Lambda})^2 \int d^2x tr_N Tr_{\hat N} (-i [\hat \partial_{\alpha}+i \hat A_{\alpha},\hat \partial_{\beta}+i \hat A_{\beta}]+ \theta^{-1}_{\alpha \beta} 1)^2 \nonumber \\
=\frac{N}{2 g^2 N_2}  \hat N  \int d^4x tr_N Tr_{\hat N} (-i [ \partial_{\alpha}+i  A_{\alpha}, \partial_{\beta}+i  A_{\beta}]+ \theta^{-1}_{\alpha \beta} 1)^2
\eea
We can interpret this result by saying that the normalization of the trace in the non-commutative $EK$ reduced theory differs by a factor of $N_2^{-1}$  from the one of a commutative
theory. Thus it may be computationally convenient to perform the calculations in the commutative theory and at the end normalize appropriately the traces. \par
Let us evaluate firstly the action of the surface operators of $Z_N$ holonomy that occur at the fixed points.
The Morita equivalent commutative \footnote{We rescale the area of the Morita equivalent theory from $\frac{L^2}{\hat N^2}$ to $L^2$ in order to perform the thermodynamic limit uniformly in
$\hat N$.} theory has gauge group $U(N \times \hat N)$ and the center of $SU(N)$ is embedded diagonally in $U(N \times \hat N)$ as $e^{\frac{i 2 \pi k}{N}} 1_{\hat N}$.
For a surface operator of $Z_N$ holonomy around the point $p$ of charge $k$, i.e. such that $M_p= e^{\frac{i 2 \pi k}{N}}$,
$N-k$ eigenvalues, $2\lambda_p$, of the $ASD$ curvature  at $p$, $F^-_{01}=2\lambda_p \delta^{(2)}(z-z_p)$, are equal to
$\frac{2 \pi k}{N}$ and $k$ eigenvalues are equal to $\frac{2 \pi (k-N)}{N}$, for the curvature to be traceless and to give rise
to the holonomy $M_p= e^{\frac{i 2 \pi k}{N}}$ \footnote{The curvature in not uniquely determined by the holonomy, since parabolic bundles admit extensions over the punctures such that the eigenvalues of the $ASD$ curvature differ by shifts of $2 \pi$. Our choice is in some sense minimal. This feature together with many others, known in the mathematical literature, is reviewed in \cite{W2}. }.
The trace of the square of the eigenvalues of the $ASD$ curvature in the fundamental representation is thus:
\bea
tr_N(4\lambda^2)=(N-k) (\frac{2 \pi k}{N})^2 + k (\frac{2 \pi (k-N)}{N})^2 
=(2 \pi)^2 \frac{k(N-k)}{N}
\eea
in such a way that the reduced action is:
\bea
S_{EK}= \frac{N \hat N (4 \pi)^2}{2 g^2}  \frac{Q}{N_2} + \frac{ \hat N^2}{2 g^2}  \sum_p 2(2 \pi)^2 k(N-k)
\eea
and the Morita equivalent one is:
\bea
S_{YM}= \frac{N \hat N (4 \pi)^2}{2 g^2}  Q + \frac{ \hat N^2}{2 g^2} N_2 \sum_p 2(2 \pi)^2 k(N-k)
\eea
From these equations it follows that, if $Q$ is finite, its contribution to the action is irrelevant with respect to the one of the $ASD$ curvature at the
parabolic singularities. \par
Once the classical quadratic divergence has been tamed by the $EK$ reduction we need to understand the logarithmic divergences that lead to a non-trivial
beta function in the Jacobian for the change of variables from the connection to the $ASD$ curvature.
These divergences have been already computed in sect.(3.4), and we can just adapt our previous calculation to the case 
of a lattice of surface operators. 
We have already observed that, as in Nekrasov localization, we should evaluate and renormalize the functional measure in a neighborhood of the fixed points and thereafter we should sit on the fixed points.
Since the fixed points have no moduli and thus no zero modes, the inverse order, first sitting on the fixed points and then renormalizing the functional measure, would not lead to the correct result. \par
But let start ignoring for the moment the zero modes and sitting on the fixed points. 
The observation that the contribution of $Q$ to the classical and the quantum effective action is irrelevant with respect to the one of the $ASD$ curvature at the parabolic singularities
resolves the issue about the different factors of $Z_Q^{-1}$ and $Z^{-1}$ that occur in sect.(3.4) as counterterms in the Jacobian of the non-$SUSY$ Nicolai map. \par
This observation allows us to ignore also issues related to the possible global non-triviality of the bundles and to the counting of the global moduli, and to concentrate only on the local
counterterms associated to the sum over the points of the parabolic divisor. \par
As in every computation involving an effective action the background field should live on a scale much larger that the quantum fluctuating field.
This may look awkward to realize for surface operators that carry delta-like singularities, that involve any momentum scale. In fact it is impossible to realize for an isolated surface
operator, but it is possible for a lattice with uniform lattice spacing \footnote{In fact it is sufficient that the lattice spacing be uniform on a scale much larger than the typical scale of the physics
involved.}. A typical example is
the following one-loop contribution to $Z^{-1}$ in Euclidean configuration space, that arises from the spin term of sect.(3.4) evaluated on surface operators:
\bea
\frac {1}{ (4 \pi^2)^2 } \sum_{ p, p' }  \int d^2u d^2v  \frac{N Tr(\mu_p \bar \mu_{p'}) }{(|z_p-z_{p'}|^2+|u-v|^2)^2}
\eea
where the sum over $p, p'$ runs over the planar lattice of the parabolic divisors of the surface operators.
There is also an orbital contribution that has the same structure. Indeed the orbital  logarithmic contribution to the beta function arises from terms of the kind \footnote{The term involving $\partial_{\alpha} A_{\alpha}$
vanishes identically around the local singularity, while the term involving $A_{\alpha}^2$ is quadratically divergent and does not contribute because of cancellations due to gauge invariance in any gauge invariant regularization of the theory. Since in our computations only functional determinants occur, the Pauli-Villars regularization suffices.}:
\bea
&&\int d^4x d^4y Tr(A_z(x) \partial_{\bar z} \frac{1}{(x-y)^2}  A_{\bar z}(y) \partial_z \frac{1}{(x-y)^2} ) \nonumber \\
&&=\int d^4x d^4y Tr(A_z(x) \frac{2 (z-w)}{(x-y)^4}  A_{\bar z}(y) \frac{2 (\bar z- \bar w)}{(x-y)^4} ) \nonumber \\
&&=\int d^2z d^2w d^2u d^2w Tr(A_z(x) A_{\bar z}(y) \frac{4 |z-w|^2}{(|z-w|^2+|u-v|^2)^4}) \nonumber \\
&&\sim- \int d^2z d^2w d^2u d^2w Tr(A_z(x) A_{\bar z}(y)  \partial_{\bar z} \partial_{w} \frac{4}{6 (|z-w|^2+|u-v|^2)^2}) \nonumber \\
&&=- \int d^2z d^2w d^2u d^2w Tr( \partial_{\bar z} \partial_{w}(A_z(x) A_{\bar z}(y))   \frac{4}{6 (|z-w|^2+|u-v|^2)^2}) \nonumber \\
&&=- \int d^2z d^2w d^2u d^2w Tr( \partial_{\bar z} A_z(x) \partial_{w} A_{\bar z}(y)  \frac{4}{6 (|z-w|^2+|u-v|^2)^2}) \nonumber \\
&&\sim \sum_{ p, p' }  \int d^2u d^2v  \frac{N Tr(\mu^0_p \mu^0_{p'}) }{(|z_p-z_{p'}|^2+|u-v|^2)^2}
\eea
where in the last line we used $\partial_{\bar z} A_z(x) \sim \sum_p \mu^0_p \delta^{(2)}(z-z_p)$ and $\partial_{w} A_{\bar z}(y) \sim -\sum_p \mu^0_p \delta^{(2)}(w-z_p)$ for the surface operators that occur at the 
fixed points, with $x=(z, \bar z, u, \bar u)$ and $y=(w, \bar w, v, \bar v)$. \par
We would like to find a regularization for which the loop
expansion of the functional determinants evaluated on surface operators satisfies the usual power counting
as in the background-field computation of the beta function. \par
To avoid quadratic divergences we restrict the sum to $p \neq p'$. Although quadratic divergences in the Morita equivalent theory correspond to finite counterterms
in the $EK$ reduced action, we would like to avoid divergences at coinciding points in higher orders of the loop expansion for the conventional power counting to hold.
We set further $\mu_p=\lambda_p=\lambda$ in such a way that :
\bea
\frac {1}{ (4 \pi^2)^2 } \sum_{ p \neq p' }  \int d^2u d^2v  \frac{N Tr(\lambda^2) }{(|z_p-z_{p'}|^2+|u-v|^2)^2}
\eea
is logarithmically divergent (both in the ultraviolet and the infrared). Indeed introducing a lattice scale $a$:
\bea
&&\frac {1}{ (4 \pi^2)^2 } \sum_{ p \neq p' }  \int d^2u d^2v  \frac{N Tr(\lambda^2) }{(|z_p-z_{p'}|^2+|u-v|^2)^2} \nonumber \\
&&\rightarrow \frac {1}{ (4 \pi^2)^2 }   \int d^2u a^{-2} d^2z a^{-2} \int d^2w d^2v  \frac{N Tr(\lambda^2) }{(|z-w|^2+|u-v|^2)^2} \nonumber \\
&&=\frac {1}{ (4 \pi^2)^2 }  N_2^2  \int d^2w d^2v  \frac{N Tr(\lambda^2) }{(|z-w|^2+|u-v|^2)^2} \nonumber \\
&&\sim N_2^2   N Tr(\lambda^2) \log \frac{\Lambda}{\mu} \nonumber \\
&&=N_2 \sum_p   N Tr(\lambda^2) \log \frac{\Lambda}{\mu}
\eea
One factor of $N_2$ is just the sum on lattice points, the other factor is the "phase space area" $N_2=(\frac{\Lambda}{2 \pi})^2 V_2$ of one surface operator.
From the preceding equation we read also that:
\bea
a^{-1}=\frac{\Lambda}{2 \pi}
\eea
To summarize, there exist a point-splitting regularization of the effective action in the background of the lattice of surface operators, that together with the implicit use of the Pauli-Villars regularization, needed to handle
quadratic tadpoles, leads to the same power counting as the usual
dimensional regularization, with the logarithmic divergences occurring as in Eq.(9.12). 
Had the contributions with $p=p'$ been included, there would appear quadratic divergences, thus spoiling
the usual power counting in higher order terms of the loop expansion. This lattice point-spitting regularization \footnote{This regularization has been found during joint work with Arthur Jaffe.}, followed by Epstein-Glaser renormalization in Euclidean configuration space (see \cite{EG} for references) is a possible starting point of a new constructive approach for the large-$N$ $YM$ theory. \par
We should understand now the contribution of the zero modes. \par 
At first we count the moduli of surface operators and then we give an argument to identify the zero modes with the moduli, as for the instantons.\par
From the computation of the $Z^{-1}$ factor that we already performed in sect.(3.4) it follows that the renormalization of the action in absence of zero modes would be:
\bea
\frac{8 \pi^2 k(N-k)}{2 g^2_W(\mu)}
 = 8 \pi^2 k(N-k)( \frac{1}{2 g^2_W(\Lambda)}- \frac{1}{(4\pi)^2} \frac{5}{3}
\log (\frac{\Lambda}{\mu}))
\eea
Adding to the action the contribution of the complex conjugate representation we get:
\bea
\frac{16 \pi^2 k(N-k)}{2 g^2_W(\mu)}
 = 16 \pi^2 k(N-k)( \frac{1}{2 g^2_W(\Lambda)}- \frac{1}{(4\pi)^2} \frac{5}{3}
\log (\frac{\Lambda}{\mu}))
\eea

\subsection{Dimension of the Lagrangian neighborhood of the fixed points}

It follows from Eq.(9.16) that, in order to get the correct one-loop beta function, the contribution of the zero modes to the renormalization of the action should be $-2k(N-k) \log (\frac{\Lambda}{\mu})$, including the fundamental and conjugate representation.
The sign is consistent with the Pauli-Villars regularization of zero modes, yet the absolute value of the coefficient of the logarithm 
is in general an even integer but not a multiple of $4$, as it would be implied by the
hyper-Kahler reduction \footnote{A hyper-Kahler manifold has necessarily a real dimension that is a multiple of $4$.}.
Thus the neighborhood of the fixed points cannot be generic. Remarkably we have already seen in sect.(7) that the fixed points sit automatically inside the Lagrangian cone of the moduli space
for which $A_u$ is nilpotent. Since in a Lagrangian neighborhood of the fixed points the dimension of the moduli space is generically one half of the dimension of a hyper-Kahler neighborhood, the correct beta function may arise. \par
We should classify now which are the Lagrangian neighborhoods of the fixed points that lead to the correct beta function. \par
One component of the Lagrangian cone  corresponds to $A_u=0$ and gives rise to unitary representations of the fundamental group.
The complex dimension of an adjoint orbit for a generic parabolic unitary bundle of rank $N$ is given by: 
\bea
\dim_{\lambda}=\frac{1}{2} (N^2-\sum_i m_i^2)
\eea
where $m_i$ are the multiplicities of the eigenvalues. Vortices of $Z_N$ holonomy have no moduli, since the holonomy lives in the center, for which the multiplicity of the eigenvalues (modulo $2 \pi$) equals the rank.
But the following slight deformation of the eigenvalues gives rise to a non-trivial adjoint orbit for the holonomy:
\bea 
\label{vorticesZN}
2 \lambda =diag(\underbrace{2\pi(k-N)/N+\epsilon}_{k},\underbrace{2\pi k/N-\epsilon k/(N-k)}_{N-k})
\eea
Thus the complex dimension of the orbit is:
\bea
\dim_{\lambda}=\frac{1}{2} (N^2-k^2-(N-k)^2)=k(N-k)
\eea
and the same holds for the complex conjugate orbit, in such a way that the real dimension of the orbit matches the number of zero modes needed for the correct beta function:
\bea
\frac{16 \pi^2 k(N-k)}{2 g^2_W(\mu)}
 = 16 \pi^2 k(N-k)( \frac{1}{2 g^2_W(\Lambda)}- \frac{1}{(4\pi)^2} (\frac{5}{3}+2)
\log (\frac{\Lambda}{\mu}))
\eea
\par
However, there is a more intrinsic characterization of the Lagrangian neighborhood of the fixed points.
Instead of deforming slightly the eigenvalues we may deform the moduli along nilpotent directions.
Hence we may require that, in a unitary gauge in the Lagrangian neighborhood, exactly the same Hitchin equations are satisfied as at the fixed points.
Therefore the eigenvalues of the $ASD$ field are precisely:
\bea 
2 \lambda =diag(\underbrace{2\pi(k-N)/N}_{k},\underbrace{2\pi k/N})
\eea
but the we allow the Higgs field, $A_u$, to have a nilpotent residue. In the Lagrangian cone not only the Higgs field has a nilpotent residue, but it is nilpotent itself \cite{S6}.
These are bundles of Hodge type \cite{SL} for which the twistor connection has a holonomy that cannot be diagonalized, but it can be set in Jordan canonical form.
In the Lagrangian cone the local moduli at a point for these bundles are parametrized by orbits of the Jordan canonical form for a real version of the complexification of the compact gauge group \cite{SL}.  
Thus in our case, for which the diagonal part of the holonomy is in $Z_N$, the local holonomy is unipotent. \par
Since the diagonal part of the holonomy is central, to compute the dimension of the orbit we need to consider only the nilpotent part.
Any such matrix is conjugate by the Jordan theorem  to a direct sum of $k$ blocks of dimension $d_i$, such that $\sum^k_{i=1} d_i=N$, where $N$ is the total rank. Each block has $d_i$ zero eigenvalues on the diagonal \footnote{A nilpotent matrix has all the eigenvalues zero.} and it is upper triangular with all $1$
on the super-diagonal. \par
The classical action is not modified at all by a deformation of the moduli along a nilpotent direction, since the nilpotent part of the holonomy is invisible in the $ASD$ curvature in a unitary gauge \cite{S6}. \par
Therefore, to get the correct beta function, we need to construct orbits of nilpotent Jordan matrices, such that the real dimension of the orbits for the action of a real version of the complexification of the gauge group is precisely the double of the complex dimension of the unitary orbits, i.e. of the flag manifolds. \par
Indeed in this case the conjugate representation describes exactly the same moduli, since the orbit is for the action of a real version of the gauge group. \par
Nilpotent orbits with the features just described, having double the dimension of a flag, are called Richardson orbits \cite{R}. \par
Here are some examples. The principal nilpotent complex orbit, i.e. the orbit of a nilpotent Jordan block of maximal rank, has precisely the same dimension as the complex orbit in the generic semisimple
case, i.e. $N^2-N=N(N-1)$, that is always even. Thus the real dimension is a multiple of $4$, as it should be. \par
By Eq.(9.17) the complex dimension is the double of the complex dimension of the maximal flag,
and thus the principal nilpotent orbit is a Richardson orbit. \par
We are looking for a nilpotent orbit with double the dimension of the partial flag in Eq.(9.19). \par
Let us first recall the general formula for the dimension of the adjoint orbit of a nilpotent \cite{Nil1}: 
\bea
&&dim O_{N}=N^2-N-2 \sum^k_{i=1}(i-1)d_i
\eea
For example for the principal nilpotent, $k=1$, $d_1=N$, and we get the aforementioned result \cite{Nil2}. For the zero nilpotent, $k=N$, $d_i=1$, and we get $0$.
The orbit of the direct sum of $k$ nilpotent Jordan
blocks of dimension $2$ and of one Jordan block of dimension $N-2k$ has double the dimension of the partial flag in Eq.(9.19).\par
Indeed in this case:
\bea
&&dim O_{N}=N^2-N-2 \sum^k_{i=1}(i-1)2-2\sum^{N-k}_{i=k+1}(i-1) \nonumber \\
&& =N^2-N +2(2k+N-2k)-2\sum^k_{i=1}i-2\sum^{N-k}_{i=k+1}i \nonumber \\
&&=N^2-N +2N-(N-k+1)(N-k)-(k+1)k  \nonumber \\
&&=N^2+N-(N-k+1)(N-k)-(k+1)k  \nonumber \\
&&= 2k(N-k)  \nonumber \\
\eea
QED \par
We should clarify in which sense the moduli that label the orbits with fixed eigenvalues occur as zero modes.
Let us consider first the unitary orbits in the Lagrangian cone.
In this case in any smooth unitary gauge the $ASD$ curvature, $\mu^-_{01}(p)$,
is conjugated to the eigenvalues by the action of the unitary group. \par
However, the classical action depends on the eigenvalues but not on the unitary matrices. Thus the unitary matrices
define flat directions in the classical action, but the holonomy of the twistor connection actually depends on the unitary matrices that label the orbit.  \par
We can associate to these moduli zero modes of the functional determinants choosing the singular gauge in which the gauge curvature is diagonal.
This gauge may be singular in the sense that may be reached by gauge transformations that are possibly singular along lines, i.e. semi-infinite strings that start at the punctures and end at infinity or with another puncture.
From the point of view of the homological localization of the holomorphic loop equation of sect.(10) these gauge transformations are allowed, provided the associated strings do not intersect the backtracking arcs
that connect the loop to a puncture, in such a way that the twistor connection is not discontinuous across the backtracking arcs. \par
In this unitary singular gauge the equations for the surface operators of unitary holonomy are:
\bea
F^-_{\alpha \beta} = \sum_p \lambda^-_{\alpha \beta}(p)  \delta^{(2)} (z-z_p(u, \bar u))
\eea
with $\lambda^-_{01}(p)$ diagonal matrices and the other $\lambda^-_{\alpha \beta}(p)$ vanishing \cite{S5}. But the connection still depends implicitly on the moduli, in such a way that zero modes occur by the standard argument at the end of sect.(3.3). \par
In the case of Richardson orbits for Hodge bundles the holonomies are conjugated to the Jordan form by a real version of the complexification of the gauge group, that factorizes into a compact and a parabolic subgroup, the compact factor being a subgroup of the unitary group.
In a unitary gauge only the orbit of the compact subgroup occurs as an adjoint orbit of the $ASD$ field at a point, while the parabolic group parametrizes the moduli of the metric. Thus it there exists a possibly singular unitary gauge in which all the moduli occur as zero modes as for orbits of unitary holonomy.

\section{Homological localization of the holomorphic loop equation}

It has been known for many years that (twisted)-$SUSY$ observables can be localized
in gauge theories with extended $SUSY$ by deformations that are trivial in the cohomology generated by the 
twisted super-charge \cite{W1,N,P}.
Since cohomology is dual to homology \cite{Bott} \footnote{The main difference between cohomology and homology is the lack of a ring structure in the latter.}, we may wonder as to whether
we can compute functional integrals by deformations that are trivial in homology  rather than in cohomology.
Were the answer be affirmative, we could get localization without supersymmetry. \par
While there is no positive answer in local field theory, gauge theories contain many non-local observables,
the Wilson loops.
Thus the natural arena for homological localization in gauge theories, as opposed to cohomological localization,
is the loop equation \cite{MM,MM1}. \par
In general the loop equation is the sum of a classical equation of motion and of a quantum term, that involves the contour integral
along the loop.
By homological localization of the loop equation we mean a deformation of the loop that is trivial in homology and for which the quantum term
vanishes, in such a way that the loop equation is reduced to a critical equation for an effective action \cite{MB1}.
Hence the needed homological deformation has to satisfy the following properties. \par 
It has to be trivial in homology. \par
It has to leave the expectation value of the loop invariant. \par
It has to imply the vanishing of the quantum term in the loop equation, i.e.
of the term that contains the contour integral along the loop. \par
In this homology framework there is a very natural analog of the operation of adding a coboundary
in cohomology,
that is based on the zig-zag symmetry of Wilson loops.
The zig-zag symmetry is the invariance of a Wilson loop by the addition of a backtracking arc ending with a cusp.
This deformation is a "vanishing boundary" in singular homology. In a regularized version the arc is the boundary of a tiny strip.
While this symmetry holds classically in most of the cases, quantum mechanically
the renormalization process may spoil it. The reason is that in general Wilson loops have perimeter and cuspidal divergences.
The perimeter divergence is linear in the cutoff scale. The cuspidal divergence is logarithmic,
with a coefficient that in turn is divergent for backtracking cusps. In $SUSY$ gauge theories with extended $SUSY$ there are examples of locally-$BPS$ Wilson loops that have no
perimeter divergence \cite{Gr}. \par
We have seen in sect.(4) that twistor Wilson loops share with their supersymmetric cousins these non-renormalization properties and, being trivial in the large-$N$ limit, they have not cuspidal divergences either.
Localization by homology leads this kind of non-renormalization properties to their extreme consequences. \par
One of the virtues of the lattice regularization of the
Nicolai map of sect.(7), from the point of view of the homological localization, is to allow identifying the cusps of the aforementioned backtracking arcs with the parabolic singularities of the reduced $EK$ theory.
We have noticed in sect.(7) that this point-like parabolic singularities are daughters of codimension-two singularities of surface operators of the parent four-dimensional theory.
The lattice version of the holomorphic loop equation of sect.(6) follows
\footnote{ The holomorphic loop equation is written in linear form since it is assumed that the loop $C_{zz}$ is simple, i.e. it has no self-intersections. } :
\bea
<Tr(\frac{\delta \Gamma}{\delta \mu(z_p, \bar z_p)'}\Psi'(L_{z_p z_p}))>=
\frac{1}{\pi} \int_{L_{z_p z_p}} \frac{ dw}{z_p -w} <Tr\Psi'(L_{z_p w})> <Tr\Psi'(L_{ w z_p})>
\eea
where $\Psi'$ is the holonomy of $B$ in the gauge $B'_{\bar z}=0$. 
The lattice points associated to the divisor of surface operators become the cusps that are the end points, $p$, of the backtracking strings, $b_p$, that
perform the deformation of the loop, $C$. Adding the backtracking strings implies the homological 
localization of the holomorphic loop equation:
\bea
<Tr(\frac{\delta \Gamma([b_p])}{\delta \mu'(z_p, \bar z_p)}\Psi'(L \cup [b_{p}]))>=0 
\eea
The homological localization can be understood in the following geometrical terms.
The existence of the regularized residue in Eq.(6.8) is the geometric obstruction to the localization
of the loop equation. The regularized residue is the line integral of a current supported on a point and computes
the de Rham cohomology of compact support in one dimension, $H^1_c(R)$ \footnote{The integral of the delta function in one dimension can be approximated by
the normalized integral of one-forms of small compact support around a point.}, that coincides with the cohomology of a point, $H^0(pt)$ \cite{Bott}.
It is interesting to display how this cohomology of a point
can be obtained by a Mayer-Vietoris argument \cite{Bott} involving partitions of unity in the right hand side of the loop equation.
For a smooth point of the arc (i.e. a point with a continuos tangent), introducing partitions of unity on the arc, $1=u_1(s)+u_2(s)$, we get:
\bea
\int dw_+(s)
\delta(z_+(s_{sm}) -w_+(s))=u_1(s_{sm}) \frac{\dot w_+(s^+_{sm})}{ |\dot w_+(s^+_{sm})|}+ u_2(s_{sm})
\frac{\dot w_+(s^-_{sm})}{|\dot w_+(s^-_{sm})|} 
=1
\eea
Let us suppose now that we try to compute the "cohomology of a backtracking cusp".
The same Mayer-Vietoris argument shows that such a cohomology does not exist in a classical
sense, since the result depends on the choice of the partition of unity:
\bea
\int dw_+(s)
\delta(z_+(s_{cusp}) -w_+(s))&=&u_1(s_{cusp}) \frac{\dot w_+(s^+_{cusp})}{ |\dot w_+(s_{cusp})|}+ u_2(s_{cusp})
\frac{\dot w_+(s^-_{cusp})}{|\dot w_+(s^-_{cusp})|} \nonumber \\
&=&u_1(s_{cusp})-u_2(s_{cusp})
\eea
This is due to the fact that we can integrate distributions on smooth manifolds,
but their extension to non-smooth ones depends on arbitrary choices in general.
In particular, if the partition of unity is symmetric, $u_1(s_{cusp})=u_2(s_{cusp})=\frac{1}{2}$, the regularized residue vanishes.
Therefore a regularization exists, that preserves the zig-zag symmetry of twistor Wilson loops, for which the holomorphic loop equation localizes. \par
Thus if every marked point of the loop equation can be transformed into a backtracking cusp we can complete our argument
about localization.
But this is precisely the effect of our lattice, since marked points of the loop contribute to
the loop equation in the lattice theory only if they coincide with the lattice points. \par
We may think that it is a change of the conformal structure around the lattice points that generates the cusps. \par
Since the Lagrangian submanifold on which twistor Wilson loops are supported on satisfies:
\bea
|dz d\bar z|=|du d\bar u|
\eea
this two-dimensional conformal transformation
lifts to a conformal rescaling of the four-dimensional metric:
\bea
ds^2 =dz d\bar z+du d\bar u
\eea
Thus it acts by adding a conformal anomaly to the effective
action:
\bea
\Gamma([b_p])=\Gamma([p])+ConformalAnomaly([b_p])
\eea
that amounts to a local counterterm, i.e. to a change of the subtraction point. \par
Therefore there is a symmetry of the $RG$ flow 
that generates the homological deformation of the loop by a vanishing boundary, i.e. by backtracking strings.
This is the analog of the action being a closed form in cohomology, since in the last case there is a symmetry of the action (i.e. the twisted supersymmetry) that generates the coboundary. \par
At this point we would like to clarify why we cannot use the Makeenko-Migdal ($MM$) loop equation to get localization. \par
We can write the $MM$ loop equation for unitary Wilson loops in the large-$N$ $YM$ theory as:
\bea
&&\int_{L_{xx}} dx_{\alpha}<\frac{N}{2 g^2} Tr(\frac{\delta S_{YM}}{\delta A_{\alpha}(x)}\Psi(x,x;A))> \nonumber \\
&&=i \int_{L_{xx}} dx_{\alpha} \int_{L_{xx}} dy_{\alpha} \delta^{(4)}(x-y) <Tr \Psi(x,y;A)>< Tr \Psi(y,x;A)> 
\eea
where
\bea
\Psi(x,y;A)=P \exp i\int_{L_{(x,y)}} A_{\alpha} dx_{\alpha}
\eea
In the case of loops without self-intersections but with cusps the $MM$ loop equation
reduces to:
\bea
&&\int_{L_{xx}} dx_{\alpha} <\frac{N}{2 g^2}Tr(\frac{\delta S_{YM}}{\delta A_{\alpha}(x)}\Psi(x,x;A))> \nonumber \\
&&=i \int_{L_{xx}} dx_{\alpha} \int_{L_{xx}} dy_{\alpha} \delta^{(4)}(x-y) <Tr \Psi(x,x;A)> <Tr1>
\eea
Performing the two contour integrations along the loop in the right hand side, we get \cite{Gr}:
\bea
&&\int_{L_{xx}} dx_{\alpha}<\frac{1}{2 g^2}Tr(\frac{\delta S_{YM}}{\delta A_{\alpha}(x)}\Psi(x,x;A))> \nonumber \\
&&\sim i(P \Lambda^3+\sum_{cusp} \frac{ \cos \Omega_{cusp}}{ \sin \Omega_{cusp}} (\pi -\Omega_{cusp}) \Lambda^2)<Tr \Psi(x,x;A)>
\eea
where $P$ is the perimeter of the loop and $\Omega_{cusp}$ the cusp angle at a cusp. For our conventions $\Omega_{cusp}=\pi$ for no cusp,
while $\Omega_{cusp}=0$ for a backtracking cusp.
The perimeter divergence arises by the double integration of the four-dimensional delta function,
i.e. of the contact term, along the loop. 
Integrating the contact term in a neighborhood of each cusp
gives rise to a sub-leading quadratic divergence, since around a cusp
there are two independent integrations instead of one, due to the two
sides of the cusp. The coefficient of the cusp 
contribution is proportional to the ratio:
\bea
\frac{\cos \Omega_{cusp}}{\sin \Omega_{cusp}} 
\eea
The numerator arises from the scalar product,
the denominator from the two independent integrations of the two-dimensional
delta function. 
In the limit in which the cusp angle $ \Omega_{cusp} $ reaches $0$
the cusp backtracks and the cusp contribution to the
contact term of the $MM$ loop equation is divergent. It turns out to be proportional to the coefficient of the logarithm in the cusp anomaly \cite{Gr}.
Therefore for backtracking cusps the cusp anomaly and the perimeter divergence mix together \cite{Gr}. Zig-zag invariance can be implemented in the $MM$ loop equation 
only by means of a subtle, regularization dependent, cancellation between the extra perimeter due to the backtracking cusp and the cusp anomaly.
This is due to the non-cohomological nature of the contact term that arises as the obstruction to localization in the $MM$ loop equation,
that is the line integral of a non-integrable distribution. \par
We can write the $MM$ loop equation also for a planar twistor Wilson loop:
\bea
&&\int_{L_{ww}} d\bar z<\frac{N}{2 g^2}Tr(\frac{\delta S_{NC}}{\delta B_z(z,\bar z)}\Psi(z,z;B))>= \nonumber \\
&&i\ \int_{L_{ww}} d\bar z\int_{L_{ww}} dz \delta^{(2)}(z-w) <Tr \Psi(w,z;B)> <Tr \Psi(z,w;B)>
\eea
Also in this case the obstruction to localization of the loop equation is of non-cohomological nature, being the line integral of a $\delta^{(2)}$, as opposed
to the $\delta^{(1)}$ that occurs in the regularized holomorphic loop equation.
For twistor Wilson loops the contact term in the $MM$
loop equation gives rise to the same divergent contribution \footnote{The missing factor of $\Lambda^2$ is in fact hidden in the normalization of $S_{NC}$ (sect.(4)).}
for backtracking cusps as for unitary Wilson loops.
Cancellations occur only in the solution of the loop equation. But even if it were possible to get cancellations at the cusps by fine-tuning, a lattice regularization of the $MM$ equation would provide links rather than points,
because the functional integral in the $MM$ equation involves the connection instead of the curvature.
\par
Hence the $MM$ loop equation cannot localize in the homological sense that we are discussing, not even for twistor
Wilson loops, because it cannot be regularized in a way that does implement the zig-zag symmetry.

\section{Canonical beta function of large-$N$ $YM$ by homological localization}

\subsection{Gluing rules for local systems}

In this section we compute the canonical beta function of the large-$N$ $YM$ theory in the scheme defined by the homological localization of the loop
equation. We take into account some global constraint that we have disregarded in our computation of sect.(9) of local counterterms in the Wilsonian scheme. \par
We have seen in the last section that, for getting homological localization of the holomorphic loop equation, we should draw backtracking strings from the loop to the lattice points
in order to transform all the marked points into cusps. \par
We now show that the cusps can be paired by the backtracking strings. \par
This is a consequence of requiring the consistency of the gluing rules for functional integrals with the localization on local systems. Indeed, to glue together two disks with twistor Wilson loops as boundaries, we need that the twistor Wilson loops on the boundaries, taken with opposite orientation, agree. Moreover, in a theory in which the twistor Wilson loops are localized on local systems, they are determined by the holonomies around the punctures inside the disks, since local systems are the same as representations of the fundamental group. \par
Therefore the precise condition for gluing is that the product of the holonomies around the points inside the two disks are the inverse of each other, in such a way that the total product is $1$. This constraint is satisfied naturally gluing a lattice of vortices
in one disk with a lattice of the same number of antivortices \footnote{By antivortices we mean surface operators that have precisely the inverse holonomy of vortices. They have the same classical action, if evaluated at the same cutoff scale, and a neighborhood with moduli space of the same dimension as for vortices.} in the other disk.
Hence in this case the number of punctures in the two disks to be glued must be equal in general. 
Another possibility is to pair vortices in one disk with vortices in the other disk and to impose the condition that the product of all the holonomies is $1$ after gluing, for example requiring that the total number of points
is an integer multiple of $N$.
This can be certainly done for disks that have the same number of punctures and the same cutoff. Afterward one of the disks is rescaled together with the cutoff to a large size in order to perform the thermodynamic limit (see below). The procedure does not spoil localization since all the punctures are paired by backtracking arcs, that can be glued on the common boundary of the disks since the punctures are equal in number. \par
This looks like a holographic correspondence \cite{MB3} that is the consequence that all the punctures can be paired by arcs connecting punctures in different disks \footnote{These are the "hairs" associated to the holographic correspondence between the ultraviolet and the infrared (see below).}. \par
The resulting configuration has not a translational invariant cutoff, as in the computation of the Wilsonian beta function, but what really matters for that computation to hold is that the lattice on which the background field lives has a very large number of punctures and it is locally uniform. \par
If the punctures of the two disks are paired, the density of the punctures in the two disks cannot be the same in the thermodynamic limit, since the area of one disk must be much larger than the area
of the other one. Thus we get two cutoff scales, $a$ and $\tilde a$, on the two disks, that we refer to as the ultraviolet and infrared scale respectively and
that we identify with the lattice spacing. Thus each puncture carries a weight that is the lattice scale. Since the weights are all equal inside the two disks, because we require at least locally uniform lattice spacing, and the punctures are equal in number, the two collections of weights are in fact projectively equal, i.e. equal up to a common rescaling factor. \par
Therefore the localization of the loop equation for twistor Wilson loops on local systems leads to a kind gluing that coincides with the one implied by a weighted arc families with projectively equal weights \cite{Pen}.  
Surprisingly these are precisely the string gluing axioms \cite{Pen,Pen1,Pen2} for weighted arc families at topological level (fig.4 I of \cite{Pen2}). Hence we may say that open strings solve 
the $YM$ loop equation for the twistor Wilson loops, in the sense that they localize the 
loop equation on a saddle point for an effective action. \par 
We have just come to the conclusion that if we combine the gluing properties
of the functional integral with the vanishing requirement for the contact term in the holomorphic loop equation, i.e.
the requirement of localization, we get the string gluing axioms at topological level.
Thus the proper graph to get localization is a weighted graph \cite{Pen} similar to a Mandelstam graph \cite{Man,Man1} (fig.4 I of  \cite{Pen2}). The weights
in this language are the sizes of the strips. \par 
A subtle point arises about the cutoff of the localized effective action.
We recall that the introduction of a lattice is essential for localization, since it allows us to transform
every non-trivial marked lattice point into a backtracking cusp.
We have seen that the gluing rules imply a different cutoff
at the cusps of the two disks of the sphere in the quantum effective action. \par
Hence the local part of the effective action, as a consequence of the stringy nature of localization in the loop
equation, is in fact bilocal with two local fields
living at different scales, one at the ultraviolet cutoff and one at the infrared cutoff, paired by the backtracking strings.
We will see momentarily that the field at the ultraviolet, but not the one at the infrared,
affects the renormalization of the Wilsonian coupling constant,
as it is expected from its very definition.
Instead the field at the infrared together with the one at the ultraviolet affects the
renormalization of the canonical coupling constant. \par
To summarize, we draw our weighted graph \cite{Pen, Pen2, Man, Man1}, that is made by two charts with the boundary loop
in common. The charts are a conformal transformation of two topological disks with punctures
(fig.(1) and fig.(2) of \cite{Pen1}).
This introduces a conformal transformation in the $EK$ two-dimensional reduced theory.
However, this transformation lifts to a conformal rescaling in the four-dimensional parent theory
because on the Lagrangian submanifold that is the support of twistor Wilson loops Eq.(10.6) holds.
Thus the four-dimensional metric changes conformally and
the effective action changes by the appropriate conformal
anomaly (Eq.(10.7)). This implies that, in addition to the explicit cutoff dependence,
the effective action on the weighted graph is related to the one
on the sphere with marked points by the addition of a divergent conformal anomaly,
because of the singularity of the conformal transformation.
This divergent conformal anomaly plays a key role to reconcile our computation of the anomalous dimension in the canonical beta function
with the general properties of the $RG$ group and, more generally, in the interpretation a posteriori of localization
as a $RG$ flow to the ultraviolet. \par 

\subsection{$Z$ factor and canonical normalization}

It is useful to write the effective action of the commutative Morita equivalent theory before the $EK$ reduction, that amounts to dividing by the factor of $N_2$ (sect.(9)).
This is convenient to define the canonical normalization of the effective action in analogy to the $SUSY$ case of sect.(3.2). \par
The contribution of one surface operator of $Z_N$ holonomy, that we call one-vortex, to the local part of the Morita equivalent Wilsonian effective action reads:
\bea
&&\exp{(-\Gamma_q(one-vortex))} \nonumber \\
&&= \exp{(- \frac{2 \pi}{H a^2} \frac{8 \pi^2 }{2 g_W^2} Z^{-1} k(N-k))}\exp{( \frac{2 \pi}{H a_T^2} \frac{k(N-k)}{2} \log(\frac{1}{H a^2}))} \nonumber \\
&&\exp{(- \frac{2 \pi}{H \tilde a^2} \frac{8 \pi^2 }{2 g_W^2} k(N-k) )} \exp{( \frac{2 \pi}{H a_T^2} \frac{k(N-k)}{2} \log(\frac{1}{H \tilde a^2}))}
\eea
where the factor of $ \frac{2 \pi}{H a_T^2} $ is the transverse measure over the zero modes two-dimensional vortex sheet that is the singular
divisor of a surface operator in four dimensions, and $H=\frac{1}{\theta}$ by definition.
The factor in the second line of the right hand side is actually the contribution of the antivortex in the infrared \footnote{Or of the vortex in the infrared.}, that is paired to the vortex by a backtracking string.
Equating the ultraviolet cutoff on the transverse and longitudinal planes, because of rotational invariance, we get $a=a_T$. Another way of formulating this condition is that the longitudinal measure on the size of vortices
that we read from the classical action and the transverse measure on the vortices two-dimensional sheet
should coincide.  
The Wilsonian beta function of sect.(9) follows,
since the contribution of the antivortex at the infrared is finite. \par
For a twistor Wilson loop in the adjoint representation 
it is possible to pair the holomorphic integral to the antiholomorphic one. In this case the counting of zero
modes corresponds to the real dimension of the orbits. Thus  we get for the bilocal part of the Wilsonian effective action:
\bea
\exp{(-\Gamma_q)}&&= \prod_p \exp{(- \frac{2 \pi}{H a^2} \frac{8 \pi^2 }{2 g_W^2} Z^{-1}
k_p(N-k_p)+c.c.)} \exp{( \frac{2 \pi}{H a_T^2} \frac{k_p(N-k_p)}{2} \log(\frac{1}{H a^2})+c.c.)} \nonumber \\
&&\exp{(- \frac{2 \pi}{H \tilde a^2} \frac{8 \pi^2 }{2 g_W^2} k_p(N-k_p)+c.c.)} 
\exp{( \frac{2 \pi}{H a_T^2} \frac{k_p(N-k_p)}{2} \log(\frac{1}{H \tilde a^2})+c.c.) }
\eea
We now come to the canonical coupling.
We observe that the fields at the ultraviolet and at the infrared are not canonically normalized
in the Wilsonian effective action. 
In order to obtain the canonical $\beta$ function, as in the $\mathcal{N}$ $=1$ $SUSY$ $YM$ case of sect.(3.2), we have to rescale the fields on the ultraviolet divisor by a factor of $g Z^{\frac{1}{2}}$ and on the infrared divisor by a factor of $g$, since $Z$ is finite in the infrared and can be normalized to $1$. This can be achieved by rescaling $a$ and $H$ as $a=a_c Z^{-\frac{1}{2}}$ and 
$H= g^{-2} H_c$ and setting $a_c=a_T$ to preserve an equal longitudinal and transverse cutoff. Rescaling $H$ is equivalent to rescale inversely $L^2$, the area of a surface operator.
The canonically normalized effective action reads:
\bea
\exp{(-\Gamma_q)}&&= \prod_p \exp{(- \frac{2 \pi}{H_c a_c^2} \frac{8 \pi^2 }{2 g_W^2} g^2
k_p(N-k_p)+c.c.)} \exp{( \frac{2 \pi}{H_c  a_c^2} g^2\frac{k_p(N-k_p)}{2} \log(\frac{1}{H_c g^{-2}Z^{-1} a_c^2})+c.c.)} \nonumber \\
&&\exp{(- \frac{2 \pi}{H_c \tilde a^2} \frac{8 \pi^2 }{2 g_W^2} g^2 k_p(N-k_p)+c.c.)} 
\exp{( \frac{2 \pi}{H_c a_c^2} g^2\frac{k_p(N-k_p)}{2} \log(\frac{1}{H_c  g^{-2} \tilde a^2})+c.c.) }
\eea
This produces the following rescaling factors at each point from the contribution of the zero modes:
\bea
[(g Z^{\frac{1}{2}})^{k(N-k)}g^{k(N-k)}]^{\frac{2 \pi}{H_c a_c^2} g^2}
\eea
Thus defining as in sect.(3.2) 
\bea 
-\frac{8\pi^2 k(N-k)g^2}{2g^2(\Lambda)}=-\frac{8\pi^2 k(N-k)g^2}{2g_W^2(\Lambda)}+2k(N-k)g^2\log g+g^2\frac{1}{2}k(N-k)\log Z
\eea
and factorizing the term $8\pi^2 k(N-k) g^2$ we get:
\bea
\frac{1}{2g_W^2(\Lambda)}=\frac{1}{2g^2(\Lambda)}+\frac{4}{(4\pi)^2}\log g+\frac{1}{(4\pi)^2}\log Z
\eea
Taking the derivative with respect to $\log\Lambda$ and using the fact that $g_W$ is 1-loop exact we obtain:
\bea
\beta_0=-\frac{1}{g^3}\frac{\partial g}{\partial\log\Lambda}+\frac{4}{(4\pi)^2}\frac{1}{g}\frac{\partial g}{\partial\log\Lambda}+\frac{1}{(4\pi)^2}\frac{\partial \log Z}{\partial\log\Lambda}
\eea
from which it follows:
\bea
\frac{\partial g}{\partial\log\Lambda}=\frac{-\beta_0 g^3+\frac{g^3}{(4\pi)^2}\frac{\partial \log Z}{\partial\log\Lambda}}{1-\frac{4}{(4\pi)^2}g^2}
\eea
Since
\bea
\frac{\partial \log Z}{\partial\log\Lambda}\sim \frac{10}{3}\frac{g_W^2}{(4\pi)^2}\frac{1}{1+\frac{10}{3}\frac{g_W^2}{(4\pi)^2}\log\frac{\Lambda}{\mu}}\sim\frac{10}{3}\frac{g_W^2}{(4\pi)^2}\sim \frac{10}{3}\frac{g^2}{(4\pi)^2}
\eea
we get:
\bea
\frac{\partial g}{\partial\log\Lambda}=\big(-\beta_0 g^3+\frac{10}{3}\frac{g^5}{(4\pi)^4}\big)\big(1+\frac{4}{(4\pi)^2}g^2+\ldots\big)=-\beta_0 g^3-\beta_1g^5+\ldots
\eea
where
\bea
\beta_1=\frac{4\beta_0}{(4\pi)^2}-\frac{10}{3}\frac{1}{(4\pi)^4}=\frac{34}{3}\frac{1}{(4\pi)^4}
\eea
that agrees with the perturbative result up to two-loops \cite{AF,AF1,2loop1,2loop2}. \par
Now we come to the computation of the anomalous dimension.
The one-loop exactness of $Z$ implies:
\bea
\frac{\partial \log Z}{\partial \log a} =-\frac{ \frac{1}{(4\pi)^2} \frac{10}{3} g_W^2}
{1-g_W^2 \frac{1}{(4\pi)^2} \frac{10}{3}
\log (\frac{\tilde a}{\sigma a})}
\eea
where now we have included the contribution of the conformal anomaly, that rescales the subtraction point by the factor of $\sigma$
to give a finite but arbitrary result for the higher order contributions to the anomalous dimension. Assuming that the beta function is independent
on the subtraction point, as required by general principles of the $RG$, the $RG$ trajectory must be followed along the line $c=-\frac{1}{(4\pi)^2} \frac{10}{3}
 \log (\frac{\tilde a}{\sigma a})=const$. We observe that it is precisely
the contribution of the conformal anomaly that allows the anomalous dimension to be a function of the coupling
only, according to the $RG$.

\section{Glueball spectrum from the localized effective action}

\subsection{Glueball propagators and anomalous dimensions in perturbation theory at large-$N$}

We start this section recalling some features and conjectures about the large-$N$ limit of pure $YM$. \par
To the leading large-$N$ order and to every order in the 't Hooft coupling constant $g$, the expectation value of a product of normalized
local gauge invariant operators factorizes \cite{Mig}. For example:
\bea
< \frac{1}{N} \sum_{\alpha \beta} tr F_{\alpha \beta}^2(x_1)...\frac{1}{N} \sum_{\alpha \beta}tr F_{\alpha \beta}^2(x_k)>
= 
< \frac{1}{N} \sum_{\alpha \beta} tr F_{\alpha \beta}^2(x_1)>...
< \frac{1}{N} \sum_{\alpha \beta} tr F_{\alpha \beta}^2(x_k)>  
\eea
Thus to this order the only information that survives is the value of the condensate.
In the case of  $<g^{-2} \sum_{\alpha \beta} \frac{1}{N} Tr F_{\alpha \beta}^2(x)>$, 
it must be proportional for a suitable regularization to the appropriate power of the renormalization
group invariant scale, $\Lambda_{YM}$, since it coincides up to numerical factors with the action density. \par
In turn $\Lambda_{YM}$ encodes
the information on the beta function of the large-$N$ theory. 
In addition it is believed that to the next to leading $\frac{1}{N}$
order the connected two-point functions of local gauge invariant operators are saturated by a sum of pure poles. \par
For example, for the scalar glueball
propagator it is conjectured that the equation holds \cite{Mig,Pol}:
\bea
\int < \frac{1}{N} \sum_{\alpha \beta} tr F_{\alpha \beta}^2(x) \sum_{\alpha \beta} \frac{1}{N} tr F_{\alpha \beta}^2(0)>_{conn}
\ e^{ip x} d^4x= 
\sum_r \frac{Z_r}{p^2+M_r^2}  
\eea
The sum of pure poles is constrained by the perturbative operator product 
expansion \cite{Narison}. It must agree asymptotically for large momentum with the "anomalous dimension"
of the glueball propagator as computed by perturbation theory plus the sum over condensates that occur in the operator product expansion  \cite{Narison}.
Indeed the scalar glueball propagator behaves in perturbation theory at large momentum, within two-loop accuracy,
up to contact terms, i.e. polynomials in the momentum squared, $p^2$, as \cite{Narison}:
\bea
 g^4(p) p^4 \log(\frac{p^2}{\mu^2}) 
\eea
\par
The logarithm explicitly displayed in the two-loop computation is necessary to reproduce the conformal behavior:
\bea
\int d^4p e^{ipx} p^4 \log(\frac{p^2}{\mu^2}) \sim \frac{1}{x^8}
\eea
The factors of $g$, the renormalized 't Hooft coupling at momentum $p$, occur because of the canonical normalization of the glueball propagator,
that involves $<\sum_{\alpha \beta} \frac{1}{N} Tr F_{\alpha \beta}^2(x)>$ rather than the action density $<g^{-2} \sum_{\alpha \beta} Tr F_{\alpha \beta}^2(x)>$, and they account for the one-loop anomalous dimension that  in this case is determined by the one-loop coefficient of the beta function. They imply logarithmic corrections to the conformal behavior.
Glueball propagators for other normalized gauge invariant operators of naive dimension $L$ involve in general one-loop anomalous dimensions that are independent on the one-loop coefficient of the beta function \cite{Zar1,Zar2} :
\bea
&&\int < \frac{1}{N} trO(x)  \frac{1}{N} trO(0)>_{conn}
\ e^{ip x} d^4x \nonumber \\
&&=\sum_r \frac{Z_r}{p^2+M_r^2}  \nonumber \\
&&= G_O(p^2) 
\sim Z^2_O(p^2) p^{2L-4} \log (\frac{p^2}{\mu^2}) 
\eea
and satisfy the following $RG$ equation:
\bea
(\frac{\partial}{\partial \log p}+ \beta(g) \frac{\partial}{\partial g}+2 \gamma_O(g))G_O(p^2)=0
\eea
where
\bea
\gamma_O(g)=\frac{\partial \log Z_O}{\partial \log p}
\eea
Sometimes it is convenient to factorize out the contribution of the anomalous dimension:
\bea
G'_O(p^2)= Z^{-2}_O(p^2)G_O(p^2)
\eea
in such a way that
the suitably normalized glueball propagator, $G'_O(p^2)$, is $RG$ invariant \footnote{We learned this suggestion by an unpublished talk of Gabriele Ferretti, \emph{"Applying the $BES$ trick to $QCD$"}, Newton Institute, Cambridge (2007).}:
\bea
(\frac{\partial}{\partial \log p}+ \beta(g) \frac{\partial}{\partial g})G'_O(p^2)=0
\eea
In the large-$N$ limit there is a sector of the theory that is integrable at one loop \cite{Zar1,Zar2}, that is made by operators of $ASD$ or $SD$ type and their covariant derivatives. The corresponding anomalous dimensions can be computed as the eigenvalues of a Hamiltonian spin chain. \par
In the $ASD$ one-loop integrable sector the anomalous dimension of $ \frac{1}{N} \sum_{\alpha \beta} Tr F_{\alpha \beta}^{-2}(x)$ is the same as the one of $ \frac{1}{N} \sum_{\alpha \beta} Tr F_{\alpha \beta}^2(x)$  since 
$<g^{-2} \sum_{\alpha \beta} \frac{1}{N} Tr (F_{\alpha \beta} \tilde F_{\alpha \beta})>$ and $<g^{-2} \sum_{\alpha \beta} \frac{1}{N} Tr F_{\alpha \beta}^2(x)>$  are both $RG$ invariant \cite{Zar1}.
Therefore the renormalization factor of $ \frac{1}{N} \sum_{\alpha \beta} Tr F_{\alpha \beta}^{-2}(x)$ is $g^2$ and the one-loop anomalous dimension coincides with $-2 \beta_0$. \par
The anomalous dimensions of a number of operators can be computed explicitly solving by the Bethe ansatz the Hamiltonian spin chain in the thermodynamic limit, that corresponds to operators of large
naive dimension $2L$ and length $L$ \footnote{We would like to thank
Konstantin Zarembo for explaining to us this result at the GGI. We would like to thank also Gabrielle Ferretti for explanations about the same subject at Chalmers University.}. In particular the anomalous dimensions of the antiferromagnetic ground states turn out of to be of the form (Eq.(27) of \cite{Zar1} and Eq.(5.23) of \cite{Zar2}):
\bea
Z_L= 1 - L g^2\frac{5}{3} \frac{1}{(4 \pi)^2} \log (\frac{\Lambda}{\mu}) +O(L^0)
\eea
The ground state of the spin chain corresponds to the operators with the most negative anomalous dimension, that turn out to be all scalars constructed by certain contractions involving only
the $ASD$ part of the curvature \cite{Zar2}. 

\subsection{Localized effective action and holomorphic/antiholomorphic fusion}

We are now ready to start our computation of the glueball spectrum. 
The large-$N$ one-loop integrable sector contains the correlators that can be computed by localization. This is perhaps not completely surprising since localization
involves the change of variables from the gauge connection to the $ASD$ part of the curvature, that lives in the one-loop integrable sector.\par
In the previous sections we have obtained the localization of trivial twistor Wilson loops, in the sense that we have reduced the loop equation for them to a critical equation for an effective action and we have renormalized the effective action in Euclidean space, in such a way that the effective action is finite. \par
Now the question arises as to which infrared information the renormalized effective action actually contains, since the effective action is naturally defined on a physical section
of space-time, as opposed to the support of the trivial twistor Wilson loops. \par
We have seen that to make sense of the holomorphic loop equation of sect.(6) after the renormalization of the Euclidean effective action it is necessary to continue analytically the Euclidean loop equation
to Minkowski space-time. Then at operator level the twistor Wilson loops become:
\bea
Tr_{\cal N} \Psi(\hat B_{\rho};L_{ww}) \rightarrow Tr_{ \cal{N}} P \exp i \int_{L_{ww}}(\hat A_{z+}+i \rho \hat D_u) dz_++(\hat A_{ z_-}+i \rho^{-1} \hat D_{\bar u}) d z_- 
\eea
as shown in sect.(4).
The support of the twistor Wilson loops analytically continued in this way becomes:
\bea
( z,  \bar z, u,  {\bar u})=(z_+, z_-, -\rho  z_+, -\rho^{-1} z_-)
\eea
For computational simplicity we set $\rho=-1$ in the following. This is just an irrelevant change of sign of the antihermitian part of $\mu_1$, since $\mu_{\rho}=\mu^0+\rho n-\rho^{-1}\bar n$.
Thus in this section $\mu=\mu_{-1}=\frac{1}{2}(\mu_{01}-i\mu_{03})$. \par
It is in the spirit and in the letter of the localization idea that the renormalized effective action analytically continued to Minkowski space-time contains information on local observables supported
on the Lagrangian submanifold displayed in Eq.(12.12) since this is the analytic continuation of the support of Euclidean twistor Wilson loops from which the effective action has been obtained by localization.
However, getting the spectrum of fluctuations exceeds by far the limited framework of localization of the homology of $1$ in the same sense in which the statement that the prepotential of $\mathcal{N}=2$ $SUSY$ $YM$
determines the low energy effective action in the Coulomb branch exceeds by far the framework of the localization of the cohomology of $1$. \par
We have seen in sect.(10) that localization by homology requires implicitly extending diagonally the action of the conformal group from two to four dimensions, an extension that can occur meaningfully only on Lagrangian submanifolds of the kind displayed in Eq.(12.12). Thus the aforementioned Lagrangian submanifold is in a sense the
only object for which we can hope that the localized effective action has a physical meaning beyond the leading large-$N$ approximation. \par
More physically homological localization requires that the $ASD$ field be singular and therefore by necessity be of magnetic type, in a theory in which smooth fields are of electric type.
Thus homological localization realizes in a technical sense long-standing ideas on the $YM$ vacuum as a dual superconductor \cite{Super1,Super2,Super3}, whose analog are the duality ideas \cite{SW} that lead to the physical justification of the prepotential as the low energy effective action in the Coulomb branch. \par
In particular 't Hooft duality leads us to hope that homological localization may capture the mass gap via the implied condensation of the magnetic charge. \par
We write now the localized effective action in the Wilsonian scheme. From the effective action we extract the glueball spectrum restricted to the Lagrangian Minkoswki section of space-time. 
To compute fluctuations we need to extend the effective action from the fixed points, characterized by $\partial_{\hat A} \hat{ \bar D}=\bar \partial_{\hat A} \hat D=0$, to fluctuations in neighborhood of them.
The fluctuations correspond generically to surface operators of semisimple type, i.e. with different eigenvalues of the (diagonalizable) holonomy and of the $ASD$ curvature, $\mu$. \par
For a twistor Wilson loop in the fundamental representation this extension can be performed only in a holomorphic way via the holomorphic loop equation. The corresponding effective action is not
Hermitian. Therefore we couple the twistor Wilson loop in the fundamental representation with the one in the complex conjugate representation\footnote{This coupling is automatic  via large-$N$ factorization if we start with twistor Wilson loops in the adjoint representation. In this representation the condensation of the magnetic charge via surface operators of $Z_N$ holonomy is compatible with triviality of twistor Wilson loops,
because of the pairwise cancellation in the holonomy of every $Z_N$ factor by its complex conjugate for any shape of the loop.}. As a result the effective action is Hermitian \footnote{The complex conjugate field and the Hermitian conjugate contain the same information since they differ by transposition of the indices.}. 
Thus we are realizing a sort of holomorphic/antiholomorphic fusion that is reminiscent and in fact technically very similar to holomorphic/antiholomorphic fusion \cite{CV,CV1} in conformal field theory. \par
Initially, thanks to large-$N$ factorization, we treat $\mu$ and $\bar \mu$  as independent variables that define two different chiral sectors. But then we choose the section in function space
where $\bar \mu$ is actually the Hermitian conjugate of $\mu$ for the effective action to be Hermitian. This enables us to compute $\frac{1}{N}$ fluctuations. \par
Yet, this $\mu / \bar \mu$ sector is only a special sector of the theory, that involves certain correlators constructed by $\mu=\frac{1}{2} (\mu^-_{01}-i  \mu^-_{03})$ and its Hermitian conjugate. These are the special correlators accessible to homological localization of the loop equation. We cannot say anything about correlators of $\mu^-_{02}$. \par
Realizing holomorphic/antiholomorphic fusion in the non-commutative Eguchi-Kawai reduced theory (sect.(4.1)) involves therefore the product of the holomporphic and antiholomorphic partition functions, extended to the holomorphic and antiholomorphic neighborhoods of the fixed points, that are thought to be each the Hermitian conjugate of the other one:
\bea
 Z&&= \int e^{-\Gamma} \big |\prod_p \delta \hat \mu'_p \big |^2   \nonumber \\
 &&=\big| \int \hat \delta A   \delta \bar{ \hat A} \delta \hat D   \delta  \hat { \bar D} 
\delta(-i F_{\hat B} - \sum_p \hat \mu_p  \delta^{(2)} (z-z_p) -\theta^{-1}1) \nonumber \\
&&\delta(\partial_{\hat A} \hat{ \bar D}+\bar \partial_{\hat A} \hat D )\delta(\partial_{\hat A} \hat{ \bar D}-\bar \partial_{\hat A} \hat D +i \sum_p  (\hat \mu_p - \hat{\bar \mu}_p)  \delta^{(2)} (z-z_p))  \nonumber \\ 
&& \exp \big(-\frac{4N \hat N}{g_W^2}  \sum_p  tr_N   \hat Tr_{\hat N} (\hat \mu_p \hat {\bar \mu}_p) \big) 
 Det (\frac{\delta \hat \mu}{\delta \hat \mu'}) \prod_p \delta \hat \mu'_p \big|^2
 \eea
We then proceed as in sect.(7). In the reduced theory the classical action is finite on surface operators because of the rescaling explained in sect.(4.1).
The $U(N)$ non-commutative theory can be compactified on a torus of large area, $L^2$.
For rational values of the dimensionless non-commutative parameter:
\bea
2 \pi \theta L^{-2 }=\frac{\hat M}{\hat N}
\eea
the non-commutative $U(N)$ theory is Morita equivalent to a $U(N \times \hat N)$ commutative gauge theory on the smaller torus, $L^2 \hat N^{-2}$,
with non-trivial 't Hooft flux through the commutative torus and with the same 't Hooft coupling constant, $g$.
Thus the large-$N$ limit of the partition function reduces to a finite dimensional inductive sequence:
\bea
Z&&= \lim_{N,\hat N \rightarrow \infty}\big| \int  
 \delta A   \delta \bar{  A}  \delta D   \delta \bar{ D} 
\delta(-i F_{B} - \sum_p  U(u,\bar u)\mu_p U(u,\bar u)^{-1} \delta^{(2)} (z-z_p)) \nonumber \\ 
&&\delta(\partial_{ A} {  D}-\bar \partial_{A}  D +i \sum_p  ( U(u,\bar u)\mu_pU(u,\bar u)^{-1} - U(u,\bar u)^{-1}{\bar \mu}_pU(u,\bar u))  \delta^{(2)} (z-z_p))  \nonumber \\ 
&&\delta(\partial_{ A} {  D}+\bar \partial_{A}  D )
 \exp \big(-\frac{4N \hat N}{g_W^2}  \sum_p  tr_N    Tr_{ \hat N} ( \mu_p  {\bar \mu}_p) \big) 
 \frac{\delta \mu}{\delta \mu'}\prod_p \delta \mu'_p \big|^2
\eea
The unitary matrices, $U(u,\bar u)$, account for the twisted boundary conditions on the commutative torus.
In the thermodynamic limit, $L^2 \rightarrow \infty$, the effect of the twist of the torus disappears since the Morita equivalent theory must coincide in the large-$N$ limit with the commutative
theory on $R^2$. The equivalence with the untwisted theory can be seen also choosing the gauge where there is no twist
in front of the delta functions. Of course this gauge is singular on the torus, in the sense that it is defined by a gauge transformation that is not single-valued on the torus, precisely because it is not continuous on the boundary
of the fundamental domain that is employed to define the torus.
But the boundary becomes irrelevant in the thermodynamic limit. \par
We require that the fields on the commutative $R^2$ so obtained in the thermodynamic limit have well defined limits at infinity, in such a way that the theory can be compactified on $S^2$.
Thus the fields on $R^2 \times R^2$ can be extended to $S^2 \times S^2$  
and the resolution of the identity reduces to the one for ordinary surface operators. Thus we get:
\bea
 Z&&=\lim_{N,\hat N \rightarrow \infty} \big| \int  
 \delta A   \delta \bar{  A}  \delta D   \delta \bar{ D} 
\delta(-i F_{B} - \sum_p  \mu_p  \delta^{(2)} (z-z_p)) \nonumber \\
&&\delta(\partial_{ A} { \bar D}+\bar \partial_{ A}  D )\delta(\partial_{ A} { \bar D}-\bar \partial_{ A} D +i \sum_p  ( \mu_p - {\bar \mu}_p)  \delta^{(2)} (z-z_p))  \nonumber \\ 
&&\exp \big(-\frac{4N \hat N}{g_W^2}  \sum_p  tr_N    Tr_{ \hat N} ( \mu_p  {\bar \mu}_p) \big) 
\frac{\delta \mu}{\delta \mu'}\prod_p \delta \mu'_p \big|^2
\eea
that in the Feynman gauge, as shown in sect.(3.3) and sect.(8), reduces to:
\bea
Z&&= \big| \int  
\big [Det^{-1/2}(-\Delta_{A}^. \delta_{\alpha \beta}-i \mu_{\alpha \beta}^{- .}) Det(-\Delta_{A}^.) \big]_{ \mu = \frac{1}{2} (  \mu^-_{01}-i  \mu^-_{03})}\nonumber \\ 
&& \Lambda^{n_b [ \mu'] }
\exp \big(-\frac{4N \hat N}{g_W^2}  \sum_p  tr_N   Tr_{\hat N} (\mu_p  {\bar \mu}_p) \big) 
\frac{\delta  \mu}{\delta \mu'} \wedge \omega'^{\frac{n_b[ \mu']}{2}} \prod_p \delta \mu'_p \big|^2
\eea
In this formula $\omega'$ should be chosen in a way compatible with the holomorphic/antiholomorphic fusion. In particular $\omega'$ should depend holomorphically on the complex
eigenvalues, $\lambda$, of $\mu$. This requires that $\omega'$ be the pullback on the moduli space of the symplectic form associated to the twistor connection for $\rho=-1$, $\hat B=\hat B_{-1}$
(sect.(7)):
\bea
\omega'=\frac{1}{2 \pi}\int' d^2 z  tr_f \hat Tr (\delta {\hat B}_{z} \wedge \delta {\hat B}_{\bar z})
\eea
The superscript in $\omega'$ refers to the version of $\omega$ defined on the punctured sphere as discussed in sect(7). This is not restrictive, since excluding the singular divisor is equivalent to omitting in $\omega$
the sum of the Kirillov forms of the adjoint orbits on the singular divisor \cite{Malkin}, because the volume form on these orbits is already taken into account by the product measure on the adjoint orbits, $|\prod_p \delta \mu_p'|$, (sect.(12.3))
that occurs by the resolution of identity. As a consequence: 
\bea
 \omega^{\frac{n_b[ \mu']}{2}} \wedge  \bar \omega^{\frac{n_b[ \mu']}{2}} \wedge \prod_p \delta \mu'_p \delta \bar \mu'_p= \omega'^{\frac{n_b[ \mu']}{2}} \wedge \bar \omega'^{\frac{n_b[ \mu']}{2}} \wedge \prod_p  \delta \mu'_p\delta \bar \mu'_p
\eea
In any case the version that involves $\omega'$ is the one that leads to the glueball potential.

\subsection{The glueball potential}

The glueball potential arises as a term in the localized effective action that is the logarithm of the modulus of the Jacobian for the change of variables of the complex field
of $ASD$ type from a unitary, $\mu$, to a holomorphic gauge, $\mu'$. \par
This term cannot arise in perturbation theory and the only way to derive it in our approach is via the holomorphic loop equation,
since the choice of the holomorphic gauge is necessary in order to produce the Cauchy kernel in the right hand side of the holomorphic loop
equation. \par
In turn the Cauchy kernel is essential, because after analytic continuation to Minkowski space-time leads to localization by homology, since
the corresponding regularized distribution is zig-zag invariant in a neighborhood of a cusp and therefore its contour integral vanishes for arcs that backtrack at the cusps. \par
The physical meaning of this Jacobian is the following. \par
For surface operators, because of the specific features of the Hitchin equations, there is a mismatch between the degrees
of freedom carried by the $ASD$ curvature at a point in a unitary and in a holomorphic gauge. \par
While at moduli level the unitary and holomorphic gauges are completely equivalent as shown in sect.(7),
the local degrees of freedom that are manifest looking at the $ASD$ curvature in the unitary and in the holomorphic gauge do not coincide completely. \par
In the holomorphic gauge there is an essentially \footnote{The moduli depend actually only on the conjugacy class of the eigenvalues of the holonomy, that determines the eigenvalues of the curvature 
only up to shifts of $2 \pi$ .} one-to-one correspondence between the moduli that occur in the holonomy, $M_p$,
of the connection, $B$, around a point, $p$, and the moduli of the local curvature, $\mu'_p$, at $p$, given by the equation:
\bea
M_p=e^{2i\mu_p'}
\eea
Thus in the holomorphic gauge $\mu_p'$ is parameterized by orbits in the complexification of the gauge group, $\mu_p'=G_p \lambda_p G_p^{-1}$, with $\lambda_p$ the (complex) eigenvalues of $\mu_p'$,
that we assume all different in the semisimple case relevant to compute fluctuations. \par
Therefore the integral over $\mu_p'=G_p\lambda_p G_p^{-1}$ is on an orbit of the complexification of the gauge group with measure:
\bea
\delta \mu_p'= Det(ad \lambda_p) \delta\lambda_p\delta G_p
\eea
where the integration on $\delta G_p$ is actually on the moduli of the orbit, that can be parametrized as $G_p= g_p P_p$, by factorizing the complexification of the gauge group into
its compact and parabolic factor \footnote{This is the Iwasawa decomposition.}. In fact it is not restrictive to project the parabolic factor, $P_p$, to its unipotent subgroup, $P_p'$, since the diagonal factor of $P_p$ acts trivially by conjugation on the eigenvalues. \par
Thus the preceding equation can be rewritten as:
\bea
\delta \mu_p'=  \Delta(\lambda_p)^2
\delta\lambda_p\delta g_p \delta P_p'
\eea
where the square of the Vandermonde determinant of the eigenvalues, $\Delta(\lambda)$:
\bea
\Delta(\lambda)=\prod_{i>j}(\lambda_i-\lambda_j)
\eea
arises by the usual Faddeev-Popov procedure, as in the holomorphic matrix models \cite{Vafa,Laz}. \par
In a unitary gauge this does not hold true.
It is a fundamental result of the theory of surface operators \cite{S5,S6} that in a unitary gauge $[\mu_p,\bar \mu_p]=0$, so that generically $\mu_p$ and $\bar \mu_p$ can be diagonalized simultaneously by a unitary gauge transformation (sect.(7)).
Thus only the moduli associated to the adjoint action of the compact unitary group, $g_p$, are manifest in the $ASD$ curvature in a unitary gauge, while the remaining degrees of freedom are hidden in the moduli of the Hermitian metric whose choice is implicit in the unitary gauge \cite{S6}.  \par
Now the non-Hermitian matrix, $\mu_p=g_p(\lambda_p + u'_p) g_p^{-1}$, that occurs in the resolution of identity that defines the non-$SUSY$ Nicolai map in a unitary gauge, is generically conjugated by a unitary transformation to a triangular matrix, $\lambda_p + u'_p$, with $u'_p$ triangular and nilpotent, in such a way that the induced measure is:
 \bea
 \delta\mu_p&&=Pf(ad \mu_p)\delta \lambda_p\delta u'_p \delta g_p \nonumber \\
 &&= \Delta(\lambda_p) \delta \lambda_p\delta u'_p \delta g_p 
 \eea
The different power of the Vandermonde determinant in Eq.(12.24) with respect to Eq.(12.22) arises by the Faddev-Popov procedure for triangularizing rather that diagonalizing $\mu_p$ \footnote{We may get the same result noticing that the integration measure $d\mu d\bar \mu$ on normal non-Hermitian matrices is $|\Delta(\lambda)|^2 d\lambda d\bar \lambda dg$ by the standard Faddeev-Popov procedure and then taking
the "square root".}. \par
But the unitary gauge transformation that sets the complex matrix $\mu_p$ in triangular form in a unitary gauge does actually diagonalize those $\mu_p$ that arise as the $ASD$ curvature on the dense locus of the surface operators in the resolution of identity of the non-$SUSY$ Nicolai map.
Therefore in the induced measure the integration over the nilpotent part, $u'_p$, actually decouples, since the $ASD$ curvature in the resolution of identity  does not depend in fact on $u'_p$ for those connections that arise by surface operators in a unitary gauge. Indeed the delta function in the resolution of identity, $\delta(-i F_{B} - \sum_p  \mu_p  \delta^{(2)} (z-z_p))$, reduces to $\delta(-i D_{B} \wedge \delta B - \sum_p  g_p u'_p g_p^{-1}  \delta^{(2)} (z-z_p))$
around the solution of $-i F_{B} - \sum_p  g_p \lambda_p g_p^{-1} \delta^{(2)} (z-z_p)=0$ and of the remaining Hitchin equations. But then the last delta function implies $u'_p=0$ because for smooth fluctuations, $\delta B$, the argument of the delta function is dominated by $\delta^{(2)} (z-z_p)$ at each point $z_p$. Therefore the integral on $u'_p$ decouples.
On the other hand the integration over the moduli, $P_p'$, of the twistor connection, $B$, contributes to the zero modes in the unitary gauge and
produces the measure:
\bea
\omega'^{\frac{n_b[ \mu']}{2}}= Pf(\omega') \prod_p \delta P_p'
\eea
in such a way that: 
\bea
\frac{\delta\mu}{\delta{\mu'}} \wedge \omega'^{\frac{n_b[ \mu']}{2}}&& =Pf (\omega') \prod_p \frac{Pf(ad \mu_p)}{Det(ad \mu_p)}\frac{\delta \lambda_p \delta P_p' \delta g_p}{\delta \lambda_p\delta G_p} \nonumber \\
&&=\frac{Pf (\omega')}{\prod_p Pf(ad \mu_p)}
\eea
where we used $G_p= g_p P_p'$. 
The same result is obtained in the singular gauge where $\mu_p$ is actually diagonal, $\mu_p=\lambda_p$, for which therefore $\delta u'_p \delta g_p$ decouples from the measure $\delta \mu_p$, since the delta function in the resolution of identity, $\delta(-i F_{B} - \sum_p  \mu_p  \delta^{(2)} (z-z_p))$, reduces to $\delta(-i D_{B} \wedge \delta B - \sum_p  u'_p  \delta^{(2)} (z-z_p))$
around the solution of $-i F_{B} - \sum_p  \lambda_p  \delta^{(2)} (z-z_p)=0$ and of the remaining Hitchin equations. 
Therefore the last delta function implies $u'_p=0$ and therefore the integral on $u'_p$ decouples.
But the measure on the moduli is now:
\bea
\omega'^{\frac{n_b[ \mu']}{2}}=  Pf(\omega') \prod_p \delta P_p' \delta g_p
\eea
Therefore the effective action reads:
\bea
\Gamma&&=\frac{8N \hat N}{g_W^2}  \sum_p  tr_N   Tr_{\hat N} (\mu_p  {\bar \mu}_p) +\sum_p \log|\Delta(\mu_p)|^2
\nonumber \\
&&- \log\big |Det^{-1/2}(-\Delta_{A}^. \delta_{\alpha \beta}-i \mu_{\alpha \beta}^{- .}) Det(-\Delta_{A}^.) \big|^2_{ \mu = \frac{1}{2} (  \mu^-_{01}-i  \mu^-_{03})} \nonumber \\
&&-{2 n_b [ \mu'] } \log \Lambda-\log|Pf(\omega')|^2
\eea
We anticipated that the second term turns out to be the glueball potential, that generates the glueball masses. The term $\log|Pf(\omega')|^2$ is irrelevant for the glueball potential, since $\omega'$ depends on the eigenvalues only through the holonomy of the connection (sect.(7)). Indeed, since the holonomy at the critical points lives in $Z_N$,
the second derivative of $\log|Pf(\omega')|^2$ at the critical points couples only to the trace of the fluctuations, $Tr(\delta \lambda_p)$, and thus decouples in the large-$N$ limit. The Pauli-Villars factors do not contribute to fluctuations. The remaining terms are considered in the following subsections. \par
It is very instructive to reinsert the factor of $N_2$ to recover the unreduced theory
and to introduce a regularization by the density of the lattice of surface operators that is more suitable for the continuum limit.
Reinserting the factor of $N_2$ we get:
\bea
\Gamma&&=\frac{8N \hat N}{g_W^2} N_2 \sum_p  tr_N   Tr_{\hat N} (\mu_p  {\bar \mu}_p) +\sum_p N_2 \log|\Delta(\mu_p)|^2
\nonumber \\
&&- \log\big |Det^{-1/2}(-\Delta_{A}^. \delta_{\alpha \beta}-i \mu_{\alpha \beta}^{- .}) Det(-\Delta_{A}^.) \big|^2_{ \mu = \frac{1}{2} (  \mu^-_{01}-i  \mu^-_{03})} \nonumber \\
&&-{2 N_2 n_b [\mu'] } \log \Lambda-\ N_2 \log|Pf(\omega')|^2 \nonumber \\
&&=\frac{8N \hat N}{g_W^2} \delta^{(2)}(0) \int d^2u \sum_p  tr_N   Tr_{\hat N} (\mu_p  {\bar \mu}_p) +\sum_p \delta^{2}(0) \int d^2u \log|\Delta(\mu_p)|^2
\nonumber \\
&&- \log\big |Det^{-1/2}(-\Delta_{A}^. \delta_{\alpha \beta}-i \mu_{\alpha \beta}^{- .}) Det(-\Delta_{A}^.) \big|^2_{ \mu = \frac{1}{2} (  \mu^-_{01}-i  \mu^-_{03})} \nonumber \\
&&-{2 \delta^{(2)}(0) \int d^2u \sum_p n_b [\mu_p'] } \log \Lambda- \delta^{(2)}(0) \int d^2u \log|Pf(\omega')|^2
\eea
with the traces that define the functional determinants properly rescaled.
We can now introduce the density of lattice points:
\bea
\rho=\sum _{p'} \delta^{(2)}(z-z_{p'}) 
\eea
normalized in such a way that
\bea
\int d^2z \sum _{p'} \delta^{(2)}(z-z_{p'})= N'_2
\eea
is the number of lattice points at the scale at which the density is $\rho$. 
Notice that the density is not normalized necessarily to $N_2$, the number of lattice points at the cutoff scale, because the primed sum is only on lattice points where the
holonomy of the surface operator is non-trivial. This allows $\rho$ to scale non-trivially with the $RG$.
Assuming translational invariance at $N=\infty$ the effective action reads:
\bea
\Gamma&&=\frac{8N \hat N}{g_W^2}  \int d^2u d^2z   \rho^2 tr_N   Tr_{\hat N} (\mu  {\bar \mu}) + \int d^2u d^2z   \rho^2 \log|\Delta(\mu) |^2
\nonumber \\
&&- \log\big |Det^{-1/2}(-\Delta_{A}^. \delta_{\alpha \beta}-i \mu_{\alpha \beta}^{- .}) Det(-\Delta_{A}^.) \big|^2_{ \mu = \frac{1}{2} (  \mu^-_{01}-i  \mu^-_{03})} \nonumber \\
&&-{2 \int d^2u d^2z   \rho^2 n_b [\mu'] } \log \Lambda- \int d^2u \rho \log|Pf(\omega')|^2
\eea
This form of the effective action is of the utmost importance, because it shows that the coefficient of the glueball potential is in fact a $RG$ invariant scale as it should be.

\subsection{The effective action is degenerate at the fixed points}

We can compute $\rho$ in terms of $\Lambda_{W}$, the $RG$ invariant scale in the Wilsonian scheme, by minimizing the renormalized effective action as a function of $\rho$ for a given $\Lambda$ and $g_W$.
In fact we know already from sect.(8) that the effect of the third and forth term in the effective action is to renormalize the coupling constant. Hence the local divergent part
of the effective action for surface operators with $Z_N$ holonomy of magnetic charge $k$ and density $\rho$ reads:
\bea
\Gamma_k&&=\frac{k(N-k) \hat N^2(4 \pi)^2}{2 g_W^2} (1 - g_W^2 \frac{10}{3} \frac{1}{(4 \pi)^2} \log \frac{\Lambda}{M} )\int d^2u d^2z   \rho^2 \nonumber \\
&&-{2 k(N-k) \hat N^2 \int d^2u d^2z   \rho^2 } \log \frac{\Lambda}{M}+... \nonumber \\
&&=k(N-k) \hat N^2(4 \pi)^2 (\frac{1}{2 g_W^2}  -  \frac{11}{3} \frac{1}{(4 \pi)^2} \log \frac{\Lambda}{M} )\int d^2u d^2z   \rho^2 
+...  \nonumber \\
&&=k(N-k) \hat N^2(4 \pi)^2 ( -  \beta_0 \log \frac{\Lambda e^{-\frac{1}{2 \beta_0 g_W^2}}  }{M} )\int d^2u d^2z   \rho^2 
+... \nonumber \\
&&= -  \beta_0 \hat N \log \frac{\Lambda e^{-\frac{1}{2 \beta_0 g_W^2}}  }{M} \int d^2u d^2z  M^4
+... \nonumber \\
&&= -  \beta_0 \hat N \log \frac{\Lambda_W } {M} \int d^2u d^2z  M^4 +...
\eea
where we have chosen the subtraction point at the scale of the action density:
\bea
M^4=\hat N k(N-k)(4 \pi)^2 \rho^2 
\eea
This condition ensures that all the sectors labelled by $k$ are degenerate in the large-$N, \hat N$ limit, in such a way that the renormalized effective action
at the subtraction scale is large and negative and equal for all of them. In this case the trivial solution with magnetic charge $k=0$ is excluded since it has greater action and therefore there is condensation
of surface operators with all the magnetic charges. Indeed the critical equation:
\bea
\frac{\delta \Gamma_k}{\delta M}= 4M^3  \log \frac{\Lambda_W } {M}-M^3=M^3(4\log \frac{\Lambda_W }{M}-1)=0
\eea
has solution:
\bea
\log \frac{\Lambda_W }{M}=\frac{1}{4}
\eea
in such a way that
the effective action reaches its  negative minimum:
\bea
-  \beta_0 \frac{\hat N}{4} \int d^2u d^2z  M^4
\eea
with $M^4$ given by:
\bea
M^4=e^{-1} \Lambda^4_W
\eea
This also means that for large $N$ the square density of surface operators scales as $\frac{1}{k}$ and as $\frac{1}{N}$. It implies also that the glueball propagators are a sum of pure poles as we will see momentarily. \par
We may wonder as to whether finite terms, i.e. not $\Lambda$ divergent, may affect this picture. It easy to see that terms proportional to $k(N-k)$ simply redefine $ \Lambda_W$.
However it is not obvious that the contributions from all finite parts have this form. Thus finite terms may affect the dependence of $\rho$ on $k$.   \par
In fact we have chosen the action density of the condensate, $ \rho^2 Tr_{\hat N} tr_{N}(\lambda^2 1_N 1_{\hat N})$, as the infrared subtraction scale, $M^4$ \footnote{ Finite changes of the
subtraction point affect the normalization of the glueball propagator but do not affect the glueball spectrum.}. This is the same prescription as for the Veneziano-Yankielowicz effective action \cite{VY} of $\cal{N}$ $=1$ $SUSY$ $YM$. This subtraction point implies that the renormalized action is the same in every sector of magnetic charge $k$,
in such a way that all the $Z_N$ magnetic charges "condense at once" with a renormalized square density that scales as $\frac{1}{k}$. \par
While this is the same prescription that occurs in the effective action of $SUSY$ theories in fact its justification at fundamental level may imply a certain fine-tuning of the finite parts in the renormalized effective action. This
fine-tuning is always possible if the surface operators  of $Z_N$ holonomy that occur at the fixed points are viewed as limit points in the closure of orbits with unipotent holonomy in the Lagrangian cone mentioned in sect.(7) and sect.(9).
Indeed in this case the relative scale of $|Pf(\omega')|^2$ can be suitably adjusted for different $k$ approaching the limit points where the nilpotent residue of the Higgs field vanishes. \par
This is essentially due to the fact that the moduli space of such orbits is non-compact and that we are suitably choosing the size of the neighborhood of the fixed points in order to satisfy
certain conditions. Indeed every "compactification" is to some extent arbitrary \footnote{In Nekrasov theory of cohomological localization a certain (to some extent arbitrary) compactification of the moduli spaces of instantons has to occur too. Such compactification turns out to be compatible with the Seiberg-Witten ideas on the electric/magnetic duality in $\mathcal{N}$ $=2$ $SUSY$ $YM$.
In our case we can argue similarly that we can choose the size of the neighborhoods of the fixed points to avoid that only a proper subgroup of $Z_N$ condenses, spoiling 't Hooft ideas on electric/magnetic duality.}. On the contrary, the unitary orbits in the Lagrangian cone, that are the other orbits that saturate the beta function (sect.(7) and sect.(9)), do not allow such fine-tuning since they are compact. \par

\subsection{The kinetic term}

We are now ready to compute the glueball propagator. It is quite clear that the classical action cannot furnish the kinetic term for the glueballs since it is ultralocal. Therefore the kinetic term must be generated by radiative corrections around surface operators. This turns out to be the case for the fluctuations of Lagrangian-embedded surface operators analytically continued to Minkowski space-time. 
They can be obtained by diagonally embedded Euclidean surface operators continued to Minkowski space-time. \par 
The diagonal embedding can be described as follows. We choose the surface $(z=u,\bar z= \bar u)$ diagonally embedded in $R^4$. Since we have defined a lattice in the $(z, \bar z)$ plane,
this defines a lattice also in the $(u, \bar u)$ plane by the diagonal map $(z_p=u_p,\bar z_p= \bar u_p)$. This lattice in the $(u, \bar u)$ plane has a set of dual plaquettes in such a way
that the $(u, \bar u)$ plane is the union of the plaquettes. Now we define the function $z_p(u, \bar u)=u$ with domain the interior of the plaquette dual to $p$ and analogously for the complex conjugate.
We also define a lattice fluctuating field supported on the plaquette dual to $p$ and locally constant as $(u, \bar u)$ vary in the support, $\delta \mu_p(u,\bar u)$. $\delta \mu_p(u,\bar u)$ is zero outside its support.
We are now ready to do computations.
We suppose that in addition to the translational invariant background of surface operators there are locally-defined fluctuating surface operators \footnote{Both the background and the fluctuations are diagonal matrices in color space.} diagonally embedded in space-time:
\bea
-i F_{B} = \sum_p  \mu  \delta^{(2)} (z-z_p)+ \sum_p \delta \mu_p(u,\bar u) \delta^{(2)} (z-z_p(u, \bar u))
\eea
Since the kinetic term for the glueball propagator must arise by the radiative corrections we examine the expansion of the functional determinants in one-loop graphs with multiple insertions
of trees. In our case these terms carry multiple insertions of the background field and of the fluctuating field supported on the lattice of surface operators. We have seen in sect.(9) and sect.(11) that the divergent parts, that contain the background field, determine the beta function.
We are now interested in the finite parts to second order, that contain the fluctuating field. \par
The justification is as follows. Every term of the loop expansion contains a trace in the adjoint representation and thus it is proportional to $N^2$, that of course diverges for large $ N$.
However, the loop expansion is in fact an expansion in powers of the density $\rho$ of surface operators.
But since the density scales as $N^{-\frac{1}{2}}$ only the leading quadratic term survives the double large-$N, \hat N$ limit. \par
The spin contribution to the effective action is:
\bea
-\frac {2N \hat N' 4}{ (4 \pi^2)^2 } \sum_{ p \neq p' }  \int d^2u d^2v  \frac{ tr_N Tr_{\hat N'} (\delta \mu_p(u, \bar u) \delta \bar \mu_{p'}(v, \bar v) ) } {(|z_p(u, \bar u)-z_{p'}(v, \bar v)|^2+|u-v|^2)^2} 
\eea
plus the complex conjugate term that we add only at the end of the computation.
The orbital contribution has the same structure and a different coefficient and sign in order to reproduce the divergent $Z^{-1}$ factor when evaluated on the
translational invariant background of surface operators \footnote{ The  $Z^{-1}$ factor, contrary to the beta function, does not depend on sitting on surface operators of $Z_N$ holonomy.} :
\bea
\frac {\frac{1}{3} N \hat N' 4}{ (4 \pi^2)^2 } \sum_{ p \neq p' }  \int d^2u d^2v  \frac{ tr_N Tr_{\hat N'} (\delta \mu_p(u, \bar u) \delta \bar \mu_{p'}(v, \bar v) ) } {(|z_p(u, \bar u)-z_{p'}(v, \bar v)|^2+|u-v|^2)^2} 
\eea
Indeed the orbital  contribution to second order, up to constant overall factors, is \footnote{The term involving $\partial_{\alpha} A_{\alpha}$
vanishes identically around the local singularity, while the term involving $A_{\alpha}^2$ is quadratically divergent and does not contribute because of cancellations due to gauge invariance.}:
\bea
&&\int d^4x d^4y Tr(A_z(x) \partial_{\bar z} \frac{1}{(x-y)^2}  A_{\bar z}(y) \partial_z \frac{1}{(x-y)^2} + A_u(x) \partial_{\bar u} \frac{1}{(x-y)^2}  A_{\bar u}(y) \partial_u \frac{1}{(x-y)^2} ) \nonumber \\
&&=\int d^4x d^4y Tr(A_z(x) \frac{2 (z-w)}{(x-y)^4}  A_{\bar z}(y) \frac{2 (\bar z- \bar w)}{(x-y)^4}+A_u(x) \frac{2 (u-v)}{(x-y)^4}  A_{\bar u}(y) \frac{2 (\bar u- \bar v)}{(x-y)^4} ) \nonumber \\
&&=\int d^2z d^2w d^2u d^2v Tr(A_z(x) A_{\bar z}(y) \frac{4 |z-w|^2}{(|z-w|^2+|u-v|^2)^4} + A_u(x) A_{\bar u}(y) \frac{4 |u-v|^2}{(|z-w|^2+|u-v|^2)^4}) \nonumber \\
&&\sim- \int d^2z d^2w d^2u d^2v Tr(A_z(x) A_{\bar z}(y)  \partial_{\bar z} \partial_{w} \frac{4}{6 (|z-w|^2+|u-v|^2)^2}+A_u(x) A_{\bar u}(y)  \partial_{\bar u} \partial_{v} \frac{4}{6 (|z-w|^2+|u-v|^2)^2}) \nonumber \\
&&=- \int d^2z d^2w d^2u d^2v Tr( \partial_{\bar z} \partial_{w}(A_z(x) A_{\bar z}(y))   \frac{4}{6 (|z-w|^2+|u-v|^2)^2}+\partial_{\bar u} \partial_{v}(A_u(x) A_{\bar v}(y))   \frac{4}{6 (|z-w|^2+|u-v|^2)^2}) \nonumber \\
&&=- \int d^2z d^2w d^2u d^2v Tr( \partial_{\bar z} A_z(x) \partial_{w} A_{\bar z}(y)  \frac{4}{6 (|z-w|^2+|u-v|^2)^2} + \partial_{\bar u} A_u(x) \partial_{u} A_{\bar u }(y)  \frac{4}{6 (|z-w|^2+|u-v|^2)^2}\nonumber \\
&&\sim \sum_{ p, p' }  \int d^2u d^2v  \frac{N Tr(\delta \mu_p \delta \bar \mu_{p'}) }{(|z_p(u,\bar u)-z_{p'}(v,\bar v)|^2+|u-v|^2)^2}
\eea
where in the last line we used $\partial_{\bar z} A_z(x) \sim \sum_p \delta \mu^0_p \delta^{(2)}(z-z_p(u,\bar u))$, $\partial_{w} A_{\bar z}(y) \sim -\sum_p \delta \mu^0_p \delta^{(2)}(w-z_p(v, \bar v))$,
$\partial_{\bar u} A_u(x) \sim \sum_p \delta n_p \delta^{(2)}(z-z_p(u,\bar u))$ and $\partial_{v} A_{\bar u}(y) \sim -\sum_p \delta {\bar n}_p \delta^{(2)}(w-z_p(v,\bar v))$ for the fluctuations of surface operators on the
diagonal Lagrangian submanifold, with $x=(z, \bar z, u, \bar u)$ and $y=(w, \bar w, v, \bar v)$.
The total result is:
\bea
-\frac {5N \hat N' 4}{3 (4 \pi^2)^2 } \sum_{ p \neq p' }  \int d^2u d^2v  \frac{ tr_N Tr_{\hat N'} (\delta \mu_p(u, \bar u) \delta \bar \mu_{p'}(v, \bar v) ) } {(|z_p(u, \bar u)-z_{p'}(v, \bar v)|^2+|u-v|^2)^2} 
\eea
The coefficient can be found without direct computation since, for the surface operators that are constant on the $(u,\bar u)$ plane and translational invariant, Eq.(12.43) must give rise to 
the logarithmic divergence that produces $Z^{-1}$.
Expressing Eq.(12.43) in terms of the density of the surface operators, $\rho$, we get:
\bea
&&-\frac {20 N \hat N'}{3 (4 \pi^2)^2 } (\int d^2v d^2u \rho^2)   \int d^2u d^2v\frac{ tr_N Tr_{\hat N'} (\delta \mu(u, \bar u) \delta \bar \mu(v, \bar v) ) } {(|u-v|^2+|u-v|^2)^2} \nonumber \\
&&=-\frac {20 N }{3 (4 \pi^2)^2 k(N-k)(4 \pi)^2} (\int d^2v d^2u e^{-1} \Lambda^4_W)   \int d^2u d^2v\frac{ tr_N Tr_{\hat N'} (\delta \mu(u, \bar u) \delta \bar \mu(v, \bar v) ) } {(|u-v|^2+|u-v|^2)^2} \nonumber \\
\eea
that for large $N$ reduces to \footnote{It is not restrictive to require $k=1,...,\frac{N}{2}$ for $N$ even, because of the symmetry of all our formulae for the exchange $k \rightarrow N-k$.}     :
\bea 
-\frac {20 e^{-1}} {3 (4 \pi^2)^2 2k (4 \pi)^2} N_2'^2   \int d^2u d^2v\frac{ tr_N Tr_{\hat N'} (\delta \mu(u, \bar u) \delta \bar \mu(v, \bar v) ) } {(u-v)^2(\bar u-\bar v)^2}  \nonumber \\
\eea
where we have set:
\bea
N_2'= \int d^2u \Lambda_W^2
\eea
This term generates the kinetic term of the glueball propagator after analytic continuation to Minkowski space-time \footnote{It is necessary to assume that $\delta \mu^-_{01}(u_+,  u_-)$ and $\delta \mu^-_{03}(u_+,  u_-)$  are the boundary values 
of holomorphic functions on the upper-half plane for each of the independent variables, $(u_+,  u_-)$, with suitable properties at infinity. }. Inserting the complex conjugate term we get:
\bea
&&-\frac {20 e^{-1}} {3 (4 \pi^2)^2 2k (4 \pi)^2} N_2'^2   \int du_+ du_- dv_+ dv_- \frac{ tr_N Tr_{\hat N'} (\delta \mu(u_+,  u_-) \delta \bar \mu(v_+,  v_-) ) } {(u_+-v_+ +i \epsilon)^2(u_- - v_-+i \epsilon)^2} + c.c.  \nonumber \\
&&=(2 \pi)^2\frac {20 e^{-1}} {3 (4 \pi^2)^2 2k (4 \pi)^2} N_2'^2   \int du_+ du_-  tr_N Tr_{\hat N'} (\delta \mu(u_+,  u_-) \partial_+\partial_- \delta \bar \mu(u_+,  u_-) ) +c.c. \nonumber \\
\eea
We notice that for obtaining this result it is crucial that fluctuations occur as surface operators, that the support of fluctuations is embedded as a Lagrangian submanifold in Euclidean space-time and that  the analytic continuation to the Minkowskian Lagrangian submanifold is performed. \par
It is interesting to compare the kinetic term just obtained with the contribution of the classical action:
\bea
&&\frac{8N \hat N' }{g_W^2}  \int d^2u d^2z   \rho^2 tr_N   Tr_{\hat N'} (\delta \mu(u, \bar u) \delta {\bar \mu}(u, \bar u)) \nonumber \\
&&=\frac{8N e^{-1 }}{g _W^2  k(N-k)(4 \pi)^2}  (\int d^2z \Lambda^2_W )  \int d^2u  \Lambda^2_W  tr_N  Tr_{\hat N'} (\delta \mu(u, \bar u) \delta \bar \mu(u, \bar u)) \nonumber \\
\eea
that at large $N$ reduces to:
\bea
\frac{8 e^{-1 }}{g _W^2  k (4 \pi)^2}  N_2'  \int d^2u   \Lambda^2_W  tr_N  Tr_{\hat N'} (\delta \mu(u, \bar u) \delta \bar \mu(u, \bar u)) \nonumber \\
\eea
Hence the classical action is irrelevant with respect to the term generated by radiative corrections for the fluctuations in the large-$N$ limit and in the thermodynamic limit in which $N_2'$ diverges.
In fact the real source of the mass term is the glueball potential whose contribution to the effective action  is:
\bea
&&\int d^2u d^2z   \rho^2 \log|\Delta(\mu) |^2 \nonumber \\
&& =  \frac{ e^{-1 }}{\hat N' k(N-k)(4 \pi)^2}      (\int d^2z \Lambda^2_W) \int d^2u  \Lambda^2_W \log|\Delta(\mu) |^2 \nonumber \\
&& =  \frac{e^{-1 }}{ \hat N' k(N-k)(4 \pi)^2}      N_2' \int d^2u  \Lambda^2_W \log|\Delta(\mu) |^2 \nonumber \\
\eea
Apparently the glueball potential in the thermodynamic limit, $N_2'  \rightarrow \infty$, is as suppressed with respect to the kinetic term as the classical action is.
Therefore the only possibility for the theory to have a mass gap is that the glueball potential is singular.
In fact these singularities of the glueball potential arise precisely when some eigenvalues coincide. This looks encouraging because it implies that the mass gap may arise only by configurations
for which the gauge group is unbroken. In addition these configurations for which the eigenvalues coincide must have infinite degeneracies in the thermodynamic limit, in order to
compensate the $\frac{1}{N_2'}$ suppression with respect to the kinetic term. The needed degeneracies follow by the ansatz
that the center $Z_N$ of $SU(N)$ occurs with multiplicity $\hat N$ in the commutative theory Morita equivalent to the non-commutative one. 
 
\subsection{The mass gap}

We are now ready to compute the mass matrix.
With our normalizations $e^{2i\mu}\in Z_N$ with $\mu \in su(N)$ and $\mu$ the translational invariant condensate. Thus:
\bea 
\label{vorticesZN}
2\mu=diag(\underbrace{2\pi(k-N)/N}_{k},\underbrace{2\pi k/N}_{N-k})
\eea
The mass matrix  in the $\mu / \bar \mu$ sector is the second derivative of the logarithm of the modulus of the Vandermonde determinant: 
\bea
\label{glueballM}
&&M^2_{ij}=\frac{\partial^2}{\partial\mu_i\partial\bar{\mu}_j}\log|\Delta(\mu)|^2=\frac{\partial^2}{\partial\mu_i\partial\bar{\mu}_j}\sum_{\alpha>\beta}\log|\mu_\alpha-\mu_\beta|^2\nonumber \\
&&=\frac{\partial}{\partial\bar{\mu}_j}\sum_{\alpha>\beta}\big(\frac{1}{\mu_\alpha-\mu_\beta} \ \delta_{\alpha i}-\frac{1}{\mu_\alpha-\mu_\beta} \ \delta_{\beta i}\big)
+ c.c. \nonumber \\
&&=\frac{\partial}{\partial\bar{\mu}_j}\big(\sum_{\beta<i}\frac{1}{\mu_i-\mu_\beta}-\sum_{\alpha>i}\frac{1}{\mu_\alpha-\mu_i}\big) + c.c.\nonumber \\
&&=\frac{\partial}{\partial\bar{\mu}_j}\big(\sum_{\beta<i}\frac{1}{\mu_i-\mu_\beta}+\sum_{\beta>i}\frac{1}{\mu_i-\mu_\beta}\big)+c.c. \nonumber \\
&&=\frac{\partial}{\partial\bar{\mu}_j}\sum_{\beta\neq i}\frac{1}{\mu_i-\mu_\beta} +c.c. \nonumber \\
&&=\pi\sum_{\beta\neq i}(\delta^{(2)}(\mu_i-\mu_\beta)\delta_{ij}-\delta^{(2)}(\mu_i-\mu_\beta)\delta_{j\beta}) + c.c.\nonumber \\
&&=\pi\sum_{\beta\neq i}\delta^{(2)}(\mu_i-\mu_\beta)\delta_{ij}-\delta^{(2)}(\mu_i-\mu_j)(1_{ij}-\delta_{ij})  +c.c.
\eea
where we used:
\bea
\frac{\partial}{\partial \bar  z_i}\frac{1}{{z}_j}=\pi\delta_{ij} \ \delta^{(2)}(z_j)
\eea
and $1_{ij}$ stands for a matrix with all entries equal to $1$,  $1_{ij}=1$, $\forall \ i,j$.\\
$\delta^{(2)}(\mu_i-\mu_j)$ is a distribution in color space. It is nonvanishing when the two eigenvalues are degenerate. \\
Let us now specialize to the $Z_N$ vortices in Eq.(12.51) with multiplicity $\hat N'$.\\
For the the first block,  $i,j=1,\ldots,k$, Eq.(12.51) gives:
\bea
M^2_{ij}=\pi\delta^{(2)}(0)((\hat N' k-1)\delta_{ij}-(1_{ij}-\delta_{ij}))=\pi\delta^{(2)}(0)(\hat N' k\delta_{ij}-1_{ij})
\eea
with $\delta^{(2)}(0)= \frac{  N \hat N'}{(2 \pi)^2 }  $ the delta function at zero in color space. \par
Since the diagonal terms scale as $\hat N'$, the non-diagonal corrections to the mass term are negligible in the large-$\hat N'$ limit,
in such a way that the theory has a mass gap \footnote{For $\hat N'=1$ and $k=1$ the theory has a massless eigenvalue in addition to the trivial diagonal
$U(1)$ that decouples. We would like to thank Daniele Dorigoni for working out this case during our course at SNS.}. \par
Similarly, for the second block,  $i,j=k+1,\ldots,\frac{N}{2}$, one gets
\bea
M^2_{ij}=\pi\delta^{(2)}(0)(\hat N' (N-k)\delta_{ij}-1_{ij})
\eea
that implies that the glueball masses are at the cutoff scale \footnote{It is not restrictive to require $k=1,...,\frac{N}{2}$ for $N$ even since our formulae are symmetric
for the exchange $k \rightarrow N-k$.} for large $N$ in this block.
Thus the glueball mass term is:
\bea
\frac{ N \hat N'e^{-1 }}{(2 \pi)^2  \hat N'  k(N-k)(4 \pi)^2}   \hat N'   N_2'  \int d^2u k  \pi  \Lambda^2_W tr_N Tr_{\hat N'}(\delta \mu \delta \bar \mu) +c.c.
\eea
to be added to the kinetic term to get:
\bea
&&\frac{e^{-1 }} { (4 \pi)^2 }\big( \frac{ (2 \pi)^2 10}{3 (4 \pi^2)^2  } N_2'^2   \frac{1}{k} \int du_+ du_-  tr_N Tr_{\hat N'} (\delta \mu \partial_+\partial_- \delta \bar \mu )+  \nonumber \\
&&+\frac{1}{(2 \pi)^2}  \hat N'   N_2'  \int du_+ du_-   \pi  \Lambda^2_W tr_N Tr_{\hat N}(\delta \mu \delta \bar \mu) \big) +c.c. \nonumber \\
&&=\frac{e^{-1 }} {(4 \pi)^2 4 \pi}
\big( \frac{10}{3  \pi  } N_2'^2   \frac{1}{k} \int du_+ du_-  tr_N Tr_{\hat N'} (\delta \mu\partial_+\partial_- \delta \bar \mu )  \nonumber \\
&&+\hat N'   N_2'  \int du_+ du_-   \Lambda^2_W tr_N Tr_{\hat N'}(\delta \mu \delta \bar \mu) \big) +c.c.\nonumber \\
\eea
Now we have the identification (Eq.(4.14)):
\bea
\hat N (\frac{2 \pi}{\Lambda})^2=2 \pi \theta 
\eea
with
\bea
2 \pi \theta = L^2 \frac{\hat M}{\hat N}
\eea
and
\bea
N_2= (\frac{\Lambda}{2 \pi})^2 L^2
\eea
Therefore:
\bea
\hat N=(\frac{\Lambda}{2 \pi})^2 L^2 \frac{\hat M}{\hat N}
\eea
at the cutoff scale and
thus for $\frac{\hat M}{\hat N} \rightarrow \frac{1}{n}$ in the large $\hat N$ limit (sect.(7)) for any finite positive integer $n \geq 2$:
\bea
 N_2= n \hat N
\eea
and at the renormalized scale:
\bea
 N_2'= n \hat N'
\eea
by the diagonal embedding. We will see in the next section that any such choice of $n$ corresponds to the choice of a different renormalization scheme
for which the preceding equation holds. \par
Finally, in the reduced theory in the sector labelled by $k$ the glueball effective action is:
\bea
&&\frac{ e^{-1 } }{(4 \pi)^2 2 \pi} 
\big( \frac{10 }{3  \pi} N_2'   \frac{1}{k} \int du_+ du_-  tr_N Tr_{\hat N'} (\delta \mu\partial_+\partial_- \delta \bar \mu )  \nonumber \\
&&+\hat N' \int  du_+ du_- \Lambda^2_W tr_N Tr_{\hat N'}(\delta \mu \delta \bar \mu) \big) + c.c. \nonumber \\
\eea

\subsection{Glueball propagators in the Wilsonian scheme}

Thus we find the following propagator
in the twistor sector of the large-$N$ theory for the Wilsonian normalization of the $EK$ reduced commutative Morita equivalent effective action:
\bea
&&\Lambda_W^6
\int < \frac{1}{N \hat N'}  tr_N Tr_{\hat N'}(\mu \bar \mu)(x_+, x_-,x_+, x_-)  \frac{1}{N \hat N'}  tr_N Tr_{\hat N'}( \mu \bar \mu)(0,0,0,0)>_{conn} e^{i(p_+x_-+ p_-x_+)} dx_+dx_-  \nonumber \\
&& \sim \frac{1}{N^2 \hat N'^2} \sum_{k=1}^{\infty} \frac{ k^2 \Lambda_W^6}{ - \alpha'  p_+ p_-+ k \Lambda_W^2} \nonumber \\
&&\sim \frac{1}{N^2 \hat N'^2} \alpha'^2  (- p_+ p_-)^2 \log \frac{ - p_+ p_-}{ \Lambda_W^2}
\eea
with:
\bea
\alpha'= \frac{ 10 }{3 \pi } n  
 \eea
Indeed it is not hard to see that, setting $k^2 \Lambda_W^4 =  [(k  \Lambda_W^2+ \alpha'  p_+ p_-)(k  \Lambda_W^2- \alpha'  p_+ p_-)+(- \alpha'  p_+ p_-)^2]$, the second line in Eq.(12.65) can be written
as a logarithmic divergent sum that reproduces the correct logarithmic behavior of perturbation theory:
\bea
&&\sum_{k=1}^{\infty} \frac{ k^2 \Lambda_W^6}{ - \alpha'  p_+ p_-+ k \Lambda_W^2} \nonumber \\
&&= \sum_{k=1}^{\infty} \frac{ \big((k  \Lambda_W^2+ \alpha'  p_+ p_-)(k  \Lambda_W^2- \alpha'  p_+ p_-)+(- \alpha'  p_+ p_-)^2 \big) \Lambda_W^2}{ - \alpha'  p_+ p_-+ k \Lambda_W^2} \nonumber \\
&&=\alpha'^2 \sum_{k=1}^{\infty} \frac{(- p_+ p_-)^2 }{ - \alpha'  p_+ p_- \Lambda_W^{-2}+k } +... ,
\eea
up to a divergent sum of condensates, proportional to a power of $\Lambda_W$, and up to a divergent sum of contact terms.
We can now define glueball composite operators in the following way.
Operators of the form:
\bea
tr_N Tr_{\hat N'}(\mu^-_{\alpha \beta} \mu^-_{\alpha \beta})^L
\eea
restricted to surface operators become:
\bea
\delta^{(2)}(0)^{2L-2}\sum_p \delta^{(2)}(0) \delta^{(2)}(z-z_p) tr_N Tr_{\hat N'}(\mu_p \bar \mu_p)^L
\eea
They can be evaluated, in the same fashion as the action density, as :
\bea
\delta^{(2)}(0)^{2L-2}  \rho^2 tr_N Tr_{\hat N'}(\mu \bar \mu)^L
\eea
We set:
\bea
\Lambda^2= \frac{N}{n} \Lambda_W^2
\eea
in such a way that the factors of $N^{2L-2}$ from $\Lambda^{2(2L-2)}$ are cancelled by the powers of $N^{-1}$ from the denominator in $tr_N Tr_{\hat N'}\delta (\mu \bar \mu)^L$ evaluated on the condensate of surface operators. This condition defines a non-perturbative scheme in which the v.e.v. of composite surface operators are actually finite in the large $(\Lambda, N)$ limit. The preceding relation implies:
\bea
\frac{1}{2 g_W^2(\Lambda_W)}=\frac{1}{2 g_W^2(\Lambda)}-\beta_0 \log \frac{N}{n}
\eea
and
\bea
N_2'= n \hat N'
\eea
in such a way that the rank $\mathcal{N}=N_2$ of the Morita equivalent gauge group stays constant along the $RG$-flow, but the degeneracy, $\hat N'$, flows into the rank, $N$, of the diagonal $SU(N)$ along the
$RG$-flow. The flow starts  with $\mathcal{N}=N_2= n \hat N$  at the cutoff scale according to Eq.(12.62), that implies $N=n$ at the cutoff scale. The flow ends with lower $\hat N'$ and larger $N$ at the renormalized scale in such a way that $n \hat N=N \hat N'$. Therefore in the large-$N_2$ limit the gauge group in the infrared is $SU(\infty)$ embedded with $ \infty$ multiplicity in $U(\infty)$,
a result consistent with the structure of the gauge group of the non-commutative gauge theory \cite{Landi,Szabo}.   
 \par
We rescale also $\rho^2$ by a factor of $N \hat N'=N_2$ in order to get a quantity on the order of $1$.
Thus our operators are:
\bea
\delta^{(2)}(0)^{2L-2}  \hat N' N \rho^2 tr_N Tr_{\hat N'}(\mu \bar \mu)^L
\eea
The corresponding glueball propagators in the Wilsonian scheme are:
\bea
&&(\frac{\Lambda}{2 \pi})^{8L-8} \hat N'^2 N^2 \rho^4   \int <tr_N Tr_{\hat N'}(\mu \bar \mu)^L(x_+, x_-) tr_N Tr_{\hat N'}(\mu \bar \mu)^L(0,0)>_{conn} e^{i(p_+x_-+ p_-x_+)}  d^4x 
 \nonumber \\
&&\sim (\frac{\Lambda_W}{2 \pi})^{8L-8} N^{4L-4}  \hat {N'}^2 N^2 \rho^4   \int <tr_N Tr_{\hat N'}(\delta \mu \mu^{L-1} \bar \mu^{L})(x_+, x_-) tr_N Tr_{\hat N'}(\mu^{L} \bar \mu^{L-1}\delta \bar \mu)(0,0)>_{conn} e^{i(p_+x_-+ p_-x_+)}  d^4x \nonumber \\
&&\sim \frac{\Lambda_W^2 L^2}{N}   \sum_{k=1}^{\infty}  \frac{ \Lambda_W^2  k^{4L-2} \Lambda_W^{8L-4}}{ - \alpha'  p_+ p_-+ k \Lambda_W^2} \nonumber \\
&&\sim  \sum_{k=1}^{\infty} \frac{  \Lambda_W^2 k^{2(2L-1)} \Lambda_W^{4(2L-1)}}{ - \alpha'  p_+ p_-+ k \Lambda_W^2} \nonumber \\
&&\sim  \sum_{k=1}^{\infty} \frac{\Lambda_W^2  \big((k  \Lambda_W^2+ \alpha'  p_+ p_-)(k  \Lambda_W^2- \alpha'  p_+ p_-)+(- \alpha'  p_+ p_-)^2 \big)^{2L-1} }{ - \alpha'  p_+ p_-+ k \Lambda_W^2} \nonumber \\
&&\sim  (- p_+ p_-)^{4L-2}  \sum_{k=1}^{\infty} \frac{ \Lambda_W^2}{ - \alpha'  p_+ p_-+ k \Lambda_W^2} \nonumber \\
&&\sim  (- p_+ p_-)^{4L-2} \log \frac{ - p_+ p_-}{ \Lambda_W^2} 
\eea
up to a sum of $RG$ invariant condensates and contact terms.
Only the leading singularity for large momentum has been displayed. In fact the subleading singularities in powers of the momentum have divergent coefficients.
We can "renormalize" these singularities, for which we have not an interpretation, as follows. 
Localization of twistor Wilson loops admits shifting the eigenvalues of surface operators by adding $2 \pi 1$ at every point \footnote{This shift may be related to the split central extension that we disregarded in Eq.(7.15) as opposed to Eq.(7.11).}. Indeed this does not modify anything in the $SU(N)$ sector, but shifts the diagonal $U(1)$
part of the action by a central term, that can be cancelled by a counterterm, in such a way that the effective action is flat for this $U(1)$. \par
We can now construct composite surface operators as in Eq.(12.74), but for the shifted curvature:
\bea 
2\mu=diag(\underbrace{2\pi k/N}_{k},\underbrace{2\pi (k+N)/N}_{N-k})
\eea
that has the same $Z_N$ holonomy. Let us call the two dimensional Fourier transform of these operators $O_{L}(p_+,p_-)$. In the Wilsonian scheme we get:
\bea
&&  <Tr_{\mathcal N} O^L(p_+, p_-)  Tr_{\mathcal N}O^L(-p_+, -p_-)>^{(W)}_{conn}  \nonumber \\
&& \sim  \sum_{k=1}^{\frac{N}{2}} \frac{  \Lambda_W^2 k^{2(2L-1)} \Lambda_W^{4(2L-1)}}{ - \alpha'  p_+ p_-+ k \Lambda_W^2}  \nonumber \\
&& +\sum_{k=1}^{\frac{N}{2}} \frac{ \frac{N^2}{k^2} \Lambda_W^2 (N+k)^{2(2L-1)} \Lambda_W^{4(2L-1)}}{ - \alpha'  p_+ p_-+ (N-k)\Lambda_W^2}  \nonumber \\
&& \sim  (- p_+ p_-)^{4L-2}  \sum_{k=1}^{\infty} \frac{ \Lambda_W^2}{ - \alpha'  p_+ p_-+ k \Lambda_W^2} + ...\nonumber \\
&& \sim  (- p_+ p_-)^{4L-2} \log \frac{ - p_+ p_-}{ \Lambda_W^2} + ...
\eea
where the dots stand for contact terms, i.e. distributions whose inverse Fourier transform is supported at coinciding points. But now there are not anymore subleading singularities in momentum with divergent coefficients. \par
The natural interpretation is that this computation in the Wilsonian scheme furnishes
the $RG$ invariant version of some glueball propagators for which the anomalous dimensions or the powers of the gauge coupling have been factored out. \par

\subsection{Glueball propagators in the canonical scheme}

If we wish to recover the anomalous dimensions we should choose a canonical scheme, in which the fields are normalized in such a way to include
the renormalization factors. \par
Localization of twistor Wilson loops is just a statement about the homology of $1$ and in principle does not provide a dictionary
to identify fluctuations of surface operators with specific glueball propagators in perturbation theory.
However, heuristically we can construct a dictionary on the following basis.
We notice that the one-loop anomalous dimensions in the ground state of the Hamiltonian spin chain coincide within one-loop accuracy with the anomalous dimensions computed by means of the localization on surface operators by the change to the $ASD$ variables:
\bea
Z_L= 1 - L g^2\frac{5}{3} \frac{1}{(4 \pi)^2} \log (\frac{\Lambda}{\mu}) +O(L^0) \sim Z^{-\frac{L}{2}}= (1 - g^2\frac{10}{3} \frac{1}{(4 \pi)^2} \log \frac{\Lambda}{\mu} )^{\frac{L}{2}}
\eea
Indeed, choosing a canonical scheme and following the definitions of sect.(11), we rescale the cutoff by a factor of $Z^{\frac{1}{2}}$ and the area $L^2$ by a factor of $g^2$. 
Thus $\Lambda=Z^{\frac{1}{2}} \Lambda_c$ and $L^2= g^2 L^2_c$.
This changes the normalization of the operators insertions
by factors of $Z$ and the normalization of the effective action of the fluctuations by a power of $g^4$, since the effective action is quadratic in the area of surface operators (Eq.(12.44)).  Hence we get:
\bea
&&(\frac{\Lambda}{2 \pi})^{8L-8} \hat N'^2 N^2 \rho^4   \int <tr_N Tr_{\hat N'}(\mu \bar \mu)^L(x_+, x_-) tr_N Tr_{\hat N'}(\mu \bar \mu)^L(0,0)>^{(W)}_{conn} e^{i(p_+x_-+ p_-x_+)} d^4x \nonumber \\
&& = (\frac{Z^{\frac{1}{2}}\Lambda_c}{2 \pi})^{8L-8} \hat N'^2 N^2 \rho^4   \int <tr_N Tr_{\hat N'}(\mu \bar \mu)^L(x_+, x_-) tr_N Tr_{\hat N'}(\mu \bar \mu)^L(0,0)>^{(W)}_{conn} e^{i(p_+x_-+ p_-x_+)} d^4x \nonumber \\
&&= g^{-4} (\frac{ Z^{\frac{1}{2}} \Lambda_c }{2 \pi})^{8L-8} \hat N'^2 N^2 \rho^4  \nonumber \\
&& \int <tr_N Tr_{\hat N'}(\delta \mu \mu^{L-1} \bar \mu^{L})(x_+, x_-) tr_N Tr_{\hat N'}(\mu^{L} \bar \mu^{L-1}\delta \bar \mu)(0,0)>^{(C)}_{conn} e^{i(p_+x_-+ p_-x_+)} d^4x 
\eea
where the canonical expectation value for the fluctuations is computed with respect to the effective action in Eq.(12.44) with $L$ replaced by $L_c$.
It follows that:
\bea
&&(\frac{\Lambda_c }{2 \pi})^{8L-8}  \hat N'^2 N^2 \rho^4   \int <tr_N Tr_{\hat N'}(\delta \mu \mu^{L-1} \bar \mu^{L})(x_+, x_-) tr_N Tr_{\hat N'}(\mu^{L} \bar \mu^{L-1}\delta \bar \mu)(0,0)>^{(C)}_{conn} e^{i(p_+x_-+ p_-x_+)} 
d^4x  \nonumber  \\
&&= g^4 Z^{-\frac{8L-8}{2}} (\frac{\Lambda_W}{2 \pi})^{8L-8} N^{4L-4} \hat N'^2 N^2 \rho^4  \nonumber \\
&&  \int <tr_N Tr_{\hat N'}(\delta \mu \mu^{L-1} \bar \mu^{L})(x_+, x_-) tr_N Tr_{\hat N'}(\mu^{L} \bar \mu^{L-1}\delta \bar \mu)(0,0)>^{(W)}_{conn} e^{i(p_+x_-+ p_-x_+)} d^4x 
\eea
Thus the perturbative anomalous dimensions for long operators in the ground state of the Hamiltonian spin chain in the thermodynamic limit \cite{Zar1,Zar2} are correctly reproduced by correlations of long surface operators in the canonical scheme. Actually they agree also for $L=1$, since in this case the anomalous dimension is determined by the beta function via the factor of $g^4$.
This suggests also that the states which surface operators factorize on by homological localization are all scalars, although in principle they may couple also to tensors since $\mu \bar \mu$ is not a scalar.
Analogously in the canonical scheme the same anomalous dimensions arise for the operators $O^L$:
\bea
&& <Tr_{\mathcal N} O^L(p_+, p_-)  Tr_{\mathcal N}O^L(-p_+, -p_-)>^{(C)}_{conn}\nonumber \\
&& = g^4(- p_+ p_- ) Z^{-\frac{8L-8}{2}}(- p_+ p_-) <Tr_{\mathcal N} O^L(p_+, p_-)  Tr_{\mathcal N}O^L(-p_+, -p_-)>^{(W)}_{conn}  \nonumber \\
&& \sim g^4(- p_+ p_-) Z^{-\frac{8L-8}{2}}(- p_+ p_-)  (- p_+ p_-)^{4L-2} \log \frac{ - p_+ p_-}{ \Lambda_W^2}
\eea

\section{ Wild local systems and Regge trajectories}

The hyper-Kahler reduction induced by the restriction to local systems can be extended to representations of the wild fundamental group \cite{S5,WW}.
Also the extension to the wild surface operators admits a gauge in which the theory is locally abelian, because of the commutativity of the coefficients 
of the higher order poles \cite{S5,WW}. 
Hence in principle we can extend the computation of the fluctuations of surface operators in Eq.(12.39) to curvatures that involve derivatives of the delta function:
\bea
-i F_{B} = \sum_p  \mu  \delta^{(2)} (z-z_p)+ \sum_p \sum_{n} \delta \mu^{(n)}_p(u,\bar u) \partial ^{n} \delta^{(2)} (z-z_p(u, \bar u)) \nonumber \\
+ \sum_p \sum_{\bar n} \delta \mu^{(\bar n)}_p(u,\bar u) \bar \partial ^{\bar n} \delta^{(2)} (z-z_p(u, \bar u))
\eea
This corresponds naturally to Regge trajectories of higher spins. We leave the computation for the future.

\section{Conclusions and outlook: $QCD$-like theories and the twistorial string theory}

The main conclusion of this paper is that there exist twistor Wilson loops that can be localized in large-$N$ pure $YM$ on local systems, i.e. on representations of the fundamental group of a punctured Riemann sphere immersed in 
space-time, i.e. on surface operators. These surface operators turn out to be connections with $Z_N$ holonomy around the punctures. The localization on surface operators leads to the one-loop exactness of the large-$N$ Wilsonian beta function and to a canonical beta function of $NSVZ$ type. \par
Some understanding of the mass gap and of the glueball spectrum occurs in a certain sector of the theory associated to twistor Wilson loops. \par
By certain changes of variables, that imply integrating on the moduli of surface operators, the loop equation for twistor Wilson loops is written in a holomorphic gauge in which a non-trivial glueball potential is generated by the change of variables. The second derivative of the glueball
potential implies a mass term for the glueballs that is non-vanishing precisely for the surface operators with degenerate eigenvalues that occur for $Z_N$ holonomy. \par
In this language glueballs arise as massive fluctuations of magnetic surface operators supported on the Lagrangian submanifold of space-time that is the support of the twistor Wilson loops. \par
This is the picture that follows from the localization of the loop equation for twistor Wilson loops and that realizes a new version of some long-standing ideas about dual superconductivity in pure $YM$ \cite{Super1,Super2,Super3}. \par
On the field theory side we may wonder as to whether the methods of this paper extend  to $\cal{N}$ $=1$ $SUSY$ $YM$, once we observe that twistor Wilson loops are not in the Parisi-Sourlas cohomology of 
the Nicolai map of $\cal{N}$ $=1$ $SUSY$ $YM$ and therefore there is no reason for which they should be localized on instantons. In fact from the point of view of the standard folklore the glueball spectrum of 
large-$N$ $\cal{N}$ $=1$ $SUSY$ $YM$ should not differ in a qualitative way from the one of $YM$. \par
Another extension would be to $QCD$-like theories, such as $YM$ minimally coupled to $N_f$ massless Dirac fermions in the fundamental representation in the large-$N$ limit, keeping the ratio $\frac{N_f}{N}$ fixed. The computation of the associated glueball spectrum would imply the determination of the lower side of the conformal window, as the point at which the mass gap disappears. \par
The basic issue involved in such extensions is the realization of the correct Wilsonian beta function around the critical points provided by the localization, and involves a crucial understanding of the fermion zero modes in a neighborhood of the critical points. \par
On the string theory side the results of this paper suggest a new string program for the $YM$ theory, if we look for exact solvability. \par
It has been known for some time that $\mathcal{N}=4$ $SUSY$ $YM$ admits a partially equivalent twistorial string \cite{W4,Mass1,Mass2,Mass3}.
The triviality of twistor Wilson loops and the fact that they are supported on Lagrangian submanifolds in twistor space suggests the existence of a stringy interpretation
of our results in terms of open topological strings ending on Lagrangian submanifolds in twistor space \cite{Vafas1,Vafas2}
in presence of surface operators. The occurrence of topological strings, as opposed to the usual strings, is due to the trivial nature of twistor Wilson loops at large $N$. \par
Since such twistorial topological string would be solvable by cohomological localization \cite{Wcs,Wcs1}, morally this conjectured topological string/gauge theory duality would provide the string cohomology 
dual to the field theory homology.
Some hints about this conjectured twistorial string of $YM$ can be found in \cite{MBs}. \par
We may wonder what such a twistorial string theory would be suited for, since it is supposed to be equivalent to the field theoretical results of this paper.
The answer is found in the old fashioned unitarization program of string theory. The field theory computation provides the free glueball spectrum in the twistor sector but does
not furnish easily information about the glueball interactions. However, in the string approach interactions are fixed by the geometry of the string world-sheet, once the free theory is known,
and thus the conjectured gauge theory/ topological string duality would open the way to computing the glueball $S$-matrix.

\section{Acknowledgments}

We would like to thank Riccardo Barbieri and Augusto Sagnotti for inviting us to give a course
at SNS in Pisa during 2010 and 2011 where many details of this paper were worked out. \par
We would like to thank Luca Lopez and Daniele Dorigoni for working out in detail the problems of our course at SNS mentioned in the footnotes (28, 40) and 81 respectively. \par 
We would like to thank the organizers and the partecipants to the GGI workshop 
on "Large-$N$ Gauge Theories" (2011) where this paper was almost completed. \par
In particular we would like to thank Adi Armoni, David Berenstein, Alex Buchel, Luigi Del Debbio, Paolo Di Vecchia, Antonio Gonzalez-Arroyo, Chris Korthals-Altes, Biagio Lucini, Yuri Makeenko, Herbert Neuberger, Vasilis Niarchos, Rodolfo Russo, Jacob Sonneschein, Michael Teper, Gabriele Veneziano, Jacek Wosiek, Konstantin Zarembo, for very interesting discussions on the physics of large-$N$ gauge theories. \par
We would like to thank the Galileo Galilei Institute for Theoretical Physics for the hospitality and the INFN for partial support during the completion of this work.


\begin{thebibliography}{99}

\bibitem{AJ}  A. Jaffe, E. Witten, {\it "Quantum Yang-Mills Theory"}, in \emph{Millennium Prize Problems},
American Mathematical Society, Providence, RI, 2006.
\bibitem{MM} Yu. M. Makeenko, A. A. Migdal, {\it Phys. Lett.} {\bf B 88} (1979) 135.
\bibitem{MM1} Yu. M. Makeenko, A. A. Migdal, {\it Nucl. Phys.} {\bf B 188} (1981) 269.
\bibitem{Mak1} Yu. Makeenko, {\it "Large-$N$ gauge theories"}, [hep-th/0001047].
\bibitem{Hooft} G. 't Hooft, {\it Nucl. Phys.} {\bf B 72} (1974) 461.
\bibitem{MB1} M. Bochicchio, {\it "Quasi $BPS$ Wilson loops, localization of loop equation by homology
and exact beta function in the large-$N$ limit of $SU(N)$ $YM$ theory"}, {\it JHEP} {\bf 0905} (2009) 116, [hep-th/0809.4662].
\bibitem{Nic1} H. Nicolai, \emph{"On a new characterization of scalar supersymmetric theories"}, \emph{Phys. Lett.} {\bf B 89} (1980) 341. 
\bibitem{Nic2} H. Nicolai, \emph{"Supersymmetry and functional integration measures"}, {\it Nucl. Phys.} {\bf B 176} (1980) 419.
\bibitem{V} V. De Alfaro, S. Fubini, G. Furlan, G. Veneziano, {\it Phys. Lett.}
{\bf B 142} (1984) 1.
\bibitem{V1} V. De Alfaro, S. Fubini, G. Furlan, G. Veneziano, {\it Nucl. Phys.}
{\bf B 255} (1985) 399.
\bibitem{Po1} A. M. Polyakov, { \it  "The wall of the cave"},
{\it Int. J. Mod. Phys.} {\bf A 14} (1999) 645, [hep-th/9809057].
\bibitem{Py} A. M. Polyakov,{ \it Nucl. Phys.} {\bf  B 164} (1980) 171.
\bibitem{Gr} D. Drukker, D. Gross, I. Ooguri, {\it Phys. Rev. }{\bf D 60} (1999) 125006, [hep-th/9904191].
\bibitem{Kr} M. Kruczenski, \emph{ "A note on twist two operators in $\mathcal {N}$ $=4$ $SYM$ and Wilson loops in Minkowski signature"}
{\it JHEP} {\bf 0212} (2002) 024, [hep-th/0210115].
\bibitem{NSVZ}  V. Novikov, M. Shifman, A. Vainshtein, V. Zakharov,
{\it Phys. Lett.} {\bf B 217} (1989) 103.
\bibitem{Super1} G. 't Hooft, {\it "On The Phase Transition Towards Permanent Quark Confinement"},
{\it Nucl. Phys.} {\bf B 138} (1978) 1.
\bibitem{Super2}G. 't Hooft, {\it "A Property Of Electric And Magnetic Flux In Nonabelian Gauge Theory"}, {\it Nucl. Phys.} {\bf  B153} (1979) 141.
\bibitem{Super3} G. 't Hooft, {\it "Topology Of The Gauge Condition And New Confinement Phases In Nonabelian Gauge Theories"},
{\it Nucl. Phys.} {\bf  B} 190 (1981) 455.
\bibitem{MB5} M. Bochicchio, {\it "Glueball propagators in large-$N$ $YM$"}, [hep-th/1111.6073].
\bibitem{Zar1}  G. Ferretti, R. Heise, K. Zarembo, \emph{"New Integrable Structures in Large-$N$ QCD"}, \emph{Phys. Rev. }{\bf D 70} (2004), [hep-th/04104187].
\bibitem{Zar2}  N. Beisert, G. Ferretti, R. Heise, K. Zarembo, \emph{"One-Loop QCD Spin Chain and its Spectrum"}, \emph{Nucl. Phys. }{\bf B 717} (2005), [hep-th/0412029].
\bibitem{MB2} M. Bochicchio,{\it "The large-$N$ limit of $QCD$ and the collective field of the Hitchin fibration"}, {\it JHEP} {\bf 9901} (1999) 006, [hep-th/9810015].
\bibitem{MB3} M. Bochicchio, {\it "Solving loop equations by Hitchin systems via holography in large-$N$ $QCD_4$"}, {\it JHEP} {\bf 0306} (2003) 026, [hep-th/0305088].
\bibitem{H2} N. J. Hitchin, {\it Duke Math. J.} {\bf 54} (1987) 91.
\bibitem{K1} E. Markman, {\it Compositio Mathematica} {\bf 93} (1994) 55.
\bibitem{K4}R. Donagi, E. Markman, {\it "Spectral covers, algebraically 
completely integrable, Hamiltonian systems, and moduli of bundles"}, 
[alg-geom /9507017].
\bibitem{S3} T. Hausel, M. Thaddeus, {\it "Mirror symmetry Langlands duality and the Hitchin systems"}, [math.AG/0205236].
\bibitem{W2} S. Gukov, E. Witten, {\it "Gauge theory, ramification and the geometric Langlands program"}, [hep-th/0612073].
\bibitem{Meyer} H. B. Meyer, {\it "Glueball Regge trajectories"}, [hep-lat/0508002].
\bibitem{AA} N. Arkani-Hamed, H. Murayama, {\it JHEP} {\bf 0006 } (2000) 030, [hep-th/9707133].
\bibitem{AF} D. Gross, F. Wilczek, {\it Phys. Rev. Lett.} {\bf 30} (1973) 1345.
\bibitem{AF1} H. Politzer, {\it Phys. Rev. Lett.} {\bf 30} (1973) 1346.
\bibitem{2loop1} W. E. Caswell, {\it Phys. Rev. Lett.} {\bf 33} (1974) 244.
\bibitem{2loop2} D. R. T. Jones, {\it Nucl. Phys.} {\bf B 75} (1974) 531.
\bibitem{DE} J. J. Duistermaat, G. J. Heckman, \emph{ Invent. Math.} {\bf 69} (1982) 259.
\bibitem{Szabo} R. J. Szabo \emph{"Equivariant Localization of Path Integrals"}, [hep-th/9608068].
\bibitem{A} M. F. Atiyah, R. Bott, \emph {Phil. Trans. Roy. Soc. London} {\bf A 308} (1982) 523. 
\bibitem{Bis1} J. Bismut, {\it Comm. Math. Phys.}
{\bf 98} (1985) 213.
\bibitem{Bis2} J. Bismut, \emph{Comm. Math. Phys.} {\bf 103} (1986) 127. 
\bibitem{W} E. Witten, \emph{J. Geom. Phys.} {\bf 9} (1992) 303, [hep-th/9204083].
\bibitem{W1} E. Witten, \emph{J. Math. Phys. } {\bf 35} (1994) 5101, [hep-th/9403195].
\bibitem{N} N. A. Nekrasov, {\it Procedings of the ICM, vol.3 p. 477}, Beijing
(2003), [hep-th/0306211].
\bibitem{SW} N. Seiberg, E. Witten,  \emph{Nucl. Phys.} {\bf B 426} (1994) 19; \emph{Erratum-ibid.} {\bf B 430} (1994) 485, [hep-th/9407087].
\bibitem{P} V. Pestun, {\it "Localization of gauge theory on a four-sphere
and supersymmetric Wilson loops"}, [hep-th/0712.2824].
\bibitem{GA0} G. Gallavotti, {\it "Constructive Quantum Field Theory",
Encyclopedia of Mathematical Physics}, Elsevier (2006), [math-ph/0510014].
\bibitem{Rev} N. Dorey, T. J. Hollowood, V. V. Khoze, M. P. Mattis, \emph{"The Calculus of Many Instantons"}, [hep-th/0206063].
\bibitem{Bian} M. Bianchi, S. Kovacs, G. C. Rossi, {\it "Instantons and Supersymmetry"},
[hep-th/0703142].
\bibitem{Parisi1} G. Parisi and N. Sourlas, \emph{"Random magnetic fields, supersymmetry and negative
dimensions"}, {\it Phys. Rev. Lett.} {\bf 43} (1979) 744.
\bibitem{Parisi2} G. Parisi, N. Sourlas, {\it "Supersymmetric field theories and stochastic differential equations"}, \emph{Nucl. Phys.} {\bf B 206} (1982) 321.
\bibitem{Dijk} R. Dijkgraaf, {\it "Les Houches Lectures on Fields, Strings and Duality"}, [hep-th/ 9703136].
\bibitem{Shif} M. Shifman {\it "Exact results in gauge theories: putting
super-symmetry at work"}, [hep-th/9906049].
\bibitem{H} N. J. Hitchin, {\it Proc. London Math. Soc.} {\bf 55} (1987) 59.
\bibitem{HKL} N. J. Hitchin, A. Karlhede, U. Lindstr\"{o}m, M. 
Roc\v{e}k, {\it Comm. Math. Phys.} {\bf 108} (1987) 535.
\bibitem{DW} R. Donagi, E. Witten, \emph{ Nucl.Phys.} {\bf B 460} (1996) 299, [hep-th/9510101].
\bibitem{Vafa} R. Dijkgraaf, C. Vafa, {\it "A perturbative window into non-perturbative
physics"}, [hep-th/0208048].
\bibitem{Laz} C. I. Lazaroiu, {\it JHEP} {\bf 0305} (2003) 044.
\bibitem{Kaw} H. Kawai, T. Kuroki, T. Morita,  \emph{"Dijkgraaf-Vafa theory as large-$N$ reduction"}, \emph{Nucl. Phys.}
{\bf B 664} (2003) 185, [hep-th/0303210].
\bibitem{WD} F. Cachazo, M. R. Douglas, N. Seiberg, E. Witten, {\it "Chiral Ring and Anomalies
in Supersymmetric Gauge Theory"}, [hep-th/0211170].
\bibitem{CV} S. Cecotti, C. Vafa, \emph{"Topological antitopological fusion"}, \emph{Nucl. Phys.} {\bf B 367} (1991) 359.
\bibitem{CV1} K. Papadodimas, \emph{"Topological Anti-Topological Fusion in Four-Dimensional Superconformal Field Theories"},
{\it JHEP} {\bf 1008} (2010) 118, [hep-th/0910.4963].
\bibitem{Pen} R. M. Kaufmann, R. C. Penner, {\it Nucl. Phys} {\bf B 748}, [math.GT/0603485].
\bibitem{Pen1} R. C. Penner, {\it "Decorated Teichmuller Theory of Bordered
Surfaces"}, [math.GT/0210326].
\bibitem{Pen2} R. K. Kaufmann, M. Livernet, R. C. Penner, {\it "Arc operads and arc algebras"},
[math/0209132].
\bibitem{Man} S. Mandelstam, \emph{"Factorization in Dual Models and Functional Integration in String Theory"}, Contribution to the volume 
{\it "The Birth of String Theory"}, eds. A. Cappelli, E. Castellani, F. Colomo, P. Di Vecchia,  [hep-th/0811.1247]. 
\bibitem{Man1} L. Hadasz, Z. Jaskolski, {\it Int. J. Mod. Phys.} {\bf A 18} (2003) 2609,
[hep-th/0202051].
\bibitem{Po2} A. Polyakov, V. Rychkov, {\it Nucl. Phys.}
{\bf B 581} (2000) 116.
\bibitem{W3} E. Witten, {\it Proc. Nato} IAS, Plenum Press, New York (1980).
\bibitem{Haa1} O. Haan, {\it Phys. Lett.} {\bf B 106} (1981) 207.
\bibitem{Haa2} O. Haan, {\it Z. Physik} {\bf C 6} (1980) 345.
\bibitem{Cv} P. Cvitanovic, P. G. Lauwers, P. N. Scharbach, {\it Nucl. Phys.}
{\bf B 203} (1982) 385.
\bibitem{Si} I. Singer, {\it "The Master Field for Two 
Dimensional Yang-Mills Theory"}, lecture at the July 1994 Mathematical Physics
Conference in Paris.
\bibitem{Douglas} M. R. Douglas, {\it Nucl. Phys. Proc. Suppl.}
{\bf 41} (1995) 66.
\bibitem{Gross} R. Gopakumar, D. J. Gross, {\it Nucl. Phys.} {\bf B 451} (1995) 379.
\bibitem{Douglas1} M. R. Douglas, {\it Phys. Lett.}
{\bf B 344} (1995) 117.
\bibitem{Voiculescu} D. V. Voiculescu, K. J. Dykema, A. Nica,
{\it "Free Random Variables"} AMS, Providence (1992).
\bibitem{Pol}  A. M. Polyakov, {\it "Gauge Fields and Strings"}, Harwood Acad.
Publ., Chur (1987).
\bibitem{Narison}  S. Narison, \emph{Nucl. Phys.} {\bf B 509} (1998) 312, [hep-ph/9612457].
\bibitem{Szabo2} R. J. Szabo, \emph{"Quantum Field Theories on non-Commutative Spaces"}, \emph{Phys. Rept.} {\bf 378} (2003), [hep-th/0109162].
\bibitem{S1} C. Simpson, {\it "The Hodge filtration
on non Abelian cohomology"}, [alg-geom/9604005].
\bibitem{S2} C. Simpson, {\it "Higgs bundles and local systems"},
{\it Publ. Math. IHES} {\bf 75} (1992) 5.
\bibitem{KM} P. B. Kronheimer, T. S. Mrowka, \emph{Topology} {\bf 32} (1993); {\bf 34} (1995).
\bibitem{Konno1} H. Konno, {\it J. Math. Soc. Japan} {\bf 45} (1993) 461. 
\bibitem{Konno2} H. U. Boden, K. Yokogawa, {\it "Moduli spaces of parabolic
 Higgs bundles and parabolic $K(D)$ pairs over smooth curves: I"}, 
 [alg-geom /9610014]. 
\bibitem{S4} J. Jost, J. Li, K. Zuo, {\it"Harmonic bundles on quasi-compact
Kahler manifolds"}, [math.AG/0108166].
\bibitem{S5} O. Biquard, P. Bloach, {\it"Wild non-Abelian Hodge theory on
curves"}, [math.DG/0111098].
\bibitem{S6} O. Biquard, O. Garcia-Prada, I. Mundet i Riera, \emph{"An introduction to Higgs bundles"}, Second International School on Geometry and Physics {\it "Geometric Langlands and Gauge Theory"}, Bellaterra (Spain) (2010).
\bibitem{Moc} T. Mochizuki, \emph{"Asymptotic behaviour of variation of pure polarized TERP structures"}, [math.DG/0811.1384].
\bibitem{BD} A. Beilinson, V. Drinfeld, {\it
"Quantization of Hitchin's integrable system and Hecke
eigensheaves" }, Drinfeld home page at Chicago Univ.
\bibitem{Frenkel} E. Frenkel, {\it "Gauge Theory and Langlands Duality"},
Bourbaki Seminar (2009),  [math.RT/0906.2747]. 
\bibitem{F} E. Frenkel, {\it "Recent advances in the Langlands program"},
 [math.AG/0303074].
\bibitem{F1} B. Feigin, E. Frenkel, N. Reshetikin, {\it Comm. Math. Phys.} {\bf
166} (1994) 27.
\bibitem{F2} E. Frenkel, {\it "Vertex algebras and algebraic curves"}, [math.QA/0007054].
\bibitem{F3} E. Frenkel, {\it "Affine algebras, Langlands duality and Bethe
ansatz"}, [q-alg/9506003].
\bibitem{F4} D. Ben-Zvi, E. Frenkel, {\it "Geometric realization of the Segal-Sugawara
construction"}, [math.AG/0301206].
\bibitem{F5} D. Ben-Zvi, E. Frenkel, {\it "Lectures on
the Wakimoto modules, opers and the center at critical
level"}, [math.QA/0210029].
\bibitem{WW} E. Witten {\it "Gauge Theory and Wild Ramification"}, Anal. Appl. (Singap.) {\bf 6} (2008), [hep-th/0710063].
\bibitem{Peter} S. Vandoren, P. van Nieuwenhuizen, {\it "Lectures on instantons"}, [hep-th/0802.1862].
\bibitem{Mas} L. J. Mason, {\it"Global anti-self-dual Yang-Mills fields in split signature and their
scattering"}, [math-phys/0505039].
\bibitem{W4} E. Witten, {\it Comm. Math. Phys. } {\bf189} (2004) 252, [hep-th/0312171].
\bibitem{DN} M. R. Douglas, N. A. Nekrasov, {\it Rev. Mod. Phys.} {\bf 73} (2001) 977,
[hep-th/0106048].
\bibitem{Twc} A. Gonzales-Arroyo, C. P. Korthals-Altes, {\it Phys. Lett.}
{\bf B 131} (1983) 396.
\bibitem{Twl1} A. Gonzales-Arroyo, M. Okawa, {\it Phys. Lett.} {\bf B 120}(1983) 174.
\bibitem{EK} T. Eguchi, H. Kawai, {\it Phys. Rev. Lett.} {\bf 48} (1982) 1063.
\bibitem{Neu} G. Bhanot, U. Heller, H. Neuberger, {\it Phys. Lett.} {\bf B 113}
 (1982) 47.
\bibitem{Twl2} T. Eguchi, R. Nakayama, {\it Phys. Lett.} {\bf B 122}
 (1983) 59.
\bibitem{Rt} G. C. Rossi, M. Testa, {\it Phys. Lett.} {\bf B 125} (1983) 476.
\bibitem{Mak2} Yu. Makeenko, {\it "The first thirty year of large-$N$ gauge theory"}, [hep-th/0407028].
\bibitem{AG} L. Alvarez-Gaume', J. L. F. Barbon, \emph{Nucl. Phys. } {\bf B 623} (2002) 165, [hep-th/0109176].
\bibitem{Wilson} K. G. Wilson, \emph{Phys. Rev.}{ \bf D10} (1974) 2445.
\bibitem{GA}  A. Gonzalez-Arroyo, \emph{Yang-Mills Fields on the 4-dimensional torus. Part I: Classical Theory 1}, [hep-th/9807108].
\bibitem{K} K. Saraikin, \emph{Comments on the Morita Equivalence}, [hep-th/0005138].
\bibitem{Local} J. Teschner, \emph{"Quantization of the Hitchin moduli spaces, Liouville theory, and the geometric Langlands correspondence I"}, [hep-th/1005.2846].
\bibitem{SL} C. Simpson, \emph{"Iterated destabilizing modifications for vector bundles with connection"}, [math.AG/0812.3472].
\bibitem{EG} C. Bergbauer, R. Brunetti, D. Kreimer, \emph{"Renormalization and resolution of singularities"}, [hep-th/0908.0633].
\bibitem{R} P. Z. Kobak, \emph{"Twistors, nilpotent orbits and harmonic maps"}.
\bibitem{Nil1} P. N. Achar, A. Henderson, B. F. Jones, {\it "Normality of Orbit Closures in the Enhanced
 Nilpotent Cone"}, [math.RT/1004.3822].
\bibitem{Nil2} S. Gukov, E. Witten, {\it "Rigid Surface Operators"}, [hep-th/0804.1561].
\bibitem{Bott} R. Bott, L. W. Tu, {\it "Differential Forms in Algebraic Topology"}, Springer-Verlag, New York (1982).                      
\bibitem{Mig} A. Migdal, {\it Ann. Phys.} {\bf 109} (1977) 365.
\bibitem{Malkin} A. Y.u. Alekseev, A. Z. Malkin, {\it "Symplectic structure of the moduli spaces of flat connections on a Riemann surface"}, \emph{Commun. Math. Phys.} {\bf169} (1995) 99, [hep-th/9312004].
\bibitem{VY} G. Veneziano, S. Yankielowicz, {\it Phys. Lett. }{\bf B 213} (1982) 231.
\bibitem{Landi} F. Lizzi , R. J. Szabo, A. Zampini, \emph{"Geometry of the Gauge Algebra in Noncommutative Yang-Mills Theory"},  {\it JHEP} {\bf 0108} (2001) 032, [hep-th/0107115].
\bibitem{Mass1} R. Boels, L. Mason, D. Skinner, \emph{"Supersymmetric Gauge Theories in Twistor Space"}, {\it JHEP} {\bf 0702} (2007) 014, [hep-th/0604040].
\bibitem{Mass2} L. Mason, D. Skinner, \emph{"The Complete Planar S-matrix of  $\mathcal{N}$ $=4$ $SYM$ as a Wilson Loop in Twistor Space"}, {\it JHEP} {\bf 1012} (2010) 018, [hep-th/1009.2225]. 
\bibitem{Mass3} M. Bullimore, D. Skinner, \emph{"Holomorphic Linking, Loop Equations and Scattering Amplitudes in Twistor Space"}, [hep-th/1101.1329]. 
\bibitem{Vafas1} A. Neitzke, C. Vafa, \emph{"N=2 strings and the twistorial Calabi-Yau"}, [hep- th/0402128].
\bibitem{Vafas2}R. Dijgraaf, S. Gukov, A. Neitzke, C. Vafa, \emph {Adv. Theor. Math. Phys. } {\bf 9} (2005) 603, [hep-th/0411073].
\bibitem{Wcs} E. Witten, {\it Commun. Math. Phys. }{\bf 118} (1988) 411.
\bibitem{Wcs1} E. Witten, {\it "Chern-Simons Gauge Theory As A String Theory"}, \emph{Prog. Math.} {\bf 133} (1995) 637, [hep-th/9207094].
\bibitem{MBs} M. Bochicchio, \emph{"The Yang-Mills String as the A-Model on the Twistor Space of the Complex Two-Dimensional Projective Space with Fluxes and Wilson Loops: the Beta Function"}, [hep-th/0811.2547].
\end{thebibliography}
\end{document}